\documentclass[12pt]{article}
\usepackage{amsmath,amssymb,amsthm,amsxtra,overpic,bbm,bm,epsfig,subfigure}
\usepackage{hyperref}
\usepackage{mathrsfs}
\usepackage{graphicx}
\usepackage{multirow}
\usepackage{color}
\usepackage{comment}
\usepackage{epstopdf}
\usepackage{float}
\numberwithin{equation}{section}
\usepackage{cite}
\usepackage{hyperref}
\usepackage{url}
\usepackage{slashed,stmaryrd}

\def\thefootnote{\fnsymbol{footnote}}
\usepackage{textcomp}

\begin{document}

\vspace{0.2cm}
\begin{center}
{\Large\bf Flavor Invariants and Renormalization-group Equations in the Leptonic Sector with Massive Majorana Neutrinos}
\end{center}
\vspace{0.2cm}

\begin{center}
{\bf Yilin Wang}~$^{a,~b}$~\footnote{E-mail: wangyilin@ihep.ac.cn},
\quad
{\bf Bingrong Yu}~$^{a,~b}$~\footnote{E-mail: yubr@ihep.ac.cn (corresponding author)},
\quad
{\bf Shun Zhou}~$^{a,~b}$~\footnote{E-mail: zhoush@ihep.ac.cn (corresponding author)}
\\
\vspace{0.2cm}
{\small
$^a$Institute of High Energy Physics, Chinese Academy of Sciences, Beijing 100049, China\\
$^b$School of Physical Sciences, University of Chinese Academy of Sciences, Beijing 100049, China}
\end{center}

\vspace{1.5cm}

\begin{abstract}
In the present paper, we carry out a systematic study of the flavor invariants and their renormalization-group equations (RGEs) in the leptonic sector with three generations of charged leptons and massive Majorana neutrinos. First, following the approach of the Hilbert series from the invariant theory, we show that there are 34 basic flavor invariants in the generating set, among which 19 invariants are CP-even and the others are CP-odd. Any flavor invariants can be expressed as the polynomials of those 34 basic invariants in the generating set. Second, we explicitly construct all the basic invariants and derive their RGEs, which form a closed system of differential equations as they should. The numerical solutions to the RGEs of the basic flavor invariants have also been found. Furthermore, we demonstrate how to extract physical observables from the basic invariants. Our study is helpful for understanding the algebraic structure of flavor invariants in the leptonic sector, and also provides a novel way to explore leptonic flavor structures.
\end{abstract}
\newpage

\def\thefootnote{\arabic{footnote}}
\setcounter{footnote}{0}

\section{Introduction}
\label{sec:intro}

One of the most important open questions in particle physics is how to understand the flavor structures of fermions~\cite{Xing2020}. In the Standard Model (SM), although quarks and charged leptons acquire their masses via the Yukawa interactions with the Higgs field after the spontaneous gauge symmetry breaking, the hierarchical fermion mass spectra, quark flavor mixing pattern and CP violation remain to be understood since the Yukawa coupling matrices are essentially arbitrary.  Furthermore, neutrino oscillation experiments have firmly established that neutrinos are massive particles and lepton flavors are significantly mixed~\cite{PDG2020}. The flavor problem is further aggravated since even the mechanism of neutrino mass generation is currently unknown.

In a class of seesaw models~\cite{Xing2020}, massive neutrinos turn out to be Majorana particles, namely, they are their own antiparticles~\cite{Majorana1937, Racah1937}. At the low-energy scale, after the heavy degrees of freedom responsible for neutrino mass generation have been integrated out, the lepton mass spectra, flavor mixing and leptonic charged-current interaction are governed by the effective Lagrangian
\begin{eqnarray}
\label{eq:effective lagrangian}
{\cal L}^{}_{\rm lepton} = - \overline{l^{}_{\rm L}} M^{}_l l^{}_{\rm R} - \frac{1}{2} \overline{\nu^{}_{\rm L}} M^{}_\nu \nu^{\rm C}_{\rm L} + \frac{g}{\sqrt{2}} \overline{l^{}_{\rm L}} \gamma^\mu \nu^{}_{\rm L} W^-_\mu + {\rm h.c.} \; ,
\end{eqnarray}
where $\nu_{\rm L}^{\rm C} \equiv {\cal C}\overline{\nu_{\rm L}^{}}_{}^{\rm T}$ has been defined with ${\cal C}\equiv {\rm i} \gamma^2 \gamma^0$ being the charge-conjugation matrix, while $M_l^{}$ and $M_\nu^{}$ denote the charged-lepton mass matrix and the Majorana neutrino mass matrix, respectively. One can diagonalize the lepton mass matrices via $V_l^{\dagger}M_l^{}V_l^\prime={\rm Diag}\left\{m_e^{},m_\mu^{},m_\tau^{}\right\}$ and $V_{\nu}^{\dagger}M_{\nu}V_{\nu}^{*}={\rm Diag}\left\{m_1^{},m_2^{},m_3^{}\right\}$, where $V_l^{}$, $V_l^{\prime}$ and $V_\nu^{}$ are $3 \times 3$ unitary matrices. In the mass basis, the leptonic flavor mixing matrix, or the Pontecorvo-Maki-Nakagawa-Sakata (PMNS) matrix~\cite{Pontecorvo1957, MNS1962}, is thus given by $V = V_l^{\dagger} V_\nu^{}$ and will appear in the charged-current interaction. It is evident that lepton masses and flavor mixing are
completely determined by the flavor structures of the lepton mass matrices $M^{}_l$ and $M^{}_\nu$.

As is well known, however, a different set of lepton mass matrices $M^{}_l$ and $M^{}_\nu$ may lead to the same physical observables, such as the lepton masses and the PMNS matrix. The reason is simply that physical observables should be independent of the basis transformations in the flavor space, whereas lepton mass matrices {\it do} depend on the basis. More explicitly, the leptonic Lagrangian in Eq.~(\ref{eq:effective lagrangian}) will be unchanged under the following unitary transformations in the flavor space
\begin{eqnarray}
\label{eq:flavor trans for field}
l_{\rm L}^{}\rightarrow l_{\rm L}^{\prime}=U_{\rm L}^{}l_{{\rm L}}^{}\;,\quad
\nu_{\rm L}^{}\rightarrow \nu_{\rm L}^{\prime}=U_{\rm L}^{}\nu_{{\rm L}}^{}\;,\quad
l_{\rm R}^{}\rightarrow l_{\rm R}^{\prime}=U_{\rm R}^{}l_{{\rm R}}^{}\;,
\end{eqnarray}
where $U_{{\rm L}}^{}$ and $U_{{\rm R}}^{}$ are two arbitrary elements in the $N$-dimensional unitary group ${\rm U}(N)$ (e.g., $N=3$ in the SM), if the lepton mass matrices transform as
\begin{eqnarray}
\label{eq:flavor trans for mass matrix}
M_{l}^{}\rightarrow M_{l}^{\prime}=U_{{\rm L}}^{}M_{l}^{}U_{{\rm R}}^{\dagger}\;,\qquad
M_{\nu}^{}\rightarrow M_{\nu}^{\prime}=U_{{\rm L}}^{}M_{\nu}^{}U_{{\rm L}}^{{\rm T}}\;.
\end{eqnarray}
One can immediately verify that the physical observables, including the lepton masses $\{m^{}_e, m^{}_\mu, m^{}_\tau\}$ and $\{m^{}_1, m^{}_2, m^{}_3\}$ and the flavor mixing parameters from the PMNS matrix $V$, are not affected by the flavor-basis transformations, whereas the flavor structures of lepton mass matrices obviously depend on the flavor basis. Consequently, it is very interesting to start with lepton mass matrices and construct the flavor-basis invariants, in which the unphysical degrees of freedom have been automatically removed.

The first flavor invariant has been constructed by Jarlskog in Refs.~\cite{Jarlskog:1985ht, Jarlskog:1985cw} in order to characterize the CP violation in the quark sector. The Jarlskog invariant is proportional to the determinant of the commutator of up- and down-type quark mass matrices and changes its sign under the CP transformation. Moreover, the flavor invariants under the joint flavor-basis and CP transformation have been systematically studied and used to derive the sufficient and necessary conditions for CP conservation in the quark or leptonic sector~\cite{Branco1986quark, Branco1986lepton, Yu2019PLB, Yu2020ICHEP, Yu2020PRD}. These invariants have also been implemented in the canonical seesaw models to make a direct connection between the CP violation at the low- and high-energy scales~\cite{Pilaftsis1997, Branco2001, CIP2006}. As the physical parameters at the low-energy scale are related to those at high-energy scales by their renormalization-group equations (RGEs), the RGEs of the flavor invariants provide an equivalent but basis-independent way to describe the running behaviors of physical parameters. The RGEs of all basic flavor invariants in the quark sector have been calculated in the Refs.~\cite{FMS2015, Talbert:2021iqn}. In the previous work~\cite{Yu2020PRD}, where the sufficient and necessary conditions for CP conservation in the leptonic sector with massive Majorana neutrinos have been investigated, two of the authors have also derived the RGEs of several CP-odd flavor invariants that are responsible for leptonic CP violation.

In this paper, we extend the previous works and perform a systematic study on the flavor invariants and their RGEs in the leptonic sector with massive Majorana neutrinos. The motivation for such an investigation is two-fold. First, as already demonstrated in Refs.~\cite{Branco1986quark, Branco1986lepton}, it is possible to construct an infinite number of flavor invariants but not all of them are independent. Strictly speaking, one has to find out all the {\it basic} flavor invariants. According to the classical invariant theory~\cite{Sturmfels2008, DK2015}, the other flavor invariants can be expressed as the {\it polynomials} of the basic ones. In the case where there are three generations of massive Majorana neutrinos, it is still unclear from the previous works how many basic invariants exist. In fact, we observe that there are totally 34 basic flavor invariants, among which 19 invariants are CP-even and the others are CP-odd. Second, after explicitly constructing the basic invariants, we derive their RGEs and show that they form a closed set of differential equations. In the conventional approach, a specific parametrization of the PMNS matrix is adopted and the RGEs of lepton masses, flavor mixing angles and CP-violating phases are utilized to examine the running effects. In terms of flavor invariants, one can achieve such a goal in a basis- and parametrization-independent manner. Moreover, the relationship between the primary flavor invariants and the physical observables will be established.

The remaining part of this paper is organized as follows. First, in Sec.~\ref{sec:flavor invariants}, we briefly recall the conventional approach to the construction of flavor invariants and the derivation of their RGEs. Then, the method of Hilbert series in the classical invariant theory is implemented to find out the basic invariants in the case of two-generation charged leptons and Majorana neutrinos in Sec.~\ref{sec:2g}. The realistic case of three-generation leptons is studied in Sec.~\ref{sec:3g}, where the complete set of RGEs for the basic flavor invariants are derived and numerically solved. We summarize our main results and conclude in Sec.~\ref{sec:summary}. Finally, we present a number of important mathematical theorems and approaches in three appendices. The famous Cayley-Hamilton theorem is collected in Appendix~\ref{appendix:CH theorem}, and a pedagogical introduction to the invariant theory and the method of Hilbert series is given in Appendix~\ref{appendix:Invariant theory}. In Appendix~\ref{appendix:syzygy}, the practically useful strategy implemented in the present work to decompose the flavor invariants into the basic ones is explained in detail, and it also offers an efficient way to derive the syzygies, which are polynomial identities among the flavor invariants.

\section{Flavor Invariants and Their RGEs}
\label{sec:flavor invariants}

First of all, we explain in this section what flavor invariants are and how to construct them based on the lepton mass matrices. Recalling the transformation rules for lepton mass matrices $M^{}_l$ and $M^{}_\nu$ under flavor-basis transformations in Eq.~(\ref{eq:flavor trans for mass matrix}), one can introduce the following matrices
\begin{eqnarray}
H_l^{}\equiv M_l^{}M_l^{\dagger}\;,\quad
H_\nu^{}\equiv M_\nu^{}M_\nu^{\dagger}\;,\quad
G_{l\nu}^{}\equiv M_\nu^{}H_l^* M_\nu^{\dagger}\;,\quad
G_{l\nu}^{(n)}\equiv M_\nu^{}(H_l^*)_{}^n M_\nu^{\dagger}\;,
\end{eqnarray}
and prove that they actually transform as the adjoint representation of the $N$-dimensional unitary group ${\rm U}(N)$, i.e.,
\begin{eqnarray}
H_{l}^{}\rightarrow U_{{\rm L}}^{}H_{l}^{}U_{{\rm L}}^{\dagger}\;,\quad
H_{\nu}^{}\rightarrow U_{{\rm L}}^{}H_{\nu}^{}U_{{\rm L}}^{{\dagger}}\;,\quad
G_{l\nu}^{}\rightarrow U_{{\rm L}}^{}G_{l\nu}^{}U_{{\rm L}}^{{\dagger}}\;,\quad
G_{l\nu}^{(n)}\rightarrow U_{{\rm L}}^{}G_{l\nu}^{(n)}U_{{\rm L}}^{{\dagger}}\;,
\end{eqnarray}
where $N$ refers to the number of lepton generations and $n \geq 2$ is a positive integer. Therefore, these matrices can serve as the ``building blocks" for constructing flavor invariants, and one can immediately write down a series of flavor invariants
\begin{eqnarray}
\label{eq:general form of invariants}
I^{abcd\cdots}_{rstu\cdots}\equiv {\rm Tr}\left\{H_l^a H_{\nu}^b G_{l\nu}^c \left[G_{l\nu}^{(n)}\right]^d H_l^r H_{\nu}^s G_{l\nu}^t \left[G_{l\nu}^{(n^{\prime})}\right]^u \cdots\right\}\;,
\end{eqnarray}
where the non-negative integers $\left\{a,b,c,d,r,s,t,u\right\}$ stand for the power indices of the corresponding matrices and the ellipses $``\cdots"$ denote the additional matrices composed of $H_l^{}$, $H_{\nu}^{}$, $G_{l\nu}^{}$ and $G_{l\nu}^{(n)}$. Thanks to the powerful Cayley-Hamilton (CH) theorem (cf. the description in Appendix \ref{appendix:CH theorem}), in the case of $N$-generation leptons, $G_{l\nu}^{(n)}$ with $n\geq N$ are no longer independent and can be expressed as a linear combination of the matrices $G_{l\nu}^{(n)}$ with $n<N$, guaranteeing a finite number of building blocks in the trace. For the same reason, the power index of each building block in Eq.~ (\ref{eq:general form of invariants}) must be smaller than the number of generations $N$, otherwise it would be reduced to those with smaller power indices by using Eq.~(\ref{eq:CH therorem general}). The above two observations lead us to the conclusion that there exist a finite number of basic flavor invariants in the generating set and the other invariants can be generated as the polynomials of the former. This conclusion is a simple and direct consequence of the general mathematical theorem in the invariant theory for the reductive group ${\rm U}(N)$.\footnote{A classical theorem in the invariant theory states that for the reductive groups, the polynomial ring constructed from the invariants under the group transformations has a finite dimension and can be generated from a finite number of basic invariants. See, e.g., Appendix~\ref{appendix:Invariant theory}, for an explanation of the relevant concepts.}

One may wonder whether the CH theorem is the unique tool that one can make use of to decompose the invariants of higher degrees into those of lower degrees. As proved in the seminal works by Processi~\cite{Processi76} and Formanek~\cite{Formanek84}, the first fundamental theorem for the invariants of $N\times N$ matrices $A^{}_i$ (for $i = 1, 2, \cdots, k$ with $k$ being a positive integer) under the ${\rm U}(N)$ group action $A^{}_i \to U^{}_{\rm L} A^{}_i U^\dagger_{\rm L}$ states that the polynomial invariant of $A^{}_i$ (for $i = 1, 2, \cdots, k$) is a polynomial of ${\rm Tr}\left[A^{}_{i_1} A^{}_{i_2} \cdots A^{}_{i_j}\right]$, where $A^{}_{i_1} A^{}_{i_2} \cdots A^{}_{i_j}$ run over all possible non-commutative monomials. Moreover, it has been found in Ref.~\cite{Processi76} that any relation among the invariants and the matrix concomitants is a consequence of the CH theorem. Hence in the subsequent discussions, we shall utilize frequently the CH theorem to reduce the flavor invariants to those in the generating set.

Then, we proceed with the RGEs of flavor invariants. As already mentioned in Sec.~\ref{sec:intro}, the RGEs of flavor invariants are helpful for investigating physical parameters running between different energy scales. Another crucial purpose for the study of RGEs is to cross-check the completeness of basic flavor invariants in the generating set. Since the derivative of any flavor invariant with respect to the energy scale must also be a flavor invariant, we shall be able to recast it into the polynomial of the basic flavor invariants. The RGEs of flavor invariants can be deduced from those of the building blocks, which are composed of lepton mass matrices by definition. At the one-loop level, the evolution of the effective Majorana neutrino mass matrix $M_{\nu}^{}$ and the charged-lepton mass matrix $M_l^{}$ are governed by~\cite{Xing2020,Chankowski:1993tx, Babu:1993qv, Antusch:2001ck, XZ2011, Ohlsson:2013xva}
\begin{eqnarray}
\frac{{\rm d} M^{}_{\nu}}{{\rm d}t} & = & \alpha^{}_{\nu} M^{}_{\nu} -\frac{3}{2} \left[ \left(Y^{}_l Y_l^{\dagger} \right) M^{}_{\nu} + M^{}_{\nu} \left( Y^{}_l Y_l^{\dagger} \right)^{\rm T} \right] \; , \label{eq:RGEMnu}\\
\frac{{\rm d}M^{}_l}{{\rm d}t} & = & \alpha^{}_l M^{}_l + \frac{3}{2} \left( Y^{}_l Y_l^{\dagger} \right) M^{}_l \; , \label{eq:RGEMl}
\end{eqnarray}
where $t\equiv {\rm ln}(\mu/\mu_0^{})/(16 \pi^2)$ has been defined with $\mu$ being the renormalization scale and $\mu_0^{}$ being the initial energy scale. In the SM framework, the relevant coefficients are
\begin{eqnarray}
\alpha^{}_{\nu} &=& -3g_2^2 + 4\lambda + 2{\rm Tr} \left[3\left(Y_{\rm u}^{} Y_{\rm u}^\dagger\right) + 3\left(Y_{\rm d}^{} Y_{\rm d}^\dagger\right) +\left(Y_l^{} Y_l^\dagger\right)\right]\;,\\
 \alpha_l^{} &=& -\frac{9}{4} g_1^2 -\frac{9}{4}g_2^2 + {\rm Tr} \left[3\left(Y_{\rm u}^{} Y_{\rm u}^\dagger\right) + 3\left(Y_{\rm d}^{} Y_{\rm d}^\dagger\right) +\left(Y_l^{} Y_l^\dagger\right)\right] \; ,
\end{eqnarray}
where $g^{}_1$ and $g^{}_2$ stand for the SM gauge couplings, $Y_{\rm u}$, $Y_{\rm d}$ and $Y_l^{}$ respectively for the Yukawa coupling matrices of up-type quarks, down-type quarks and charged-leptons, and $\lambda$ for the quartic Higgs coupling~\cite{Xing2020}. Starting with Eqs.~(\ref{eq:RGEMnu})-(\ref{eq:RGEMl}) and recalling the definitions of $H^{}_l \equiv M^{}_l M^\dagger_l$, $H^{}_\nu \equiv M^{}_\nu M^\dagger_\nu$ and $G^{}_{l\nu} \equiv M^{}_\nu H^{*}_l M^\dagger_\nu$, one easily obtains
\begin{eqnarray}
\frac{{\rm d} H^{}_l}{{\rm d}t} & = & 2\alpha^{}_l H^{}_l + 6H_l^2 /v_{}^2\; ,
\label{eq:RGEHl} \\
\frac{{\rm d} H^{}_{\nu}}{{\rm d}t} & = & 2 \alpha_{\nu} H^{}_{\nu} -3\left( \left\{H_l^{},H_\nu^{}\right\} +2 G^{}_{l\nu}\right)/v_{}^2 \; , \label{eq:RGEHnu} \\
\frac{{\rm d}G^{}_{l\nu}}{{\rm d}t} & = & 2(\alpha^{}_{\nu} + \alpha^{}_{l}) G^{}_{l\nu} -3\left\{H_l^{},G_{l\nu}^{}\right\}/v_{}^2\; , \label{eq:RGEGlnu}
\end{eqnarray}
where the relation $Y^{}_l = \sqrt{2} M^{}_l/v$ has been used with $v \approx 246~{\rm GeV}$ being the vacuum expectation value of the Higgs field and the anti-commutator of two matrices $\left\{A,B\right\}\equiv AB+BA$ has been defined. With the help of Eqs.~(\ref{eq:RGEHl})-(\ref{eq:RGEGlnu}), for any given flavor invariants, it is then straightforward to calculate their RGEs. Notice that the RGEs of $G^{(n)}_{l\nu}$ for $n\geq 2$ can be derived in a similar manner to that of $G^{}_{l\nu}$, as they are all built upon $M^{}_\nu$ and $H^*_l$.

Though the flavor invariants have been studied for a long time, it is only until recently~\cite{JM2009} realized that the ring of flavor invariants is finitely generated. The main task is to make clear the number of basic invariants and the syzygies among them, and find  out the way to construct the basic invariants and syzygies explicitly. In the subsequent two sections, we concentrate on the flavor invariants and their RGEs in the cases of two- and three-generation leptons, respectively.

\section{Leptonic Flavor Invariants: Two Generations}
\label{sec:2g}

In this section, we consider the case of two-generation leptons. Although this case is unrealistic, it is less trivial than the toy model discussed in Appendix~\ref{appendix:Invariant theory}, whose invariant ring is simply free. In contrast, the corresponding invariant ring for two-generation leptons is a complete intersection (but not free). Therefore, we can take it as an excellent example to illustrate how to read off basic flavor invariants and syzygies from the plethystic logarithm (PL). After calculating the Hilbert series (HS) and the PL in the case of two-generation leptons by using the Molien-Weyl (MW) formula,\footnote{The readers who are unfamiliar with the invariant theory are encouraged to first look into Appendix~\ref{subapp:MW formula} for a brief introduction to the MW formula and its simple application to the toy model of one-generation leptons.} we explicitly construct all the basic flavor invariants and the syzygies. Then comes the computation of the RGEs of all the basic flavor invariants.

\subsection{Hilbert Series}
\label{subsec:2g HS}
In the two-generation case with $N = 2$, the building blocks for flavor invariants transform as below
\begin{eqnarray}
H_l^{} \rightarrow U_{\rm L}^{} H_l^{} U_{\rm L}^{\dagger}\;,\quad
M_{\nu}^{} \rightarrow U_{\rm L}^{} M_{\nu}^{} U_{\rm L}^{\rm T}\;,\quad
U_{\rm L} \in {\rm U}(2)\;,
\end{eqnarray}
from which we can see that $H_l^{}$ belongs to the adjoint representation of ${\rm U}(2)$ while $M_{\nu}^{}$ to the rank-two symmetric tensor representation, i.e.,
\begin{eqnarray}
H_l^{}: {\bf 2} \otimes {\bf 2^*}\;,\quad
M_{\nu}^{}: \left({\bf 2} \otimes {\bf 2}\right)_{\rm s}\;,\quad
M_{\nu}^{\dagger}: \left({\bf 2^*} \otimes {\bf 2^*}\right)_{\rm s}\;,
\end{eqnarray}
where ${\bf 2}$ and ${\bf 2^*}$ denote respectively the fundamental and anti-fundamental representation of ${\rm U}(2)$ and the subscript ``s" refers to the symmetric part. As the character functions of ${\bf 2}$ and ${\bf 2^*}$ are  $z_1^{}+z_2^{}$ and $z_1^{-1}+z_2^{-1}$ respectively, we get the character functions of $H_l^{}$ and $M_{\nu}^{}$,
\begin{eqnarray}
\chi_{l}^{}(z_1^{},z_2^{})=\left(z_1^{}+z_2^{}\right)\left(z_1^{-1}+z_2^{-1}\right)\;,\quad
\chi_{\nu}^{}(z_1^{},z_2^{})=z_1^2+z_1^{}z_2^{}+z_2^2+z_1^{-2}+z_1^{-1}z_2^{-1}+z_2^{-2}\;,
\end{eqnarray}
where $z_i^{}$ (for $i=1,2$) are the coordinates on the maximum torus of ${\rm U}(2)$ (see Appendix~\ref{subapp:MW formula} for more details). By labeling the degree of $M_l^{}$ and $M_{\nu}^{}$ with $q^{}_{l}$ and $q_{\nu}^{}$ respectively, we find the plethystic exponential (PE), namely,
\begin{eqnarray}
{\rm PE}\left(z_1^{},z_2^{};q_l^2,q_{\nu}\right)
&=&{\rm PE}\left[\chi_{l}^{}(z_1^{},z_2^{})q_l^2+\chi_{\nu}^{}(z_1^{},z_2^{})q_{\nu}^{}\right]
={\rm exp}\left(\sum_{k=1}^{\infty}\frac{\chi_{l}^{}(z_1^{k},z_2^{k})q_l^{2k}+\chi_{\nu}^{}(z_1^{k},z_2^{k})q_{\nu}^{k}}{k} \right)\nonumber\\
&=&\left[\left(1-q_l^2\right)^2\left(1-z_2^{}z_1^{-1}q_l^2\right)\left(1-z_1^{}z_2^{-1}q_l^2\right)\left(1-q_{\nu}^{}z_1^2\right)\left(1-q_{\nu}^{}z_2^2\right)\left(1-q_{\nu}^{}z_1^{}z_2^{}\right) \right. \nonumber\\
&&\left.\times \left(1-q_{\nu}^{}z_1^{-2}\right)\left(1-q_{\nu}^{}z_2^{-2}\right)\left(1-q_{\nu}^{}z_1^{-1}z_2^{-1}\right) \right]^{-1}.
\end{eqnarray}

Substituting the ${\rm PE}$ and the Haar measure of ${\rm U}(2)$ group in Eq.~(\ref{eq:Haar measure U(2)}) into the MW formula in Eq.~(\ref{eq:Molien-Weyl formula}), we obtain the multi-graded HS
\begin{eqnarray}
{\cal H}\left(q_l^{},q_{\nu}^{}\right)
&=& \int \left[{\rm d}\mu \right]_{{\rm U}(2)}{\rm PE}\left(z_1^{},z_2^{};q_l^2,q_{\nu}^{}\right)\nonumber\\
&=&\frac{1}{\left(1-q_{l}^2\right)^2}\frac{1}{2\left(2\pi{\rm i}\right)^2}\oint_{\left|z_1\right|^{}=1}\frac{{\rm d}z_1^{}}{z_1^{}}\oint_{\left|z_2\right|^{}=1}\frac{{\rm d}z_2^{}}{z_2^{}}\left(2-\frac{z_1^{}}{z_2^{}}-\frac{z_2^{}}{z_1^{}}\right)\left[\left(1-z_2^{}z_1^{-1}q_l^2\right)\left(1-z_1^{}z_2^{-1}q_l^2\right) \right. ~~~ \nonumber\\
&& \left.\times \left(1-q_{\nu}^{}z_1^2\right)\left(1-q_{\nu}^{}z_2^2\right)\left(1-q_{\nu}^{}z_1^{}z_2^{}\right)\left(1-q_{\nu}^{}z_1^{-2}\right)\left(1-q_{\nu}^{}z_2^{-2}\right)\left(1-q_{\nu}^{}z_1^{-1}z_2^{-1}\right) \right]^{-1}\;. ~~~
\end{eqnarray}
After completing the contour integrals by virtue of the residue theorem, one arrives at
\begin{eqnarray}
\label{eq:HS multi-graded in 2 generations}
{\cal H}\left(q_l^{},q_{\nu}^{}\right)=\frac{1+q_l^4q_{\nu}^4}{\left(1-q_l^2\right)\left(1-q_l^4\right)\left(1-q_{\nu}^2\right)\left(1-q_{\nu}^4\right)\left(1-q_l^2q_{\nu}^2\right)\left(1-q_l^4q_{\nu}^2\right)}\;,
\end{eqnarray}
where $q_l^{}$ and $q_{\nu}^{}$ denote the degree of $M_l^{}$ and $M_{\nu}^{}$, respectively. Starting with the multi-graded HS in Eq.~(\ref{eq:HS multi-graded in 2 generations}), we can calculate the PL
\begin{eqnarray}
\label{eq:PL 2 generations}
{\rm PL}\left[{\mathcal H}\left(q_l^{},q_{\nu}^{}\right) \right]=q_1^2+q_{\nu}^2+q_l^4+q_l^2q_{\nu}^2+q_{\nu}^4+q_l^4 q_{\nu}^2+q_l^4q_{\nu}^4-q_l^8q_{\nu}^8\;,
\end{eqnarray}
and the ungraded HS
\begin{eqnarray}
\label{eq:HS ungraded in 2 generations}
{\mathscr H}\left(q\right)\equiv {\cal H}\left(q,q\right)=\frac{1+q^8}{\left(1-q^2_{}\right)^2\left(1-q^4_{}\right)^3\left(1-q^6_{}\right)}\;,
\end{eqnarray}
where the last identity has been obtained by identifying $q^{}_l = q^{}_\nu \equiv q$ in Eq.~(\ref{eq:HS multi-graded in 2 generations}). Some comments on the results in Eqs.~(\ref{eq:PL 2 generations}) and (\ref{eq:HS ungraded in 2 generations}) are in order.
\begin{itemize}
    \item From the denominator on the right-hand side of Eq.~(\ref{eq:HS ungraded in 2 generations}), we can observe that there are in total 6 algebraically-independent invariants (also called primary invariants), corresponding to 6 physical observables in the model (i.e., two charged-lepton masses, two neutrino masses, one flavor mixing angle and one CP-violating phase). Furthermore, the degrees of the primary invariants can be read off from the power indices of $q$ in the parentheses, while the number of invariants of the same degree is indicated by the power index of the corresponding parenthesis in the denominator. More explicitly, in our case, there are two primary invariants of degree 2, three of degree 4 and one of degree 6.

    \item On the other hand, Eq.~(\ref{eq:PL 2 generations}) shows that there are totally 7 invariants in the generating set and their degrees are the same as those primary ones, except for one invariant with the degree of $M_l^{}$ and $M_{\nu}^{}$ to be both four. This invariant, which is not algebraically-independent of the other six ones in the generating set, shares a syzygy with them at the degree of $(8,8)$. This is exactly what the last term with a minus sign in Eq.~(\ref{eq:PL 2 generations}) points to.
\end{itemize}
As we shall see shortly, these observations are confirmed by the explicit construction of all the basic invariants in the next subsection.

\subsection{Construction of Flavor Invariants}
\label{subsec:2g construction}

\begin{table}[t!]
\centering
\begin{tabular}{l|c|c|c}
\hline \hline
flavor invariants & $(q_l^{}, q_{\nu}^{})$ & $q_l^{}+q_{\nu}^{}$ & CP parity \\
\hline \hline
$J_{1}^{}\equiv {\rm Tr}\left(H_{l}^{}\right)$ & $(2,0)$ & 2 & + \\
\hline
$J_{2}^{}\equiv {\rm Tr}\left(H_{\nu}^{}\right)$ & $(0,2)$ & 2 & +\\
\hline
$J_{3}^{}\equiv {\rm Tr}\left(H_{l}^{2}\right)$ & $(4,0)$ & 4 &+\\
\hline
$J_{4}^{}\equiv {\rm Tr}\left(H_{l}^{}H_{\nu}^{}\right)$ & $(2,2)$&4 &+\\
\hline
$J_{5}^{}\equiv {\rm Tr}\left(H_{\nu}^{2}\right)$ & $(0,4)$ & 4 &+\\
\hline
$J_{6}^{}\equiv {\rm Tr}\left(H_{l}^{}G_{l\nu}^{}\right)$ & $(4,2)$ & 6 &+\\
\hline
$J_{7}^{-}\equiv {\rm Tr}\left(\left[H_{l}^{}, H_{\nu}^{}\right] G_{l\nu}^{}\right)$ & $(4,4)$ & 8 & $-$\\
\hline
\hline
\end{tabular}
\vspace{0.5cm}
\caption{Summary of the basic flavor invariants in the generating set along with their degrees and CP parities in the case of two-generation leptons, where $q_l^{}$ and $q_{\nu}^{}$ denote the degree of $M_l^{}$ and $M_{\nu}^{}$, respectively. Note that the commutator $\left[A,B\right]\equiv AB-BA$ of two matrices has been defined.}
\label{table:2 generations}
\end{table}
In the two-generation case, only $H_l^{}$, $H_{\nu}^{}$ and $G_{l\nu}^{}$ are the building blocks of flavor invariants and the power index of each building block is at most two. In addition, the degrees $(q_l^{},q_\nu^{})$ of each basic invariant can be easily read off from Eq.~(\ref{eq:PL 2 generations}). Thus all the basic invariants in the generating set can be explicitly constructed using these building blocks. The final results are summarized in Table~\ref{table:2 generations}, together with their degrees and CP parities.

It is worthwhile to mention that there is one unique CP-odd invariant $J^-_7\equiv{\rm Tr}\left(\left[H_{l}^{}, H_{\nu}^{}\right] G_{l\nu}^{}\right)$ in the generating set, whereas the corresponding CP-even invariant $J_7^{+}\equiv{\rm Tr}\left(\left\{H_{l}^{}, H_{\nu}^{}\right\} G_{l\nu}^{}\right)$ can actually be decomposed into the polynomial of the primary invariants $\left\{J_1^{}, J_2^{},..., J_6^{}\right\}$, i.e.,
\begin{eqnarray}
\label{eq:J7P decomposition}
J_7^+=-\frac{1}{2}\left(J_1^2 J_2^2-J_1^2J_5^{}-2J_2^{}J_6^{}-2J_4^2\right) \;.
\end{eqnarray}
This should be the case as expected. Additionally, $\left(J_7^-\right)^2$ can be written as the polynomial of $\left\{J_1^{}, J_2^{},..., J_6^{}\right\}$, which corresponds to a syzygy at the degree of $(8,8)$, as what the negative term in Eq.~(\ref{eq:PL 2 generations}) shows,\footnote{A systematic method to decompose an arbitrary invariant into the polynomials of basis invariants and to find all the syzygies at a certain degree can be found in Appendix~\ref{appendix:syzygy}.}
\begin{eqnarray}
\label{eq:syzygy 2g}
\left(J_7^-\right)^2&=&\frac{1}{2}J_2^2J_3^{}\left(J_2^2 J_3^{}-3J_3^{}J_5^{}+4J_4^2\right)+\left(J_3^{}J_5^{}-J_4^2\right)^2-J_6^{}\left(2J_2^{}J_4^2+J_2^2J_6^{}-2J_5^{}J_6^{}\right)-\frac{1}{4}J_1^{}\nonumber\\
&\times&\left(J_2^2-J_5^{}\right) \left[J_1^3\left(J_2^2+J_5^{}\right)+8J_4^{}\left(J_2^{}J_3^{}-2J_6^{}-J_1^2J_2^{}\right)+4J_1^{}\left(J_2^{}J_6^{}-J_3^{}J_5^{}+3J_4^2\right)\right]\;.\nonumber\\
\end{eqnarray}
Thus far we have explicitly constructed all 7 basic invariants, as shown in Table~\ref{table:2 generations}, and the primary ones are $\{J^{}_1, J^{}_2, \cdots, J^{}_6\}$.

Finally, we demonstrate that the basic invariants can be expressed in terms of the physical observables. In the basis where the mass matrix of charged-leptons is real and diagonal, we have
\begin{eqnarray}
M_l^{}={\rm Diag}\left\{m_e^{}, m_{\mu}^{}\right\}\;, \qquad
M_{\nu}^{}=V \cdot {\rm Diag}\left\{m_1^{},m_2^{}\right\} \cdot V^{\rm T}\;,
\end{eqnarray}
where the PMNS matrix $V$ can be parametrized as
\begin{eqnarray}
V=\left(
\begin{matrix}
\cos\theta&\sin\theta \cr
-\sin\theta&\cos\theta
\end{matrix}
\right)\cdot
\left(
\begin{matrix}
e^{{\rm i}\varphi}&0\cr
0&1\\
\end{matrix}
\right) \; ,
\end{eqnarray}
with $\theta$ being the flavor mixing angle and $\varphi$ being the Majorana-type CP phase. Adopting this standard parametrization, one can write down the explicit forms of flavor invariants in terms of all the six physical parameters, namely,
\begin{eqnarray}
\label{eq:parameterization of invariants in two generations}
J_1^{}&=&m_e^2+m_{\mu}^2\;,\quad J_2^{}=m_1^2+m_2^2\;,\quad J_3^{}=m_e^4+m_{\mu}^4\;,\nonumber \\
J_4^{}&=&(m_1^2 m_e^2+m_2^2 m_{\mu}^2)\cos^2_{}\theta+\left(m_2^2 m_e^2+m_1^2 m_{\mu}^2\right)\sin^2_{}\theta\;, \quad J_5^{}=m_1^4+m_2^4\;,\nonumber \\
J_6^{}&=& \left(m_1^2 m_e^4+m_2^2 m_{\mu}^4\right)\cos^4_{}\theta+\left(m_2^2 m_e^4+m_1^2 m_{\mu}^4\right)\sin^4_{}\theta \nonumber \\
&&+\frac{1}{2}\left[\left(m_1^2+m_2^2\right)m_e^2 m_{\mu}^2+m_1^{}m_2^{}\left(m_{\mu}^2-m_e^2\right)^2 \cos 2\varphi \right]\sin^2_{}2\theta \;,\nonumber\\
J_7^{-}&=& -\frac{{\rm i}}{2}m_1^{}m_2^{}\left(m_2^2-m_1^2\right)\left(m_{\mu}^2-m_e^2\right)^2\sin^2_{} 2\theta \sin 2\varphi \;.
\end{eqnarray}
On the other hand, the physical observables, which by definition are the directly measurable quantities in experiments, should be independent of the flavor basis transformation. So it is useful to express them completely in terms of the flavor invariants. In the two-generation case, all the physical parameters can be easily extracted from flavor invariants using Eq.~(\ref{eq:parameterization of invariants in two generations}),
\begin{eqnarray}
m_{e,\mu}^2&=&\frac{1}{2}\left(J_1^{}\mp \sqrt{2J_3^{}-J_1^2}\right)\;,\nonumber \\
m_{1,2}^2&=&\frac{1}{2}\left(J_2^{}\mp\sqrt{2J_5^{}-J_2^2}\right)\;,\nonumber\\
\cos 2\theta &=& \frac{2J_4^{}-J_1^{}J_2^{}}{\sqrt{2J_3^{}-J_1^2}\sqrt{2J_5^{}-J_2^2}}\;,\nonumber\\
\cos 2\varphi &=&\frac{\left(J_1^2J_2^{}-4J_1^{}J_4^{}+2J_6^{}\right)\left(J_2^2-J_5^{}\right)+2\left(J_2^{}J_4^2-J_5^{}J_6^{}\right)}{\sqrt{2}\sqrt{J_2^2-J_5^{}}\left[J_3^{}\left(J_2^2-J_5^{}\right)+J_5^{}\left(J_1^2-J_3^{} \right) -2J_4^{}\left(J_1^{} J_2^{}-J_4^{}\right)\right]} \;,
\end{eqnarray}
where the upper and lower signs in the first (second) identity refer respectively to $m^{}_e$ and $m^{}_\mu$ ($m^{}_1$ and $m^{}_2$). In the hierarchical limit $m_{\mu}^2 \gg m_e^2$, one obtains $m_e^{2}\approx (J_1^2-J_3^{})/(2\sqrt{J_3^{}})$ and $m_{\mu}^{2}\approx \sqrt{J_3^{}}$.

\subsection{RGEs of Flavor Invariants}
\label{subsec:2g RGE}
Starting from the RGEs of the building blocks in Eqs.~(\ref{eq:RGEHl})-(\ref{eq:RGEGlnu}), one can derive the RGEs of all the basic flavor invariants in the generating set
\begin{eqnarray}
\label{eq:J1}
\frac{{\rm d} J_1^{}}{{\rm d}t}&=&2\alpha_{l}^{}J_{1}^{}+6J_{3}^{}/v_{}^2\;,\\
\label{eq:J2}
\frac{{\rm d} J_2^{}}{{\rm d}t}&=&2\alpha_{\nu}^{}J_{2}^{}-12J_{4}^{}/v_{}^2\;,\\
\label{eq:J3}
\frac{{\rm d} J_3^{}}{{\rm d}t}&=&4\alpha_{l}^{}J_{3}^{}+6J_{1}^{}\left(3J_3^{}-J_1^{2}\right)/v_{}^2\;,\\
\label{eq:J4}
\frac{{\rm d} J_4^{}}{{\rm d}t}&=&2\left(\alpha_{l}^{}+\alpha_{\nu}^{}\right)J_{4}^{}-6J_{6}^{}/v_{}^2\;,\\
\label{eq:J5}
\frac{{\rm d} J_5^{}}{{\rm d}t}&=&4\alpha_{\nu}^{}J_{5}^{}-12\left[2J_{2}^{}J_{4}^{}-J_{1}^{}\left(J_{2}^2-J_5^{}\right)\right]/v_{}^2\;,\\
\label{eq:J6}
\frac{{\rm d} J_6^{}}{{\rm d}t}&=&2\left(2\alpha_{l}^{}+\alpha_{\nu}^{}\right)J_{6}^{}\;,\\
\label{eq:J7}
\frac{{\rm d} J_7^{-}}{{\rm d}t}&=&4\left(\alpha_{l}^{}+\alpha_{\nu}^{}\right)J_{7}^{-}\;,
\end{eqnarray}
which form a closed system of differential equations, implying the completeness of the generating set. Moreover, notice that the RGEs of all the CP-even flavor invariants are by themselves closed, while the derivative of the unique CP-odd flavor invariant $J_7^{-}$ is proportional to itself. This can be understood by carrying out the CP transformation on both sides and noting the fact that there is only one CP phase in the two-generation case.

The RGEs in Eqs.~(\ref{eq:J1})-(\ref{eq:J7}) are in general difficult to solve analytically, except for those in the last two equations. However, if we neglect the second terms on the right-hand sides of Eqs.~(\ref{eq:J1})-(\ref{eq:J5}) that are actually suppressed by the small ratios $m^2_i/v^2$ (for $i = 1, 2$) and $m^2_\alpha/v^2$ (for $\alpha = e, \mu$) as compared to the first terms, then the approximate analytical solutions turn out to be very simple, viz.
\begin{eqnarray}
J_1^{}(t)&\approx& J_1^{}(0)\exp\left\{2\int^t_0  \alpha^{}_l(t^\prime) {\rm d}t^\prime \right\} \; ,\nonumber\\
J_2^{}(t)&\approx& J_2^{}(0)\exp\left\{2\int^t_0  \alpha^{}_\nu(t^\prime)  {\rm d}t^\prime \right\} \; ,\nonumber\\
J_3^{}(t)&\approx& J_3^{}(0)\exp\left\{4\int^t_0  \alpha^{}_l(t^\prime) {\rm d}t^\prime \right\} \; ,\nonumber\\
J_4^{}(t)&\approx& J_4^{}(0)\exp\left\{2\int^t_0 \left[ \alpha^{}_l(t^\prime) + \alpha^{}_\nu(t^\prime)\right] {\rm d}t^\prime \right\} \; ,\nonumber\\
J_5^{}(t)&\approx& J_5^{}(0)\exp\left\{4\int^t_0  \alpha^{}_\nu(t^\prime)  {\rm d}t^\prime \right\} \; ,\nonumber\\
J_6^{}(t)&=& J_6^{}(0)\exp\left\{2\int^t_0 \left[ 2\alpha^{}_l(t^\prime) + \alpha^{}_\nu(t^\prime)\right] {\rm d}t^\prime \right\} \; ,\nonumber\\
J_7^{-}(t)&=& J_7^{-}(0)\exp\left\{4\int^t_0 \left[ \alpha^{}_l(t^\prime) + \alpha^{}_\nu(t^\prime)\right] {\rm d}t^\prime \right\} \; .
\end{eqnarray}
These approximate solutions are actually trivial in the sense that no flavor mixing is assumed and the running effects of the flavor mixing angle $\theta$ and the CP phase $\varphi$ are entirely ignored.

\section{Leptonic Flavor Invariants: Three Generations}
\label{sec:3g}
Now we consider the case of three-generation leptons, where the algebraic structure of the invariant ring is much more complicated than that in the former case. The main reason is that the invariant ring is a non-complete intersection for the three-generation case, in contrast to the free ring for the one-generation case and the complete intersection ring for the two-generation case. The number of basic invariants grows very quickly with the number of generations. Moreover, the existence of the Majorana neutrino mass matrix further complicates the situation.\footnote{In the quark sector, where all the building blocks
reside in the adjoint representation of ${\rm U}(N)$, the invariant ring is still a complete intersection even in the three-generation case. In the leptonic sector, the mass matrix of charged leptons transforms similarly to those of quarks, while the Majorana neutrino mass matrix transforms as the rank-2 symmetric tensor representation. These very different transformation features of the building blocks (i.e., the charged-lepton mass matrix and the Majorana neutrino mass matrix) lead to a much more complicated algebraic structure of the invariant ring in the leptonic sector.}

In the remaining parts of this section, we first compute the HS and PL by using the MW formula, and explain their main features. The explicit construction of all the basic flavor invariants is then carried out, and a brief comparison with the results in Ref.~\cite{JM2009} is made. It is important to notice that the explicit construction is helpful for us to understand the algebraic structure of the invariant ring. Then, we show that all the physical observables can be analytically extracted from basic flavor invariants, which provides a basis-independent way to describe the running behaviors of physical observables. Finally we calculate the RGEs of all the basic flavor invariants and observe that they form a closed system of differential equations. The numerical solutions to the RGEs of the basic invariants are also given.

\subsection{Hilbert Series}
\label{subsec:3g HS}
In the three-generation case with $N = 3$, the building blocks for constructing flavor invariants transform as
\begin{eqnarray}
H_l^{} \rightarrow U_{\rm L}^{} H_l^{} U_{\rm L}^{\dagger}\;,\quad
M_{\nu}^{} \rightarrow U_{\rm L}^{} M_{\nu}^{} U_{\rm L}^{\rm T}\;,\quad
U_{\rm L} \in {\rm U}(3)\;,
\end{eqnarray}
from which we can see that $H_l^{}$ belongs to the adjoint representation of ${\rm U}(3)$ while $M_{\nu}^{}$ to the rank-2 symmetric tensor representation, i.e.,
\begin{eqnarray}
H_l^{}: {\bf 3} \otimes {\bf 3^*}\;,\quad
M_{\nu}^{}: \left({\bf 3} \otimes {\bf 3}\right)_{\rm s}\;,\quad
M_{\nu}^{\dagger}: \left({\bf 3^*} \otimes {\bf 3^*}\right)_{\rm s}\;,
\end{eqnarray}
where ${\bf 3}$ and ${\bf 3^*}$ denote respectively the fundamental and anti-fundamental representation of ${\rm U}(3)$, and the subscript ``s" refers to the symmetric part. With the help of the character functions $z_1^{}+z_2^{}+z_3^{}$ and $z_1^{-1}+z_2^{-1}+z_3^{-1}$ for ${\bf 3}$ and ${\bf 3^*}$, we can derive the character functions of $H_l^{}$ and $M_{\nu}^{}$,
\begin{eqnarray}
\chi_{l}^{}(z_1^{},z_2^{},z_3^{})&=&\left(z_1^{}+z_2^{}+z_3^{}\right)\left(z_1^{-1}+z_2^{-1}+z_3^{-1}\right)\;,\nonumber\\
\chi_{\nu}^{}(z_1^{},z_2^{},z_3^{})&=&z_1^2+z_2^2+z_3^2+z_1^{}z_2^{}+z_1^{}z_3^{}+z_2^{}z_3^{}\nonumber\\
&~& + z_1^{-2}+z_2^{-2}+z_3^{-2}+z_1^{-1}z_2^{-1}+z_1^{-1}z_3^{-1}+z_2^{-1}z_3^{-1}\;,
\end{eqnarray}
where $z_i^{}$ (for $i=1,2,3$) are the coordinates on the maximum torus of ${\rm U}(3)$. By labeling the degree of $M_l^{}$ and $M_{\nu}^{}$ as $q^{}_{l}$ and $q_{\nu}^{}$ respectively, one can find
\begin{eqnarray}
{\rm PE}\left(z_1^{},z_2^{},z_3^{};q_l^2,q_{\nu}\right)
&=&{\rm PE}\left[\chi_{l}^{}(z_1^{},z_2^{},z_3^{})q_l^2+\chi_{\nu}^{}(z_1^{},z_2^{},z_3^{})q_{\nu}^{}\right] \nonumber\\
&=&{\rm exp}\left(\sum_{k=1}^{\infty}\frac{\chi_{l}^{}(z_1^{k},z_2^{k},z_3^{k})q_l^{2k}+\chi_{\nu}^{}(z_1^{k},z_2^{k},z_3^{k})q_{\nu}^{k}}{k} \right)\nonumber\\
&=&\left[\left(1-q_l^2\right)^3\left(1-q_l^2z_2^{}z_1^{-1}\right)\left(1-q_l^2z_1^{}z_2^{-1}\right)\left(1-q_l^2z_3^{}z_1^{-1}\right)\left(1-q_l^2z_1^{}z_3^{-1}\right)\right.\nonumber\\
&&\left.\times \left(1-q_l^2z_2^{}z_3^{-1}\right)\left(1-q_l^2z_3^{}z_2^{-1}\right)\left(1-q_{\nu}^{}z_1^2\right)\left(1-q_{\nu}^{}z_2^2\right)\left(1-q_{\nu}^{}z_3^2\right)\right.\nonumber\\
&& \left.\times \left(1-q_{\nu}^{}z_1^{}z_2^{}\right)\left(1-q_{\nu}^{}z_1^{}z_3^{}\right)\left(1-q_{\nu}^{}z_2^{}z_3^{}\right)\left(1-q_{\nu}^{}z_1^{-2}\right)\left(1-q_{\nu}^{}z_2^{-2}\right)\right. \nonumber \\
&& \left.\times \left(1-q_{\nu}^{}z_3^{-2}\right)\left(1-q_{\nu}^{}z_1^{-1}z_2^{-1}\right)\left(1-q_{\nu}^{}z_1^{-1}z_3^{-1}\right)\left(1-q_{\nu}^{}z_2^{-1}z_3^{-1}\right)
\right]^{-1}\;.
\end{eqnarray}

Substituting the ${\rm PE}$ and the Haar measure of the ${\rm U}(3)$ group in Eq.~(\ref{eq:Haar measure U(3)}) into the MW formula in Eq.~(\ref{eq:Molien-Weyl formula}), we get the multi-graded HS, i.e.,
\begin{eqnarray}
{\cal H}\left(q_l^{},q_{\nu}^{}\right)
&=& \int \left[{\rm d}\mu \right]_{{\rm U}(3)}{\rm PE}\left(z_1^{},z_2^{},z_3^{};q_l^2,q_{\nu}^{}\right)\nonumber\\
&=& \frac{1}{6\left(2\pi {\rm i}\right)^3}\oint_{\left|z_1^{}\right|=1}\frac{{\rm d}z_1^{}}{z_1^{}}\oint_{\left|z_2^{}\right|=1}\frac{{\rm d}z_2^{}}{z_2^{}}\oint_{\left|z_3^{}\right|=1}\frac{{\rm d}z_3^{}}{z_3^{}}\nonumber\\
&& \times \left[-\frac{\left(z_2^{}-z_1^{}\right)^2\left(z_3^{}-z_1^{}\right)^2\left(z_3^{}-z_2^{}\right)^2}{z_1^2z_2^2z_3^2} \right]\times {\rm PE}\left(z_1^{},z_2^{},z_3^{},q_l^2,q_{\nu}\right)\;.
\end{eqnarray}
Applying the residue theorem to the contour integrals, one obtains
\begin{eqnarray}
\label{eq:HS multi-graded in 3 generations}
{\cal H}\left(q_l^{},q_{\nu}^{}\right)=\frac{{\cal N}\left(q_l^{},q_{\nu}^{}\right)}{{\cal D}\left(q_l^{},q_{\nu}^{}\right)}\;,
\end{eqnarray}
where
\begin{eqnarray*}
{\cal N}\left(q_l^{},q_{\nu}^{}\right)
&=&-q_l^{24} q_{\nu}^{18} - 2 q_l^{20} q_{\nu}^{14} - 2 q_l^{20} q_{\nu}^{12} - q_l^{20} q_{\nu}^{10} - 2 q_l^{18} q_{\nu}^{14} - 3 q_l^{18} q_{\nu}^{12} -  q_l^{18} q_{\nu}^{10} - 3 q_l^{16} q_{\nu}^{14}\nonumber\\
&&- 3 q_l^{16} q_{\nu}^{12} - 3 q_l^{16} q_{\nu}^{10}- q_l^{16} q_{\nu}^{8} - q_l^{16} q_{\nu}^{6} - q_l^{14} q_{\nu}^{14} - q_l^{14} q_{\nu}^{12} - q_l^{14} q_{\nu}^{10} -  2 q_l^{14} q_{\nu}^{8} - q_l^{14} q_{\nu}^{6}\nonumber\\
&&- q_l^{12} q_{\nu}^{14} +  q_l^{12} q_{\nu}^{4} + q_l^{10} q_{\nu}^{12}+ 2 q_l^{10} q_{\nu}^{10} + q_l^{10} q_{\nu}^{8} + q_l^{10} q_{\nu}^{6} + q_l^{10} q_{\nu}^{4} + q_l^{8} q_{\nu}^{12} + q_l^{8} q_{\nu}^{10}\nonumber\\
&&+ 3 q_l^{8} q_{\nu}^{8} + 3 q_l^{8} q_{\nu}^{6} + 3 q_l^{8} q_{\nu}^{4} + q_l^{6} q_{\nu}^{8} + 3 q_l^{6} q_{\nu}^{6} + 2 q_l^{6} q_{\nu}^{4} + q_l^{4} q_{\nu}^{8} + 2 q_l^{4} q_{\nu}^{6} + 2 q_l^{4} q_{\nu}^{4} + 1\;,\nonumber\\
{\cal D}\left(q_l^{},q_{\nu}^{}\right)
&=&\left(1 - q_l^2\right) \left(1 - q_l^4\right) \left(1 - q_l^6\right) \left(1 - q_{\nu}^2\right) \left(1 - q_{\nu}^4\right) \left(1 - q_{\nu}^6\right) \left(1 -q_l^2 q_{\nu}^2\right) \left(1 - q_l^4 q_{\nu}^2\right)^2\nonumber\\
&&\times \left(1 - q_l^2 q_{\nu}^4\right) \left(1 - q_l^6 q_{\nu}^2\right) \left(1 - q_l^4 q_{\nu}^4\right) \left(1 - q_l^8 q_{\nu}^2\right)\;.
\end{eqnarray*}
Given the multi-graded HS, we can calculate the PL
\begin{eqnarray}
\label{eq:PL 3 generations}
{\rm PL}\left[{\cal H}\left(q_l^{},q_{\nu}^{}\right) \right]&=&q_l^2 +q_{\nu}^2+q_l^4+q_l^2 q_{\nu}^2+q_{\nu}^4+q_l^6+2q_l^4 q_{\nu}^2+q_l^2 q_{\nu}^4+q_{\nu}^6+q_l^6 q_{\nu}^2+3q_l^4 q_{\nu}^4+q_l^8 q_{\nu}^2\nonumber \\
&&+2q_l^6 q_{\nu}^4
+2q_l^4 q_{\nu}^6+3q_l^8 q_{\nu}^4+3q_l^6 q_{\nu}^6+q_l^4 q_{\nu}^8+q_l^{10} q_{\nu}^4+3q_l^8 q_{\nu}^6+q_l^6 q_{\nu}^8+q_l^{12}q_{\nu}^4\nonumber\\
&&+q_l^{10}q_{\nu}^6-{\cal O}(\left[q_l^{}q_{\nu}^{}\right]^{18})\;,
\end{eqnarray}
and the ungraded HS
\begin{eqnarray}
\label{eq:HS ungraded in 3 generations}
{\mathscr H }\left(q\right)\equiv{\cal H}\left(q,q\right)=\frac{{\mathscr N}\left(q\right)}{{\mathscr D}\left(q\right)}\;,
\end{eqnarray}
where
\begin{eqnarray*}
{\mathscr N}\left(q\right)&=&1 + q^6 + 2 q^8 + 4 q^{10} + 8 q^{12} + 7 q^{14} + 9 q^{16} + 10 q^{18} +  9 q^{20} + 7 q^{22} + 8 q^{24} + 4 q^{26}\\&& + 2 q^{28} + q^{30} + q^{36}\;,\nonumber\\
{\mathscr D}\left(q\right)&=& (1 - q^2)^2 (1 - q^4)^3 (1 - q^6)^4 (1 - q^8)^2 (1 - q^{10})\;.
\end{eqnarray*}

Here we give some helpful comments on the multi-graded HS in Eq.~(\ref{eq:HS multi-graded in 2 generations}) and that in Eq.~(\ref{eq:HS multi-graded in 3 generations}), and on the results for two- and three-generation cases obtained in Ref.~\cite{JM2009}. First, those results in Ref.~\cite{JM2009} are derived mainly by making observations, whereas we have implemented a systematic approach here and reproduced the same results. Second, from the denominator of Eq.~(\ref{eq:HS ungraded in 3 generations}) we find that there are totally 12 algebraically-independent invariants, corresponding to 12 physical observables in the model (i.e., three charged-lepton masses, three neutrino masses, three flavor mixing angle and three CP-violating phases).  Third, Eq.~(\ref{eq:PL 3 generations}) tells us the total number and corresponding degrees of invariants in the generating set. As a result, there are totally 33 basic invariants\footnote{However, there are actually 34, rather than 33, basic invariants in the generating set. See Sec.~\ref{subsec:3g construction} for the reasoning and explanation for this discrepancy.} in the generating set with the highest degree of $(12,4)$ and $(10,6)$. This is, however, different from the conclusion drawn in Ref.~\cite{JM2009}, where the authors assert that the highest degree of invariants in the generating set is $(12,10)$.

\subsection{Construction of Flavor Invariants}
\label{subsec:3g construction}
In the three-generation case, only $H_l^{}$, $H_{\nu}^{}$, $G_{l\nu}^{}$ and $G_{l\nu}^{(2)}$ serve as the building blocks for constructing flavor invariants and the power index of each building block can be at most three. The invariants in the generating set constructed using these building blocks with the help of Eq.~(\ref{eq:PL 3 generations}) are summarized in Table~\ref{table:3 generations}, together with their degrees and CP parities.

\renewcommand\arraystretch{0.95}
\begin{table}[H]
\centering
\begin{tabular}{l|c|c|c}
\hline \hline
flavor invariants & $(q_l^{}, q_{\nu}^{})$ & $q_l^{}+q_{\nu}^{}$ & CP parity \\
\hline \hline
$I_{1}^{}\equiv {\rm Tr}\left(H_{l}^{}\right)$ & $(2,0)$ & 2 & $+$ \\
\hline
$I_{2}^{}\equiv {\rm Tr}\left(H_{\nu}^{}\right)$ & $(0,2)$ & 2 & $+$ \\
\hline
$I_{3}^{}\equiv {\rm Tr}\left(H_{l}^{2}\right)$ & $(4,0)$ & 4 & $+$ \\
\hline
$I_{4}^{}\equiv {\rm Tr}\left(H_{l}^{}H_{\nu}^{}\right)$ & $(2,2)$&4 & $+$ \\
\hline
$I_{5}^{}\equiv {\rm Tr}\left(H_{\nu}^{2}\right)$ & $(0,4)$ & 4 & $+$ \\
\hline
$I_{6}^{}\equiv {\rm Tr}\left(H_{l}^{3}\right)$ & $(6,0)$ & 6 & $+$ \\
\hline
$I_{7}^{}\equiv {\rm Tr}\left( H_{l}^{2} H_{\nu}^{}\right)$ & $(4,2)$ & 6 & $+$ \\
\hline
$I_{8}^{}\equiv{\rm Tr}\left(H_{l}^{}G_{l\nu}^{}\right)$ & $(4,2)$ & 6 & $+$ \\
\hline
$I_{9}^{}\equiv {\rm Tr}\left(H_{l}^{}H_{\nu}^2\right)$ & $(2,4)$ & 6 & $+$ \\
\hline
$I_{10}^{}\equiv {\rm Tr}\left(H_{\nu}^3\right)$ & $(0,6)$ & 6 & $+$ \\
\hline
$I_{11}^{}\equiv{\rm Tr}\left(H_{l}^2 G_{l\nu}^{}\right)$ & $(6,2)$ & 8 & $+$ \\
\hline
$I_{12}^{}\equiv {\rm Tr}\left(\left\{H_{l}^{},H_{\nu}^{}\right\}G_{l\nu}^{}\right)$ & $(4,4)$ & 8 & $+$ \\
\hline
$I_{13}^{}\equiv {\rm Tr}\left(\left[H_{l}^{},H_{\nu}^{}\right]G_{l\nu}^{}\right)$ & $(4,4)$ & 8 & $-$ \\
\hline
$I_{14}^{}\equiv {\rm Tr}\left(H_{l}^{2}H_{\nu}^{2}\right)$  & $(4,4)$ & 8 & $+$ \\
\hline
$I_{15}^{}\equiv{\rm Tr}\left(H_{l}^{2}G_{l\nu}^{(2)}\right)$ & $(8,2)$ & 10 & $+$ \\
\hline
$I_{16}^{}\equiv {\rm Tr}\left(\left\{H_{l}^{2},H_{\nu}^{}\right\}G_{l\nu}^{}\right)$ & $(6,4)$ & 10 & $+$ \\
\hline
$I_{17}^{}\equiv{\rm Tr}\left(\left[H_{l}^{2},H_{\nu}^{}\right]G_{l\nu}^{}\right)$ & $(6,4)$ & 10 & $-$ \\
\hline
$I_{18}^{}\equiv {\rm Tr}\left(\left\{H_{l}^{},H_{\nu}^{2}\right\}G_{l\nu}^{}\right)$ & $(4,6)$ & 10 & $+$ \\
\hline
$I_{19}^{}\equiv {\rm Tr}\left(\left[H_{l}^{},H_{\nu}^{2}\right]G_{l\nu}^{}\right)$ & $(4,6)$ & 10 & $-$ \\
\hline
$I_{20}^{}\equiv {\rm Tr}\left(\left\{H_{l}^{2},H_{\nu}^{}\right\}G_{l\nu}^{(2)}\right)$ & $(8,4)$ & 12 & $+$ \\
\hline
$I_{21}^{}\equiv {\rm Tr}\left(\left[H_{l}^{2},H_{\nu}^{}\right]G_{l\nu}^{(2)}\right)$ & $(8,4)$ & 12 & $-$ \\
\hline
$I_{22}^{}\equiv {\rm Tr}\left(H_{l}^2H_{\nu}^{}H_{l}^{}G_{l\nu}^{}\right)-{\rm Tr}\left(H_{l}^2G_{l\nu}^{}H_{l}^{}H_{\nu}^{}\right)$ & $(8,4)$ & 12 & $-$ \\
\hline
$I_{23}^{}\equiv {\rm Tr}\left(\left\{H_{l}^{},H_{\nu}^{2}\right\}G_{l\nu}^{(2)}\right)$ & $(6,6)$ & 12 & $+$ \\
\hline
$I_{24}^{}\equiv {\rm Tr}\left(\left[H_{l}^{},H_{\nu}^{2}\right]G_{l\nu}^{(2)}\right)$ & $(6,6)$ & 12 & $-$ \\
\hline
$I_{25}^{}\equiv {\rm Tr}\left(H_{l}^2H_{\nu}^{2}H_{l}^{}H_{\nu}^{}\right)-{\rm Tr}\left(H_{l}^2H_{\nu}^{}H_{l}^{}H_{\nu}^{2}\right)$ & $(6,6)$ & 12 & $-$ \\
\hline
$I_{26}^{}\equiv{\rm Tr}\left(H_{l}^{}H_{\nu}^{2}G_{l\nu}^{}H_{\nu}^{}\right)-{\rm Tr}\left(H_{l}^{}H_{\nu}^{}G_{l\nu}^{}H_{\nu}^{2}\right)$ & $(4,8)$ & 12 & $-$ \\
\hline
$I_{27}^{}\equiv{\rm Tr}\left(H_{l}^2H_{\nu}^{}H_{l}^{}G_{l\nu}^{(2)}\right)-{\rm Tr}\left(H_{l}^2G_{l\nu}^{(2)}H_{l}^{}H_{\nu}^{}\right)$ & $(10,4)$ & 14& $-$ \\
\hline
$I_{28}^{}\equiv{\rm Tr}\left(\left\{H_l^{2},H_{\nu}^{}\right\}G_{l\nu}^{2}\right)$ & $(8,6)$ & 14 & $+$ \\
\hline
$I_{29}^{}\equiv {\rm Tr}\left(\left[H_l^{2},H_{\nu}^{}\right]G_{l\nu}^{2}\right)$ & $(8,6)$ & 14 & $-$ \\
\hline
$I_{30}^{}\equiv {\rm Tr}\left(H_{l}^2H_{\nu}^{2}H_{l}^{}G_{l\nu}^{}\right)-{\rm Tr}\left(H_{l}^2G_{l\nu}^{}H_{l}^{}H_{\nu}^{2}\right)$ & $(8,6)$ & 14 & $-$ \\
\hline
$I_{31}^{}\equiv{\rm Tr}\left(H_{l}^2H_{\nu}^{2}G_{l\nu}^{}H_{\nu}^{}\right)-{\rm Tr}\left(H_{l}^2H_{\nu}^{}G_{l\nu}^{}H_{\nu}^{2}\right)$ & $(6,8)$ & 14 & $-$ \\
\hline
$I_{32}^{}\equiv {\rm Tr}\left(H_{l}^2G_{l\nu}^{}H_{l}^{}G_{l\nu}^{(2)}\right)-{\rm Tr}\left(H_{l}^2G_{l\nu}^{(2)}H_{l}^{}G_{l\nu}^{}\right)$ & $(12,4)$ & 16 & $-$ \\
\hline
$I_{33}^{}\equiv{\rm Tr}\left(H_{l}^2H_{\nu}^{}H_{l}^{}G_{l\nu}^{2}\right)-{\rm Tr}\left(H_{l}^2G_{l\nu}^{2}H_{l}^{}H_{\nu}^{}\right)$ & $(10,6)$ & 16 & $-$ \\
\hline
$I_{34}^{}\equiv{\rm Tr}\left(H_l^2 H_\nu^2 G_{l\nu}^2\right)-{\rm Tr}\left(H_l^2 G_{l\nu}^2 H_\nu^2\right)$ & $(8,8)$ & 16 & $-$ \\
\hline
\hline
\end{tabular}
\vspace{0.5cm}
\caption{Summary of the basic flavor invariants in the generating set along with their degrees and CP parities in the three-generation case, where $q_l^{}$ and $q_{\nu}^{}$ denote the degree of $M_l^{}$ and $M_{\nu}^{}$, respectively. Note that the commutator $\left[A,B\right]\equiv AB-BA$ and the anti-commutator $\left\{A,B\right\}\equiv AB+BA$ of two matrices have been defined.}
\label{table:3 generations}
\end{table}
\renewcommand\arraystretch{1}

It should be noted that according to Eq.~(\ref{eq:PL 3 generations}) there are only 33 basic invariants in the generating set with the highest degree of $(12,4)$ and $(10,6)$, and they correspond to the first 33 invariants constructed in Table~\ref{table:3 generations}. However, during the calculations of their RGEs, we find another new flavor invariant at degree of (8,8), i.e.,
\begin{eqnarray}
I_{34}^{}\equiv{\rm Tr}\left(H_l^2 H_\nu^2 G_{l\nu}^2\right)-{\rm Tr}\left(H_l^2 G_{l\nu}^2 H_\nu^2\right)\;,
\end{eqnarray}
which \emph{cannot} be written as the polynomial of the above 33 basic invariants by using the general decomposition method introduced in Appendix~\ref{appendix:syzygy}. Furthermore, we have checked that the constructed 33 basic invariants are independent, namely, none of them can be expressed as the polynomial of other 32 basic invariants. Consequently, it is reasonable to claim that the generating set with only 33 basic invariants are incomplete. After adding $I_{34}^{}$ into the generating set, we have verified that any flavor invariants can be written as the polynomials of the 34 basic invariants, so a complete generating set includes 34 generators in total, as summarized in Table~\ref{table:3 generations}.

But how can we understand the discrepancy about the number of the generators between the PL in Eq.~(\ref{eq:PL 3 generations}) and the generating set? Why does the PL in Eq.~(\ref{eq:PL 3 generations}) not have a positive term of degree $(8,8)$? Such a discrepancy can be solved, though in a bit subtle way, by investigating the syzygies at this degree. Using the method developed in Appendix~\ref{appendix:syzygy}, one can prove that there is one and only one syzygy at the degree of $(8,8)$, which can be explicitly written as
\begin{eqnarray}
\label{eq:syzygy 3g}
&&I_1^4 \left(I_2^{} I_{10}^{} - I_5^2\right) + \frac{4}{3} I_1^3 \left(2I_2^3 I_4^{} - 3I_2^2 I_9^{} - 3I_2^{} I_4^{} I_5^{} - 2I_4^{} I_{10}^{} + 6I_5^{} I_9^{}\right) + \frac{2}{3}I_1^2\left\{I_2^4 I_3^{} - 2I_2^3\left(I_7\right.\right.\nonumber\\
&& \left. +2I_8\right)- 3I_2^2\left(I_3^{} I_5^{} + 5I_4^2 - 2I_{12}^{} - 2I_{14}^{}\right) - I_2^{} \left[I_3^{} I_{10}^{}-30 I_4^{} I_9^{} - 6I_5^{} \left(I_7^{} + I_8^{}\right) + 6I_{18}^{}\right] + 3I_3^{} I_5^2 \nonumber\\
&&\left. + 3I_5^{}\left(3I_4^2 - 2I_{12}^{} -4I_{14}^{}\right) -18 I_9^2 +2I_{10}^{} \left(I_7^{} +2I_8^{}\right)\right\} - \frac{4}{3} I_1^{} \left\{I_2^3\left(2I_3^{} I_4^{} -I_{11}^{}\right) - 3I_2^2 \left[I_3^{} I_9^{} +2I_4^{}\right.\right.\nonumber\\
&& \left.\times\left(I_7^{} + I_8^{}\right)-I_{16}^{}\right]-3 I_2^{} \left[I_3^{} I_4^{} I_5^{}+2I_4^3 -2 I_4^{} \left(I_{12}^{} +2I_{14}^{}\right) -I_5^{} I_{11}^{} -2\left(I_7^{} +I_8^{}\right)I_9^{} +I_{23}^{}\right]\nonumber\\
&& \left. -2 I_3^{} \left(I_4^{} I_{10}^{} - 3I_5^{} I_9^{}\right)  + 12I_4^2 I_9^{}+ 6I_4^{}\left[I_5^{}\left(I_7^{} + I_8^{}\right) - I_{18}^{}\right] - 6I_5^{} I_{16}^{} -6I_9^{} \left(I_{12}^{} +2I_{14}^{}\right) + 4I_{10}^{} I_{11}^{}\right\}\nonumber\\
&& -\frac{2}{3} I_2^4 I_3^2 + \frac{2}{3}I_2^3\left(2 I_3^{} I_7^{}+ 3I_{15}^{}\right)+ 2I_2^2\left[ I_3^2 I_5^{} +I_3^{} \left(I_4^2 -2I_{14}^{}\right) - 4I_4^{} I_{11}^{} - I_8^2\right] -\frac{1}{3} I_2^{} \left[ I_3^2 I_{10}^{}+ 12I_3^{}\right.\nonumber\\
&& \left.\times\left(I_5^{} I_7^{} + I_4^{} I_9^{} \right) + 12I_4^2\left(2I_7^{} + I_8^{}\right) - 36I_4^{} I_{16}^{} + 6I_5^{} I_{15}^{} - 24 I_7^{} I_{14}^{} -12I_8^{} I_{12}^{} -24I_9^{} I_{11}^{} +12 I_{28}^{}\right]\nonumber\\
&&  - I_3^2 I_5^2 - \frac{2}{3} I_3^{} \left( 3I_4^2 I_5^{} - 12 I_5^{} I_{14}^{} + 2 I_7^{} I_{10}^{} - 6I_9^2\right)+8I_4^2 I_{14}^{}+ 8I_4^{} \left[I_5^{} I_{11}^{} + \left(I_7^{} + I_8^{}\right)I_9^{} - I_{23}^{}\right] \nonumber\\
&& + 2I_5^{}\left(2I_7^2 + I_8^2 -2I_{20}^{}\right) - 4\left(I_8^{} I_{18}^{} +2I_9^{}I_{16}^{} - I_{10}^{} I_{15}^{}\right) - I_{12}^2+ I_{13}^2 - 8I_{14}^2=0\;.
\end{eqnarray}
Note that this syzygy is independent of $I_{34}^{}$ and should correspond to the negative term $-q_l^8 q_\nu^8$ in the PL. However, we can observe from Eq.~(\ref{eq:PL 3 generations}) that the negative terms or the syzygies start from the total degree of 18, without the $-q_l^8 q_\nu^8$ term. Therefore, the form of PL in Eq.~(\ref{eq:PL 3 generations}) should be understood as follows
\begin{eqnarray}
{\rm PL}\left[{\cal H}\left(q_l^{},q_{\nu}^{}\right) \right]
=\cdots +q_l^{12}q_{\nu}^4+q_l^{10}q_{\nu}^6+q_l^8q_{\nu}^8-q_l^8q_{\nu}^8-{\cal O}(\left[q_l^{}q_{\nu}^{}\right]^{18})\;.
\end{eqnarray}
In other words, both a basic invariant and a syzygy at the same degree of $(8,8)$ exist, but the HS is blind to such a situation because these two terms cancel each other out in the PL.\footnote{We have also checked in other cases with total degree 16 and have confirmed that there is indeed no other cancellation anywhere other than at the degree of $(8,8)$.}

To the best of our knowledge, the subtle cancellation between the basic invariant and syzygy in the PL has rarely been discussed in the literature and a strict mathematical elaboration is still lacking. However, such a cancellation may happen, for example, in the construction of the generators of the gauge-invariant operators if the invariant ring is complicated enough.\footnote{If one just needs to count the number of the linearly-independent gauge-invariant operators of a \emph{certain} mass dimension, as one usually does in the effective field theory, then there is no worry about the subtle cancellation. This is because the number of independent operators in such cases can be directly read off from the coefficients of the HS rather than the PL. The former by definition are all positive numbers.} It is also worthwhile to mention that the study of the flavor invariants in the case of three-generation leptons has drawn the attention of mathematicians to a related computation of the invariants and the HS for a trigraded Cohen-Macaulay ring~\cite{Wallach2009}.

\subsection{Physical Observables from Flavor Invariants}
\label{subsec:extraction}

In this subsection we shall establish the relations between physical observables and basic flavor invariants. In the basis where the mass matrix of charged leptons is real and diagonal, we have
\begin{eqnarray}
\label{eq:parameterization of Ml and Mnu}
M_l^{}={\rm Diag}\left\{m_e^{}, m_{\mu}^{},m_{\tau}^{}\right\}\;,\qquad
M_{\nu}^{}=V \cdot {\rm Diag}\left\{m_1^{},m_2^{},m_3^{}\right\} \cdot V^{\rm T}_{}\;,
\end{eqnarray}
where the standard parametrizaiton of the PMNS matrix is adopted, i.e.,
\begin{eqnarray}
V = \left( \begin{matrix} c^{}_{13} c^{}_{12} & c^{}_{13} s^{}_{12} & s^{}_{13} e^{-{\rm i}\delta} \cr -s_{12}^{} c_{23}^{} - c_{12}^{} s_{13}^{} s_{23}^{} e^{{\rm i}\delta}_{} & + c_{12}^{} c_{23}^{} - s_{12}^{} s_{13}^{} s_{23}^{} e^{{\rm i}\delta}_{} & c_{13}^{} s_{23}^{} \cr + s_{12}^{} s_{23}^{} - c_{12}^{} s_{13}^{} c_{23}^{} e^{{\rm i}\delta}_{} & - c_{12}^{} s_{23}^{} - s_{12}^{} s_{13}^{} c_{23}^{} e^{{\rm i}\delta}_{} & c_{13}^{} c_{23}^{} \end{matrix} \right) \cdot \left(\begin{matrix} e^{{\rm i}\rho} & 0 & 0 \cr 0 & e^{{\rm i}\sigma} & 0 \cr 0 & 0 & 1\end{matrix}\right) \; ,
\end{eqnarray}
with $c_{ij}^{}\equiv\cos\theta_{ij}^{}$ and $s_{ij}^{}\equiv\sin\theta_{ij}^{}$ (for $ij=12, 13, 23$). Here $\left\{\theta_{12}^{}, \theta_{13}^{}, \theta_{23}^{}  \right\}$ are three flavor mixing angles, $\delta$ is the Dirac-type CP phase and $\left\{\rho,\sigma\right\}$ are two Majorana-type CP phases. Using the explicit expressions of $M_l^{}$ and $M_{\nu}^{}$ in Eq.~(\ref{eq:parameterization of Ml and Mnu}), one can also obtain the explicit forms of all the flavor invariants in terms of all the 12 physical observables, just as what we have done in the case of two-generation leptons.

Although the calculations are tedious and some tricks are needed, we find that it is possible to extract all these 12 physical observables analytically from flavor invariants. This provides a basis-independent way to describe the running behaviors of physical observables. For later convenience, we introduce some useful working invariants, namely,
\begin{eqnarray}
k_1&\equiv & 3 I_1^{}/v^2\;,\quad
k_2\equiv I_1^2-I_3^{}\;,\quad
k_3\equiv \left(I_1^{3}-3I_1^{} I_3^{}+2I_6^{}\right)/3\;,\nonumber\\
p_1&\equiv & 3 I_2^{}/v^2\;,\quad
p_2\equiv I_2^2-I_5^{}\;,\quad
p_3\equiv \left(I_2^{3}-3I_2^{} I_5^{}+2I_{10}^{}\right)/3\;.
\end{eqnarray}
Under the hierarchical condition $m_\tau^{}\gg m_{\mu}^{}\gg m_e^{}$, one can extract the masses of three charged leptons from flavor invariants in a simple manner
\begin{eqnarray}
	\label{eq:lep mass using inv}
	m_e^2\approx \frac{k_3^{}}{k_2^{}}\;,\quad
	m_{\mu}^2\approx \frac{k_2^{}}{2I_6^{1/3}}\;,\quad
	m_{\tau}^2 \approx I_6^{1/3}\;.
\end{eqnarray}
As for the neutrino masses, in the case of normal mass ordering with $m_3^{}\gg m_2^{}> m_1^{}$, we have
\begin{eqnarray}
	\label{eq:neu mass using inv NO}
	m_{1,2}^2\approx \frac{1}{2}\left[\frac{p_2^{}}{2I_{10}^{1/3}}\mp\sqrt{\left(\frac{p_2^{}}{2I_{10}^{1/3}}\right)^2_{}-4\left(\frac{p_3^{}}{2 I_{10}^{1/3}} \right)} \right]\;,\quad
	m_3^{2}\approx I_{10}^{1/3}\;,
\end{eqnarray}
while in the case of inverted mass ordering with $m_2^{}>m_1^{}\gg m_3^{}$,
\begin{eqnarray}
	\label{eq:neu mass using inv IO}
	m_{1,2}^2  \approx  \frac{1}{2}\left[I_2^{}-m_3^2\mp\sqrt{2I_5^{}-I_2^2+2I_2^{}m_3^2}\right] \; , \quad 	m_3^2 \approx  \frac{p_3^{}}{p_2^{}}\;.
\end{eqnarray}

In order to express the three mixing angles $\{\theta_{12}^{}, \theta_{23}^{}, \theta_{13}^{}\}$ and three CP-violating phases $\{\delta, \rho, \sigma\}$ in terms of flavor invariants, we find that it is convenient to choose the basis where the Majorana neutrino mass matrix is diagonal and present the mixing angles and CP phases in terms of the elements of $H_l^{}$ and the masses of leptons. The latter can in turn be expressed in terms of flavor invariants. In the chosen basis, we have $H_\nu^{} = {\rm Diag}\, \{ m_1^{2}, m_2^{2}, m_3^{2} \}$ and the Hermitian matrix $H_l^{}$ can be generally written as
\begin{eqnarray}
	H_l^{} = \left( \begin{matrix} H_{11}^{} & H_{12} e^{{\rm i} h_{12}^{}} & H_{31} e^{-{\rm i} h_{31}^{}} \\ H_{12} e^{-{\rm i} h_{12}^{}} & H_{22}^{} & H_{23} e^{{\rm i} h_{23}^{}} \\ H_{31} e^{{\rm i} h_{31}^{}} & H_{23} e^{-{\rm i} h_{23}^{}} & H_{33}^{} \end{matrix} \right)\;,
\end{eqnarray}
which is related to the physical parameters via $H_l^{} = V^\dagger_{}\, {\rm Diag}\, \{ m_e^{2}, m_\mu^{2}, m_\tau^{2} \}\, V$ with $V$ being the PMNS matrix. Note that the off-diagonal elements of $H^{}_l$ in the upper-right corner have been written as $H^{}_{12} e^{{\rm i}h^{}_{12}}$, $H^{}_{31} e^{-{\rm i}h^{}_{31}}$ and $H^{}_{23} e^{{\rm i}h^{}_{23}}$, where $H^{}_{ij}$ (for $ij = 12, 23, 31 $) are the moduli.

First, we express the elements of $H_l^{}$ in terms of the flavor invariants and neutrino masses, where the latter have already be given in terms of flavor invariants using Eq.~(\ref{eq:neu mass using inv NO}) and Eq.~(\ref{eq:neu mass using inv IO}). By noticing that
\begin{eqnarray*}
	{\rm Tr}(H_l^{} H_\nu^{}) &=& I_4^{} = H_{11}^{} m_1^2 + H_{22}^{} m_2^2 + H_{33}^{} m_3^2\;, \\
	{\rm Tr}(H_l^{} H_\nu^2) &=& I_9^{} = H_{11}^{} m_1^4 + H_{22}^{} m_2^4 + H_{33}^{} m_3^4\;, \\
	{\rm Tr}(H_l^{} H_\nu^3) &=& I_2^{} I_9^{} - \frac{1}{2} p_2^{} I_4^{} + \frac{1}{2} p_3^{} I_1^{} = H_{11}^{} m_1^6 + H_{22}^{} m_2^6 + H_{33}^{} m_3^6\;,	
\end{eqnarray*}
where the determinant of the coefficient matrix is nonzero, namely, $m_1^2m_2^2m_3^2 \Delta_{12}\Delta_{23}\Delta_{31}\neq 0$ with $\Delta_{ij}^{} \equiv m_i^2 - m_j^2$ (for $i,j = 1,2,3$), we obtain
\begin{eqnarray}
	H_{ii}^{} &=& \frac{2\left(I_9^{}m_i^2+I_4^{}m_j^2 m_k^2 \right)+I_1^{}p_3^{}-I_4^{}p_2^{}}{2 m_i^2 \Delta_{ji} \Delta_{ki}}\;,
\end{eqnarray}
where $(i,j,k)=(1,2,3)$ or $(2,3,1)$ or $(3,1,2)$. On the other hand, making use of the following equations
\begin{eqnarray*}
	{\rm Tr}(H_l^{2} H_\nu^{}) &=& I_7^{} = H_{11}^{2} m_1^2 + H_{22}^{2} m_2^2 + H_{33}^{2} m_3^2\nonumber\\ &+& (m_1^2 + m_2^2) H_{12}^2 + (m_2^2 + m_3^2) H_{23}^2 + (m_3^2 + m_1^2) H_{31}^2\;, \\
	{\rm Tr}(H_l^{2} H_\nu^2) &=& I_{14}^{} = H_{11}^{2} m_1^4 + H_{22}^{2} m_2^4 + H_{33}^{2} m_3^4\nonumber\\ &+& (m_1^4 + m_2^4) H_{12}^2 + (m_2^4 + m_3^4) H_{23}^2 + (m_3^4 + m_1^4) H_{31}^2 \;, \\
	{\rm Tr}(H_l^{2} H_\nu^3) &=& I_2^{} I_{14}^{} - \frac{1}{2} p_2^{} I_7^{} + \frac{1}{2} p_3^{} I_3^{} = H_{11}^{2} m_1^6 + H_{22}^{2} m_2^6 + H_{33}^{2} m_3^6\nonumber\\ &+& (m_1^6 + m_2^6) H_{12}^2 + (m_2^6 + m_3^6) H_{23}^2 + (m_3^6 + m_1^6) H_{31}^2 \;,		
\end{eqnarray*}
with a nonzero determinant of the coefficient matrix $2m_1^2m_2^2m_3^2 \Delta_{12}\Delta_{23}\Delta_{31}\neq 0$, we have
\begin{eqnarray}
	H_{ij}^2 &=&\left\{n_1^{}\left[\left(m_j^4+m_k^4\right)m_i^4+\left(m_i^2+m_j^2 \right)m_k^4\Delta_{jk}^{}\right]-n_2^{}\left(m_i^2+\Delta_{jk}^{} \right)\left(m_j^2+m_k^2\right)\left(m_k^2+m_i^2\right)\right.\nonumber\\
&&\left. +n_3^{}\left[\Delta_{jk}^{}m_k^2+m_i^2\left(m_j^2+m_k^2\right)\right]
\right\} / \left(2m_i^2 m_j^2 m_k^2 \Delta_{jk}^{} \Delta_{ki}^{}\right)\;,
\end{eqnarray}
where $(i,j,k)=(1,2,3)$ or $(2,3,1)$ or $(3,1,2)$ and
\begin{eqnarray*}
	n_1^{} &\equiv& I_7^{} - \left(H_{11}^2 m_1^2 + H_{22}^2 m_2^2 + H_{33}^2 m_3^2\right)\;,\\
	n_2^{} &\equiv& I_{14}^{} - \left(H_{11}^2 m_1^4 + H_{22}^2 m_2^4 + H_{33}^2 m_3^4\right)\;,\\
	n_3^{} &\equiv& I_2^{} I_{14}^{} - \frac{1}{2} p_2^{} I_7^{} + \frac{1}{2} p_3^{} I_3^{} - \left(H_{11}^2 m_1^6 + H_{22}^2 m_2^6 + H_{33}^2 m_3^6\right)\;.
\end{eqnarray*}
As for the phases in $H_l^{}$, we need to consider some CP-odd flavor invariants. If we choose
\begin{eqnarray*}
	\frac{{\rm i}}{2}I_{13}^{} &=&  H_{12}^2 m_1^{} m_2^{} \Delta_{12}^{} \sin\left(2 h_{12}^{}\right) + H_{23}^2 m_2^{} m_3^{} \Delta_{23}^{} \sin\left(2 h_{23}^{}\right)+ H_{31}^2 m_3^{} m_1^{} \Delta_{31}^{} \sin\left(2 h_{31}^{}\right) \;, \\
	\frac{{\rm i}}{2}I_{19}^{} &=& H_{12}^2 m_1^{} m_2^{} \left(m_1^{2} + m_2^{2}\right) \Delta_{12}^{} \sin\left(2 h_{12}^{}\right) + H_{23}^2 m_2^{} m_3^{} (m_2^{2} + m_3^{2}) \Delta_{23}^{} \sin\left(2 h_{23}^{}\right) \nonumber\\ &&+ H_{31}^2 m_3^{} m_1^{} (m_3^{2} + m_1^{2}) \Delta_{31}^{} \sin\left(2 h_{31}^{}\right)\;, \\
	\frac{{\rm i}}{2}I_{26}^{} &=& H_{12}^2 m_1^{3} m_2^{3} \Delta_{12}^{} \sin\left(2 h_{12}^{}\right) + H_{23}^2 m_2^{3} m_3^{3} \Delta_{23}^{} \sin\left(2 h_{23}^{}\right) + H_{31}^2 m_3^{3} m_1^{3} \Delta_{31}^{} \sin\left(2 h_{31}^{}\right)\;,
\end{eqnarray*}
with a nonzero determinant $H_{12}^2H_{23}^2H_{31}^2m_1^2m_2^2m_3^2\Delta_{12}^2\Delta_{23}^2\Delta_{31}^2\neq0$ of the coefficient matrix,\footnote{Here we assume that there are no zero elements in the matrix $H_l^{}$ in general.} then the phases are determined by
\begin{eqnarray}
	\sin(2 h_{ij}^{}) &=& -\frac{{\rm i}}{2\Delta_{ij}^{}\Delta_{jk}^{}\Delta_{ki}^{}}\frac{I_{26}^{} -  I_{19}^{} m_k^2 + I_{13}^{} m_k^4}{H_{ij}^2m_i^{}m_j^{}}\;,
\end{eqnarray}
where $(i,j,k)=(1,2,3)$ or $(2,3,1)$ or $(3,1,2)$.

Second, using the relations $H_l^{} = V^\dagger_{}\, {\rm Diag}\, \{ m_e^{2}, m_\mu^{2}, m_\tau^{2} \}\, V$ and $H_l^{2} = V^\dagger_{}\, {\rm Diag}\, \{ m_e^{4}, m_\mu^{4}, m_\tau^{4} \}\, V$, one can directly express the matrix elements of $V$ in terms of those of $H_l^{}$ and $H_l^{2} $,
\begin{eqnarray}
\label{eq:extract PMNS from Hl}
	\left|V_{ei}\right|^2 &=& \frac{\left(H_l^2\right)_{ii}^{} - \left(H_l^{}\right)_{ii}^{} \left(m_\mu^2 + m_\tau^2\right) + m_\mu^2 m_\tau^2}{\Delta_{\mu e} \Delta_{\tau e}}\;,\\
	\left|V_{\mu i}\right|^2 &=& \frac{\left(H_l^2\right)_{ii}^{} - \left(H_l^{}\right)_{ii}^{} \left(m_\tau^2 + m_e^2\right) + m_\tau^2 m_e^2}{\Delta_{\tau\mu } \Delta_{e \mu}}\;,\\
	\left|V_{\tau i}\right|^2 &=& \frac{\left(H_l^2\right)_{ii}^{} - \left(H_l^{}\right)_{ii}^{} \left(m_e^2 + m_\mu^2\right) + m_e^2 m_\mu^2}{\Delta_{e \tau} \Delta_{\mu\tau  }}\;,
\end{eqnarray}	
with $i=1,2,3$ and
\begin{eqnarray}
	V_{ei}^* V_{ej}^{} &=& \frac{\left(H_l^2\right)_{ij}^{} - \left(H_l^{}\right)_{ij}^{} \left(m_\mu^2 + m_\tau^2\right)}{\Delta_{\mu e}\Delta_{\tau e}}\;,\\
	V_{\mu i}^* V_{\mu j}^{} &=& \frac{\left(H_l^2\right)_{ij}^{} - \left(H_l^{}\right)_{ij}^{} \left(m_\tau^2 + m_e^2\right)}{\Delta_{\tau\mu}\Delta_{e \mu}}\;,\\
	V_{\tau i}^* V_{\tau j}^{} &=& \frac{\left(H_l^2\right)_{ij}^{} - \left(H_l^{}\right)_{ij}^{} \left(m_e^2 + m_\mu^2\right)}{\Delta_{e\tau}\Delta_{\mu\tau}}\;,
\end{eqnarray}	
with $i,j=1,2,3$ $(i\neq j)$ and $\Delta_{\alpha \beta}^{} \equiv m_\alpha^2 - m_\beta^2$ (for $\alpha,\beta = e,\mu,\tau$). Note that $(H_l^{})_{ij}^{}$ and $(H_l^2)_{ij}^{}$ denote the $(i,j)$-element of $H_l^{}$ and $H_l^2$, respectively.

Finally, we can extract three mixing angles and three phases directly from the matrix elements of $V$ in the standard way
\begin{eqnarray}
\label{eq:extract physical observables from PMNS}
	s_{13}^2 = |V_{e3}|^2\;,\quad s_{12}^2 = \frac{|V_{e2}|^2}{1-|V_{e3}|^2}\;,\quad s_{23}^2 = \frac{|V_{\mu 3}|^2}{1-|V_{e3}|^2}\;,\quad
	\sin \delta =\frac{{\rm Im}\left(V_{e2}^{}V_{e3}^{*}V_{\mu 2}^{*}V_{\mu 3}\right)}{s_{12}^{} c_{12}^{} s_{23}^{} c_{23}^{} s_{13}^{} c_{13}^2}\;,
\end{eqnarray}
and
\begin{eqnarray}
\label{eq:Majorana phases}
\rho = - \delta - {\rm Arg}\left(\frac{V_{e1}^* V_{e3}^{}}{c_{12}^{} c_{13}^{} s_{13}^{}}\right)\;,\quad
	\sigma = - \delta - {\rm Arg}\left(\frac{V_{e2}^* V_{e3}^{}}{s_{12}^{} c_{13}^{} s_{13}^{}}\right)\;.
\end{eqnarray}
Note that we have used the identities $V_{e1}^* V_{e3}^{} = {\rm e}^{-{\rm i}(\delta + \rho)}_{} c_{12}^{} c_{13}^{} s_{13}^{}$ and $V_{e2}^* V_{e3}^{} = {\rm e}^{-{\rm i}(\delta + \sigma)}_{} s_{12}^{} c_{13}^{} s_{13}^{}$ to get the Majorana CP phases. Thus we have explicitly expressed all the physical observables in terms of flavor invariants, with which we can describe the running behaviors of physical observables in a basis-independent way.

To summarize, we first use $\left\{I_2^{},I_5^{},I_{10}^{}\right\}$ to obtain three neutrino masses. Then in the basis where the neutrino mass matrix is diagonal, we utilize $\left\{I_1^{},I_4^{},I_9^{} \right\}$ and $\left\{I_3^{},I_7^{},I_{14}^{} \right\}$ to derive the absolute values of the diagonal and off-diagonal elements of $H_l^{}$, respectively. Finally, the three phases in $H_l^{}$ can be extracted with three CP-odd flavor invariants $\left\{I_{13}^{},I_{19}^{},I_{26}^{}\right\}$. The masses of charged-leptons can be obtained by calculating the eigenvalues of $H_l^{}$ while the three flavor mixing angles and three CP phases can be extracted from the elements of $H_l^{}$ using Eqs.~(\ref{eq:extract PMNS from Hl})-(\ref{eq:Majorana phases}). As a result, in the three-generation case, the 12 physical observables in the leptonic sector
\begin{eqnarray*}
\left\{m_1^{},m_2^{},m_3^{},m_e^{},m_\mu^{},m_\tau^{},\theta_{12}^{},\theta_{13}^{},\theta_{23}^{},\delta,\rho,\sigma \right\}
\end{eqnarray*}
are equivalent to the following 12 flavor invariants
\begin{eqnarray*}
\left\{I_1^{},I_2^{},I_3^{},I_4^{},I_5^{},I_7^{},I_9^{},I_{10}^{},I_{13}^{},I_{14}^{},I_{19}^{},I_{26}^{}\right\}\;,
\end{eqnarray*}
three of which are CP-odd and the others are CP-even, reflecting the fact that there are totally three CP phases.

\subsection{RGEs of Flavor Invariants}
\label{subsec:3g RGE}

Starting from the RGEs of the building blocks, i.e., Eqs.~(\ref{eq:RGEHl})-(\ref{eq:RGEGlnu}), together with that of $G_{l\nu}^{(2)}$
\begin{eqnarray}
\label{eq:RGEG2}
\frac{{\rm d}G^{(2)}_{l\nu}}{{\rm d}t}=2\left(2\alpha_l^{}+\alpha_{\nu}^{} +k_1^{}\right)G_{l\nu}^{(2)}-\frac{3}{v^2}\left[\left\{H_l^{},G_{l\nu}^{(2)} \right\}+k_2^{}G_{l\nu}^{}-k_3^{}H_{\nu}^{}\right]\;,
\end{eqnarray}
one can directly calculate the RGEs of the flavor invariants in the generating set. First, the RGEs of all the CP-even flavor invariants read

{\allowdisplaybreaks
\begin{eqnarray}
\label{eq:I1}
\frac{{\rm d}I_1^{}}{{\rm d}t} &=& 2 \alpha_{l}^{} I_1^{} + 6 I_{3}^{} / v_{}^2 \;,\\
\frac{{\rm d}I_2^{}}{{\rm d}t} &=& 2 \alpha_{\nu}^{} I_2^{} - 12 I_{4}^{} / v_{}^2 \;,\\
\frac{{\rm d}I_3^{}}{{\rm d}t} &=& 4 \alpha_{l}^{} I_3^{} + 12 I_{6}^{} / v_{}^2 \;,\\
\frac{{\rm d}I_4^{}}{{\rm d}t} &=& 2 \left( \alpha_{l}^{} + \alpha_{\nu}^{} \right) I_4^{} - 6 I_8^{} / v^2_{} \;,\\
\frac{{\rm d}I_5^{}}{{\rm d}t} &=& 4 \alpha_{\nu}^{} I_5^{} - 24 I_9^{} / v^2_{}\;,\\
\frac{{\rm d}I_6^{}}{{\rm d}t} &=& 2 \left( 3 \alpha_l^{} + 4 k_1^{} \right) I_6^{} + 3 \left( I_1^4 - 6 I_1^2 I_3^{} + 3 I_3^2 \right) / v^2_{} \;,\\
\frac{{\rm d}I_{7}^{}}{{\rm d}t} &=& 2 \left( 2 \alpha_l^{} + \alpha_{\nu}^{} + k_1^{} \right) I_7^{} - 3 \left( k_2^{} I_4^{} - k_3^{} I_2^{} + 2 I_{11}^{} \right) / v^2_{} \;,\\
\frac{{\rm d}I_{8}^{}}{{\rm d}t} &=& 2 \left( 2 \alpha_l^{} + \alpha_{\nu}^{} \right) I_8^{} \;,\\
\frac{{\rm d}I_{9}^{}}{{\rm d}t} &=& 2 \left( \alpha_{l}^{} + 2 \alpha_{\nu}^{} - 2 k_1^{} \right) I_9^{} - 3 \left[ 2 \left( I_{12}^{} - 2 I_{14}^{} \right) + k_2^{} p_2^{} + 4 I_2^{} \left( I_7^{} - I_1^{} I_4^{} \right) + 2 I_4^2 \right] / v^2_{} \;,\\
\frac{{\rm d}I_{10}^{}}{{\rm d}t} &=& 2 \left( 3 \alpha_{\nu}^{} - 2 k_1^{} \right) I_{10}^{} - 6 \left[ 3 \left( 2 I_2^{} I_9^{} - p_2^{} I_4^{} \right) + I_1^{} \left( I_2^3 - 3 I_2^{} I_5^{} \right) \right] / v^2_{} \;,\\
\frac{{\rm d}I_{11}^{}}{{\rm d}t} &=& 2 \left( 3 \alpha_{l}^{} + \alpha_{\nu}^{} + k_1^{} \right) I_{11}^{} - 3 \left( k_2^{} I_8^{} - k_3^{} I_4^{} \right) / v^2_{} \;,\\
\frac{{\rm d}I_{12}^{}}{{\rm d}t} &=& 4 \left( \alpha_l^{} + \alpha_{\nu}^{} \right) I_{12}^{} - 12 \left[ I_1^2 \left( I_2^{} I_4^{} - I_9^{} \right) + 2 I_1^{} \left( I_{12}^{} - I_2^{} I_8^{} - I_4^2 \right) \right. \nonumber\\
&&\left. + 2 I_2^{} I_{11}^{} - I_2^{} I_3^{} I_4^{} + I_3^{} I_9^{} + 2 I_4^{} \left( I_7^{} + I_8^{} \right) - 2 I_{16}^{} \right] / v^2_{} \;,\\
\frac{{\rm d}I_{14}^{}}{{\rm d}t} &=& 4 \left( \alpha_{l}^{} + \alpha_{\nu}^{} \right) I_{14}^{} - \left[ 6 I_{16}^{} - I_1^3 \left( 3 I_5^{} - 2 I_2^2 \right) - 3 I_1^2 \left( 2 I_2^{} I_4^{} - 3 I_9^{} \right) \right. \nonumber\\
&&\left. - 3 I_1^{} \left( I_2^2 I_3^{} + 4 I_{14}^{} - 2 I_2^{} I_7^{} - 2 I_3^{} I_5^{} \right) + I_2^2 I_6^{} - 3 \left( I_3^{} I_9^{} + I_5^{} I_6^{} - 2 I_4^{} I_7^{} \right) \right] / v^2_{} \;,\\
\frac{{\rm d}I_{15}^{}}{{\rm d}t} &=& 2 \left( 4 \alpha_{l}^{} + \alpha_{\nu}^{} + 2 k_1^{} \right) I_{15}^{} - 6 \left( k_2^{} I_{11}^{} - k_3^{} I_7^{} \right) / v^2_{} \;,\\
\frac{{\rm d}I_{16}^{}}{{\rm d}t} &=& 2 \left( 3 \alpha_{l}^{} + 2 \alpha_{\nu}^{} \right) I_{16}^{} + 3 \left[ I_1^3 \left( I_9^{} - I_2^{} I_4^{} \right) + 2 I_1^2 \left( I_2^{} I_8^{} + I_4^2 - I_{12}^{} + I_{14}^{} \right) \right. \nonumber\\
&& \left. + I_1^{} \left( I_2^{} I_3^{} I_4^{} + 2 I_2^{} I_{11}^{} + 2 I_4^{} I_7^{} - I_3^{} I_9^{} \right) - I_2^{} \left( 2 k_2^{} I_7^{} + k_3^{} I_4^{} + 4 I_{15}^{} \right) \right. \nonumber\\
&& \left. - 2 I_3^{} I_{14}^{} - 6 I_4^{} I_{11}^{} - 2 I_7^{} \left( 2 I_7^{} + I_8^{} \right) + 3 k_3^{} I_9^{} - k_2^{} I_{12}^{} + 4 I_{20}^{} \right] / v^2_{}\;,\\
\frac{{\rm d}I_{18}^{}}{{\rm d}t} &=& 2 \left( 2 \alpha_{l}^{} + 3 \alpha_{\nu}^{} - 2 k_1^{} \right) I_{18}^{} - 3 \left\{ I_1^3 \left( 3 I_2^{} I_5^{} + 4 p_3^{} - 2 I_{10}^{} - I_2^3 \right) + I_1^2 \left( 7 I_2^2 I_4^{} - 2 I_2^{} I_9^{} \right. \right. \nonumber\\
&& \left.\left. - 3 I_4^{} I_5^{} - 4 I_4^{} p_2^{} \right) + I_1^{} \left[ I_2^3 I_3^{} - 4 I_2^2 \left ( I_7^{} + I_8^{} \right) + I_2^{} \left( 4 I_{12}^{} - 3 I_3^{} I_5^{} - 8 I_4^2 \right) + 2 I_3^{} \left( I_{10}^{} \right. \right. \right. \nonumber\\
&& \left.\left.\left. - 2 p_3^{} \right) + 4 I_5^{} \left( I_7^{} - I_8^{} \right) + 4 I_7^{} p_2^{} - k_2^{} p_3^{} \right] + I_2^2 \left( 4 I_{11}^{} - 3 I_3^{} I_4^{} \right) + 2 I_2^{} \left( I_3^{} I_9^{} + 4 I_4^{} I_7^{} \right. \right. \nonumber\\
&& \left.\left. - k_2^{}I_9^{} - 2I_{16}^{} \right) - I_4^{} \left( I_3^{} I_5^{} - 4 I_{12}^{} - k_2^{} p_2^{} \right) + 4 \left( I_5^{} I_{11}^{} + 2 I_8^{} I_9^{} - I_{23}^{} \right) \right\} / v^2_{}\;,\\
\frac{{\rm d}I_{20}^{}}{{\rm d}t} &=& 4 \left( 2 \alpha_l^{} + \alpha_{\nu}^{} + 2 k_1^{} \right) I_{20}^{} - 6 \left[ I_1^3 \left( I_2^{} I_7^{} - I_{14}^{} \right) + I_1^2 \left( 2 I_{16}^{} - I_2^{} I_{11}^{} - 2 I_4^{} I_7^{} \right) + I_1^{} \left( 2 I_2^{} I_{15}^{} \right.\right.\nonumber\\
&&\left.\left.- I_2^{} I_3^{} I_7^{} + I_3^{} I_{14}^{} + 2 I_7^2 \right) + I_2^{} \left( k_3^{} I_7^{} - I_1^2 I_{11}^{} \right) + 2 \left( I_4^{} I_{15}^{} + I_7^{} I_{11}^{} \right) - 3 k_3^{} I_{14}^{} \right. \nonumber\\
&& \left. + k_{2}^{} I_{16}^{} \right] / v^2_{} \;,\\
\frac{{\rm d}I_{23}^{}}{{\rm d}t} &=& 2 \left( 3 \alpha_l^{} + 3 \alpha_{\nu} + k_1^{} \right) I_{23}^{} - 3 \left\{ I_1^3 I_2^{} \left( I_2^{} I_4^{} - I_9^{} \right) + 2 I_1^2 \left[ I_2^2 \left( 2 I_7^{} - I_8^{} \right) + I_2^{} \left( I_{12}^{} + 2 I_{14}^{} \right.\right.\right.\nonumber\\
&& \left.\left.\left.- I_4^2 \right) + 2 \left( p_3^{} I_3^{} - p_2^{} I_7^{} \right) - I_5^{} I_7^{} \right]
+ I_1^{} \left[ I_2^{} \left( I_3^{}I_9^{} + 2I_{16}^{} - 2 I_2^{} I_{11}^{} - I_2^{} I_3^{} I_4^{} - 6I_4^{} I_7^{} \right) \right.\right.\nonumber\\
&& \left.\left.+ 4 I_7^{} I_9^{}  - 2 k_3^{} p_3^{} \right] + I_2^2 \left( 4 I_{15}^{} - 4 I_3^{} I_7^{} + k_3^{} I_4^{} \right) - I_{2}^{} \left[ 2 \left( 2 I_3^{} I_{14}^{} + I_4^{}  I_{11}^{} \right) - 2 I_7^{} \left( 4 I_7^{} + I_8^{} \right)  \right. \right. \nonumber\\
&& \left.\left.+ 3 \left( k_3^{} I_9^{} + 2 k_2^{} I_{14}^{} \right) +  k_2^{} I_{12}^{} + 4 I_{20}^{} \right] + I_3^{} \left( 2 I_5^{} I_7^{} + 4 p_2^{} I_7^{} - 4 p_3^{} I_3^{} - 3 k_2^{} p_3^{} \right) + I_4^{} \left( k_3^{} p_2^{} \right. \right.\nonumber\\
&& \left. \left. + 6 I_{16}^{} \right) + I_7^{} \left( 3 k_2^{} p_2^{} - 4 I_{14}^{} \right) + k_2^{} p_2^{} I_8^{} + 2 \left( 4 I_9^{} I_{11}^{} + k_2^{} I_{18}^{} - I_{28}^{} \right) \right\} /v^2_{}\;,\\
\frac{{\rm d}I_{28}^{}}{{\rm d}t} &=& 2 \left( 4 \alpha_l^{} + 3 \alpha_{\nu}^{} + k_1^{} \right) I_{28}^{} - 3 \left\{ 2 I_1^3 \left( I_2^{} I_4^2 - I_4^{} I_9^{} + k_2^{} p_3^{} \right) + I_1^2 \left[ - 4 \left( I_4^3 + I_7^{} I_9^{} \right) + 2 p_2^{} I_{11}^{} \right. \right. \nonumber\\
&& \left. \left. + I_4^{} \left( 4 I_{12}^{} - 4 I_{14}^{} + 8 I_2^{} I_7^{} - 4 I_2^{} I_8^{} + k_2^{} p_2^{} \right) + k_3^{} \left( I_2^3 + 4 I_{10}^{} - 5 I_2^{} I_5^{} - 8 p_3^{} \right) \right] - 2 I_1^{}  \right. \nonumber\\
&& \left. \times \left[ 2 I_2^2 \left( k_2^{} I_8^{} + k_3^{} I_4^{} \right) + I_2^{} \left( I_3^{} I_4^{2} + 6 I_4^{} I_{11}^{} + 4 I_7^{} I_8^{} - 2 k_2^{} I_{12}^{} + 2 k_2^{} I_4^2 + 2 k_3^{} I_9^{} \right) + 6 I_4^2 I_7^{} \right.\right.\nonumber\\
&& \left.\left.- I_4^{} \left( I_3^{} I_9^{} + 2 I_{16}^{} + 2 k_2^{} I_9^{} + 4 k_3^{} I_5^{} + 4 k_3^{} p_2^{} \right) + k_2^{} p_3^{} I_3^{} - 2 \left( 2 I_7^{} I_{12}^{} + 2 I_9^{} I_{11}^{} - k_2^{} I_{18}^{} \right. \right. \right. \nonumber\\
&& \left. \left. + k_2^{} I_5^{} I_8^{} \right) \right] - k_3^{} I_2^3 I_3^{} + I_2^2 \left( k_2^{} I_3^{} I_4^{} + 4 k_3^{} I_7^{} - 2 I_3^{} I_{11}^{} \right) + I_2^{} \left[ 2 k_3^{} I_4^2 + 4 I_4^{} \left( 2 I_{15}^{} - 2 I_3^{} I_7^{} \right. \right.\nonumber\\
&& \left. \left. + k_2^{} I_7^{} \right) + 5 k_3^{} I_3^{} I_5^{} +4\left( 4I_7^{} I_{11}^{} + k_3^{} I_{14}^{} - k_2^{} I_{16}^{} \right) \right] + I_3^{} \left[ I_4^{} \left( 4 I_{14}^{} - k_2^{} I_5^{} - 2k_2^{} p_2^{} \right) + 2 I_5^{} I_{11}^{}  \right.\nonumber\\
&& \left. + 4 \left( I_7^{} I_9^{} - k_3^{} I_{10}^{} \right) + 10 k_3^{} p_3^{} \right] + 12 I_4^2 I_{11}^{} + 2 I_4^{} \left( 8 I_7^2 + 2 I_7^{} I_8^{} + k_2 I_{12} - 3 k_3^{} I_9^{} - 2 k_2^{} I_{14}^{}\right.\nonumber\\
&& \left.\left. - 4 I_{20}^{} \right) - 8 I_7^{} \left( I_{16}^{} + k_3^{} I_5^{} + k_3^{} p_2^{} \right) + 4 \left( k_2^{} p_2^{} I_{11}^{} + I_8^{} I_{16}^{} + k_2^{} I_{23}^{} - 2 I_{11}^{} I_{14}^{} \right) \right\} / \left( 2 v^2_{} \right)\,.
\end{eqnarray}}
It should be noted that the RGEs of all the CP-even flavor invariants form a closed system of differential equations, which is independent of any CP-odd flavor invariants. This feature is similar to the two-generation case and manifests the fact that if there is no CP violation at the initial scale (i.e., all the CP phases take trivial values) then CP will be conserved all the way during the RGE running~\cite{Yu2020PRD}. As for the CP-odd flavor invariants
{\allowdisplaybreaks
\begin{eqnarray}
\frac{{\rm d}I_{13}^{}}{{\rm d}t} &=& 4 \left( \alpha_l^{} + \alpha_{\nu}^{} \right) I_{13}^{} \;,\\
\frac{{\rm d}I_{17}^{}}{{\rm d}t} &=& 2 \left( 3 \alpha_l^{} + 2 \alpha_{\nu}^{} + k_1^{} \right) I_{17}^{} - 3 \left( k_2^{} I_{13}^{} + 2 I_{22}^{} \right) / v^2_{} \;,\\
\frac{{\rm d}I_{19}^{}}{{\rm d}t} &=& 2 \left( 2 \alpha_l^{} + 3 \alpha_{\nu}^{} - 2 k_1^{} \right) I_{19}^{} - 12 \left[ I_2^{} I_{17}^{} - \left( I_1^{} I_2^{} - I_4^{} \right) I_{13}^{} - I_{24}^{} \right] / v^2_{} \;,\\
\frac{{\rm d}I_{21}^{}}{{\rm d}t} &=& 4 \left( 2 \alpha_l^{} + \alpha_{\nu}^{} + k_1^{} \right) I_{21}^{} - 6 \left( k_2^{} I_{17}^{} + 2 I_{27}^{} \right) / v^2_{} \;,\\
\frac{{\rm d}I_{22}^{}}{{\rm d}t} &=& 2 \left( 4 \alpha_l^{} + 2 \alpha_{\nu}^{} + k_1^{} \right) I_{22}^{} - 3 k_3^{} I_{13}^{} / v^2_{} \;,\\
\frac{{\rm d}I_{24}^{}}{{\rm d}t} &=& 2 \left( 3 \alpha_l^{} + 3 \alpha_{\nu}^{} + k_1^{} \right) I_{24}^{} - 3 \left[ I_2^{} \left( 2 I_{22}^{} - k_2^{} I_{13}^{} \right) + 2 \left( I_4^{} I_{17}^{} + k_2^{} I_{19}^{} + I_{29}^{} \right) \right] / v^2_{} \;,\\
\frac{{\rm d}I_{25}^{}}{{\rm d}t} &=& 2 \left( 3 \alpha_l^{} + 3 \alpha_{\nu}^{} + k_1^{} \right) I_{25}^{} - 6 \left[ I_1^{} \left( I_2^{} I_{17}^{} - I_{24}^{} \right) - I_2^{} \left( I_3^{} I_{13}^{} + 2 I_{22}^{} \right) + I_3^{} I_{19}^{} - I_4^{} I_{17}^{} \right. \nonumber\\
&& \left. + I_7^{} I_{13}^{} + 3 I_{30}^{} \right] / v^2_{} \;,\\
\frac{dI_{26}^{}}{dt} &=& 4 \left( \alpha_l^{} + 2 \alpha_{\nu}^{} - 2 k_1^{} \right) I_{26}^{} - 12 \left[ I_1^{} \left( I_5^{} I_{13}^{} - I_2^{} I_{19}^{} \right) + I_2^{} I_{24}^{} + I_4^{} I_{19}^{} - I_5^{} I_{17}^{} - I_9^{} I_{13}^{} \right. \nonumber\\
&& \left. - 2 I_{31}^{} \right] / v^2_{} \;,\\
\frac{{\rm d}I_{27}^{}}{{\rm d}t} &=& 2 \left( 5 \alpha_l^{} + 2 \alpha_{\nu}^{} + 2 k_1^{} \right) I_{27}^{} - 3 \left( 2 I_{32}^{} + k_2^{} I_{22}^{} + k_3^{} I_{17}^{} \right) / v^2_{} \;,\\
\frac{{\rm d}I_{29}^{}}{{\rm d}t} &=& 2 \left( 4 \alpha_l^{} + 3 \alpha_{\nu}^{} + k_1^{} \right) I_{29}^{} - 3 \left[ 2 k_2^{} I_1^{} \left( I_{19}^{} - I_2^{} I_{13}^{} \right) + 2 k_2^{} I_2^{} I_{17}^{} + I_4^{} \left( k_2 I_{13}^{} + 2 I_{22}^{} \right) \right. \nonumber\\
&&\left. + 2 I_8^{} I_{17}^{} - 2 k_2^{} I_{24}^{} \right] / v^2_{} \;,\\
\frac{{\rm d}I_{30}^{}}{{\rm d}t} &=& 2 \left( 4 \alpha_l^{} + 3 \alpha_{\nu}^{} + k_1^{} \right) I_{30}^{} - 3 \left[ 2 I_1^{} \left( I_2^{} I_3^{} I_{13}^{} - I_3^{} I_{19}^{} - I_4^{} I_{17}^{} + I_{29}^{} \right) - I_2^{} \left( 2 I_3^{} I_{17}^{}  \right. \right. \nonumber\\
&&\left. \left. + k_3^{} I_{13}^{} \right) + 2 \left( I_3^{} I_{24}^{} + I_8^{} I_{17}^{} - I_{11}^{} I_{13}^{} + k_3^{} I_{19}^{} + 3 I_4^{} I_{22}^{} - 2 I_{33}^{} \right) \right] / v^2_{} \;,\\
\frac{{\rm d}I_{31}^{}}{{\rm d}t} &=& 2 \left( 3 \alpha_l^{} + 4 \alpha_{\nu}^{} + k_1^{} \right) I_{31}^{} - \left\{ 3 I_1^2 ( - I_2^{} I_{19}^{} + I_5^{} I_{13}^{} + 3 I_{26}^{})+ I_1^{} \left(-3 I_2^2 I_{17}^{} +4 I_2^{} I_{24}^{} \right.\right.\nonumber\\
&& \left. - I_5^{} I_{17}^{} - 2I_{31}^{}\right)+ 3 I_2^2 \left(I_3^{} I_{13}^{} + I_{22}^{} \right) - I_2^{} \left(I_3^{} I_{19}^{} - 12 I_4^{} I_{17}^{} + 6 I_7^{} I_{13}^{}+ 6 I_{29}^{}+ 2 I_{30}^{}\right) \nonumber\\
&&  - I_3^{}\left(2I_5^{} I_{13}^{}+7I_{26}^{}\right) + 2 I_4^{} (-2I_{24}^{} + I_{25}^{})- I_5^{} I_{22}^{} + 4 I_7^{} I_{19}^{} - 14 I_9^{} I_{17}^{}  +2I_{13}^{} I_{14}^{} \nonumber\\
&&  \left. + 12 I_{34}^{} \right\}/v^2_{}\;,\\
\frac{{\rm d}I_{32}^{}}{{\rm d}t} &=& 4 \left( 3 \alpha_l^{} + \alpha_{\nu}^{} + k_1^{} \right) I_{32}^{} - 6 k_3^{} I_{22}^{} / v^2_{} \;,\\
\frac{{\rm d}I_{33}^{}}{{\rm d}t} &=& 2 \left( 5 \alpha_{l}^{}+3\alpha_{\nu}^{}\right)I_{33}^{}-\left\{I_1^3\left(I_2^{}I_{17}^{}-2I_{24}^{}\right)-2I_1^2\left[I_2^{}\left(I_3^{}I_{13}^{}+I_{22}^{} \right)+I_7^{}I_{13}^{}-I_3^{}I_{19}^{}\right.\right.\nonumber\\
&&\left.\left. - 2 I_4^{} I_{17}^{} + I_{29}^{} \right] + I_1^{} \left[ I_2^{} \left( I_3^{} I_{17}^{} - 6 k_3^{} I_{13}^{} \right) + k_2^{} I_{24}^{} + 6 \left( I_{11}^{} I_{13}^{} + k_3^{} I_{19}^{} - I_8^{} I_{17}^{} \right) \right] \right. \nonumber \\
&&\left. + I_2^{} \left[ 2 \left( I_3^{} I_{22}^{} + I_{32}^{} \right) + 6 k_3^{} I_{17}^{} - k_2^{} \left( I_3^{} I_{13}^{} + I_{22}^{} \right) \right] + I_3^{} \left[ k_2^{} I_{19}^{} + 2 \left( I_7^{} I_{13}^{} - I_4^{} I_{17}^{} \right) \right] \right. \nonumber\\
&& + I_4^{} \left( 3 k_3^{} I_{13}^{} - 2 k_2^{} I_{17} - 4 I_{27}^{} \right) + 2 I_7^{} \left( k_2^{} I_{13}^{} - I_{22}^{} \right) + 2 I_8^{} \left( I_{21}^{} + 3 I_{22}^{} \right) + 2 \left( I_{11}^{} I_{17}^{} \right. \nonumber\\
&& \left.\left. - 2 I_{13}^{} I_{15}^{} \right) + 3 k_2^{} I_{30}^{} - 3k_3^{} \left( 2 I_{24}^{} + I_{25}^{} \right)
\right\}/v^2\;,\\
\label{eq:I34}
\frac{{\rm d}I_{34}^{}}{{\rm d}t} &=& 8 \left( \alpha_l^{} + \alpha_\nu^{} \right) I_{34}^{} - \left\{ 3 I_1^3 \left( 6 I_{26}^{} - p_2^{} I_{13}^{} \right) + 2 I_1^2 \left[ 3 I_2^2 I_{17}^{} - I_2^{} \left(3 I_4^{} I_{13}^{} + 4 I_{24}^{}\right) - 3 I_4^{} I_{19}^{}\right.\right.\nonumber\\
&& \left. + I_5^{} I_{17}^{}+ 9 I_9^{} I_{13}^{}-7I_{31}^{} \right] - I_1^{} \left[-I_2^2 I_3^{} I_{13}^{} +2I_2^{} \left(4 I_4^{} I_{17}^{} - 2 I_7^{} I_{13}^{} + 2 I_{29}^{} - 5 I_{30}^{} \right)  \right.\nonumber\\
&&\left.+ I_3^{} \left( I_5^{} I_{13}^{} + 20 I_{26}^{} \right) - 2 I_4^{} \left(17 I_{24}^{}-5 I_{25}^{}\right) + 10 I_5^{} I_{22}^{} - 24 I_7^{} I_{19}^{} + 26 I_9^{} I_{17}^{} + 28 I_{13}^{} I_{14}^{} \right. \nonumber\\
&& \left. + 8I_{34}^{} \right]- 4 I_2^2 I_3^{} I_{17}^{} + 2 I_2^{} \left[I_3^{}\left( 3 I_4^{} I_{13}^{} +7 I_{24}^{} \right)- 2 I_4^{} I_{22}^{} - 3 I_6^{} I_{19}^{} + 4 \left( I_7^{} + I_8^{} \right) I_{17}^{} \right. \nonumber\\
&& \left. -4 I_{11}^{} I_{13}^{} + 4I_{33}^{}\right] -2I_3^{} \left(6 I_4^{} I_{19}^{} + 5 I_5^{} I_{17}^{} + I_{31}^{}\right) - 4 I_4^2 I_{17}^{} + 2 I_4^{} \left(14 I_{29}^{} - 5 I_{30}^{}\right) \nonumber\\
&&  + 6 I_5^{} I_6^{} I_{13}^{} + 18 I_6^{} I_{26}^{} - 44 I_7^{} I_{24}^{} + 2 I_8^{} \left(2 I_{24} + 9 I_{25}^{}\right)+26 I_9^{} I_{22}^{} + 8 I_{11}^{} I_{19}^{} - 4 I_{12}^{} I_{17}^{}  \nonumber\\
&& \left. + 4 I_{13}^{} I_{16}^{} + 36 I_{14}^{} I_{17}^{}\right\}/\left(2v^2_{}\right) \;,
\end{eqnarray}}
where the CP-even and CP-odd flavor invariants are entangled. This means if one of the three CP phases is nontrivial at the beginning, then other CP phases will be generated by radiative corrections as the CP-odd flavor invariants evolve~\cite{Yu2020PRD}.

The right-hand sides of the RGEs of all 34 flavor invariants have been written as polynomials of the 34 flavor invariants themselves. This confirms our previous claim that there are totally 34, instead of 33, flavor invariants in the generating set.

\subsection{Numerical Solutions}
\begin{figure}[H]
	\centering
	\includegraphics[scale=0.65]{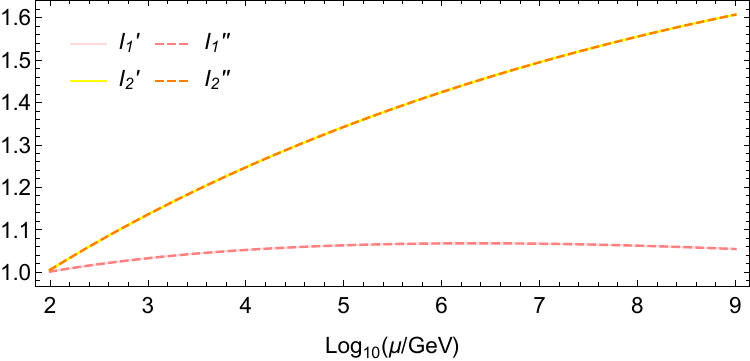}\qquad
	\vspace{0.7cm}
	\includegraphics[scale=0.65]{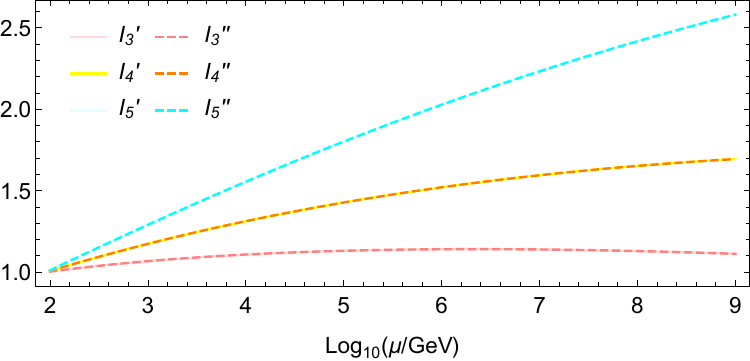}
	\vspace{0.7cm}
	\includegraphics[scale=0.65]{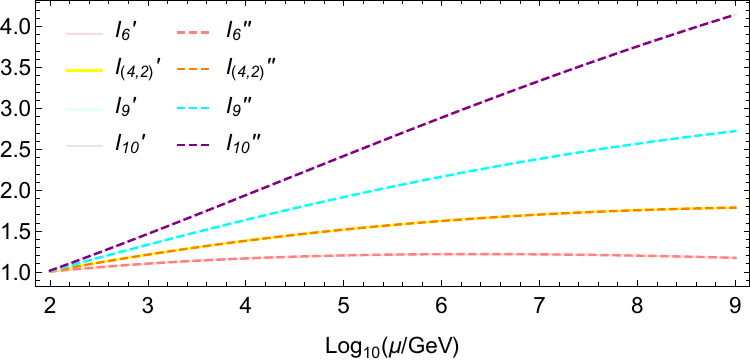}\qquad
	\includegraphics[scale=0.65]{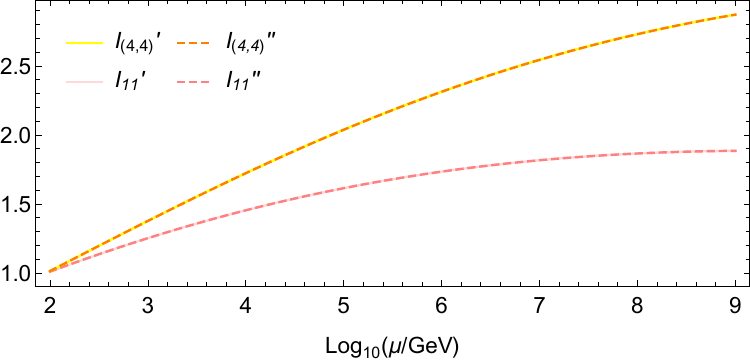}
	\vspace{0.7cm}
	\includegraphics[scale=0.65]{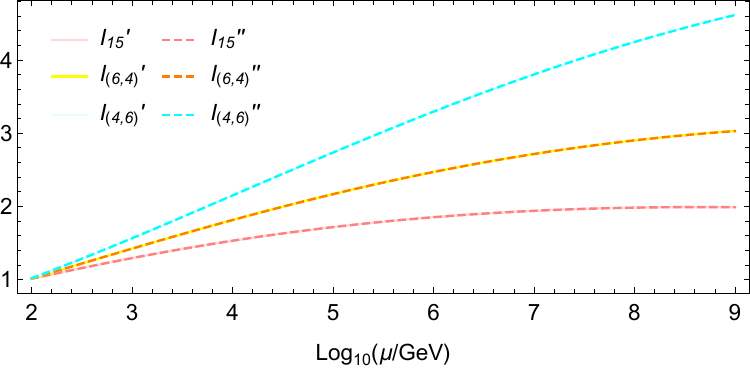}\qquad
	\includegraphics[scale=0.65]{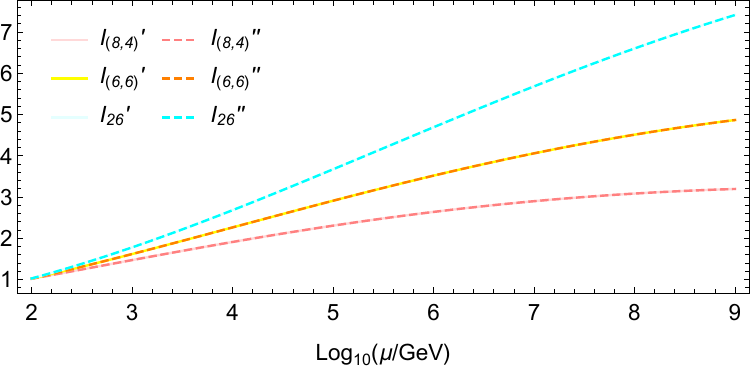}
	\vspace{0.5cm}
	\includegraphics[scale=0.65]{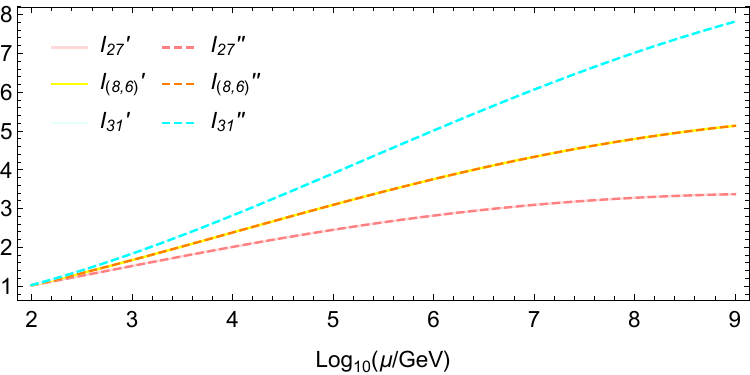}\qquad
	\includegraphics[scale=0.65]{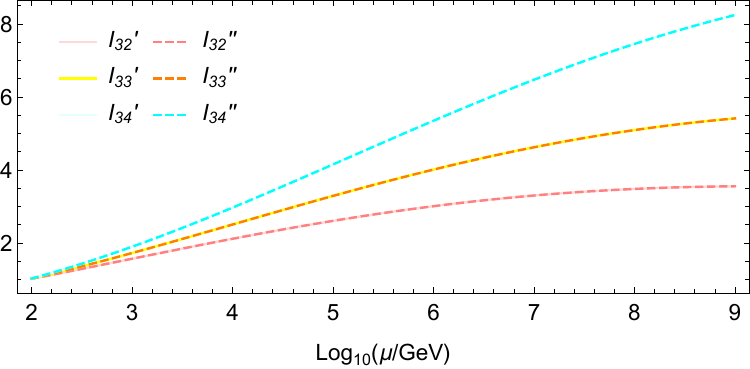}
	\caption{The evolution of the 34 basic flavor invariants in the generating set from  $\mu=10_{}^2\,{\rm GeV}$ to $\mu=10_{}^9\,{\rm GeV}$. Note that all the curves in the figure correspond to the invariants normalized by their initial values, i.e., $I_j^{\prime(\prime \prime)}(\mu)\equiv I_j(\mu)/I_j(\mu_0^{})$ (for $j=1,2,...,34$). Note also that invariants at the same degree have been represented by one curve and labeled by their common degrees since their running behaviors are quite similar. For example, $I_{(4,2)}^{\prime (\prime \prime)}$ stands for $I_7^{\prime(\prime \prime)}$ and  $I_8^{\prime(\prime \prime)}$, $I_{(4,4)}^{\prime (\prime \prime)}$ for $I_{12}^{\prime(\prime \prime)}$, $I_{13}^{\prime(\prime \prime)}$ and  $I_{14}^{\prime(\prime \prime)}$, and so on. $I_j^{\prime}(\mu)$ are calculated by directly solving the RGEs of basic invariants in Eqs.~(\ref{eq:I1})-(\ref{eq:I34}) numerically. As a comparison and cross-check, $I_j^{\prime\prime}(\mu)$ are derived by firstly solving the RGEs of SM parameters numerically and then substituting the running SM parameters into the definitions of the basic flavor invariants.}
	\label{fig:fig1}
\end{figure}

The RGEs of the 34 basic flavor invariants in Eqs.~(\ref{eq:I1})-(\ref{eq:I34}) are impossible to solve analytically, so we proceed with numerical solutions. For this purpose, we have to specify the values of the invariants at the initial energy scale. At the energy scale $\mu_0^{}=M_Z^{}=91.19\,{\rm GeV}$, the input values of all relevant SM parameters are summarized in Table~\ref{table:initial values}~\cite{PDG2020,NuFit2020,HZ2020}, from which one can compute the initial values of all the basic invariants. Then we can solve Eqs.~(\ref{eq:I1})-(\ref{eq:I34}) numerically and the results are shown in Fig.~\ref{fig:fig1}.

To make a cross-check, we also solve the RGEs of the SM physical parameters directly with their initial values given in Table~\ref{table:initial values} and substitute the results into the definition of the basic flavor invariants to obtain their values at any running scale.

\begin{table}[t!]
	\centering
	\begin{tabular}{c|c|c|c|c|c}
		\hline \hline
		$m_u^{}/{\rm MeV}$ & 1.230 & $m_e^{}/{\rm MeV}$ & 0.4831 & $g_1^{}$ & 0.4612 \\
		\hline
		$m_d^{}/{\rm MeV}$ & 2.670 & $m_\mu^{}/{\rm GeV}$ & 0.1018 & $g_2^{}$ & 0.6510 \\
		\hline
		$m_s^{}/{\rm MeV}$ & 53.16 & $m_\tau^{}/{\rm GeV}$ & 1.729 & $g_3^{}$ & 1.210 \\
		\hline
		$m_c^{}/{\rm GeV}$ & 0.620 & $m_1^{}/{\rm eV}$ & 0.05000 & $\lambda$ & 0.1395 \\
		\hline
		$m_b^{}/{\rm GeV}$ & 2.839 & $m_2^{}/{\rm eV}$ & 0.05074 & $\delta_{}^{\rm q}/{\rm rad}$ & 1.196 \\
		\hline
		$m_t^{}/{\rm GeV}$ & 168.3 & $m_3^{}/{\rm eV}$ & 0.07081 & $\delta_{}/{\rm rad}$ & 3.403 \\
		\hline
		$\sin \theta_{12}^{\rm q}$ & 0.2265 &	$\sin \theta_{12}^{}$ & 0.5514 & $\rho/{\rm rad}$ & 0\\
		\hline
		$\sin \theta_{23}^{\rm q}$ & 0.04053 & $\sin \theta_{23}^{}$ & 0.7550 & $\sigma/{\rm rad}$ & 0 \\
		\hline
		$\sin \theta_{13}^{\rm q}$ & 0.003610 & $\sin \theta_{13}^{}$ & 0.1490 &    &   \\
		\hline
		\hline
	\end{tabular}
	\vspace{0.5cm}
	\caption{The input values of the relevant SM physical parameters at the initial energy scale $\mu_0^{}=M_Z^{}=91.19\,{\rm GeV}$. The flavor mixing parameters $\{\theta^{\rm q}_{12}, \theta^{\rm q}_{13}, \theta^{\rm q}_{23}, \delta^{\rm q}\}$ refer to those in the quark sector~\cite{PDG2020}, while $\{\theta^{}_{12}, \theta^{}_{13}, \theta^{}_{23}, \delta, \rho, \sigma\}$ to those in the leptonic sector~\cite{NuFit2020}.
		The initial values of the running quark and charged-lepton masses, the gauge couplings $g_i^{}$ (for $i=1,2,3$) and the quartic Higgs coupling $\lambda$ come from Ref.~\cite{HZ2020}. For neutrino masses, we assume the normal mass ordering and take $m_1^{}=0.05\,{\rm eV}$. Furthermore, the Majorana CP phases $\rho$ and $\sigma$ are set to be zero.}
	\label{table:initial values}
\end{table}

At the one-loop level, the evolution of the physical parameters in the SM are governed by~\cite{Xing2020}
\begin{eqnarray}
\label{eq:RGEg1}
\frac{{\rm d}g_1^{}}{{\rm d} t} &=& \frac{41}{10} g_1^3\;,\qquad
\frac{{\rm d}g_2^{}}{{\rm d} t} = -\frac{19}{6} g_2^3\;,\qquad
\frac{{\rm d}g_3^{}}{{\rm d} t} = -7 g_3^3\;,\nonumber\\
\frac{{\rm d}Y_{\rm u}^{}}{{\rm d} t} &=&\left[\alpha_{\rm u}^{} + \frac{3}{2} \left(Y_{\rm u}^{} Y_{\rm u}^\dagger\right) - \frac{3}{2} \left(Y_{\rm d}^{} Y_{\rm d}^\dagger\right)\right] Y_{\rm u}^{}\;,\nonumber\\
\frac{{\rm d}Y_{\rm d}^{}}{{\rm d} t}&=&\left[\alpha_{\rm d}^{} - \frac{3}{2} \left(Y_{\rm u}^{} Y_{\rm u}^\dagger\right) + \frac{3}{2} \left(Y_{\rm d}^{} Y_{\rm d}^\dagger\right)\right] Y_{\rm d}^{}\;,\nonumber\\
\frac{{\rm d} \lambda}{{\rm d} t} &=& 24\lambda_{}^2 - 3\lambda\left(\frac{3}{5} g_1^2 + 3g_2^2\right) + \frac{3}{8} \left(\frac{3}{5}g_1^2 + g_2^2\right)^2_{} +\frac{3}{4}g_2^4\nonumber\\
	&&+ 4 \lambda {\rm Tr}\left[3 \left(Y_{\rm u}^{} Y_{\rm u}^\dagger\right) + 3\left(Y_{\rm d}^{} Y_{\rm d}^\dagger\right) + \left(Y_l^{} Y_l^\dagger\right)\right]\nonumber\\
	&&- 2{\rm Tr}\left[3 \left(Y_{\rm u}^{} Y_{\rm u}^\dagger\right)^2_{} + 3\left(Y_{\rm d}^{} Y_{\rm d}^\dagger\right)^2_{} + \left(Y_l^{} Y_l^\dagger\right)^2_{}\right]\;.
\end{eqnarray}
In addition, we have the RGEs of $M_\nu^{}$ and $M_l^{}$ in Eqs.~(\ref{eq:RGEMnu})-(\ref{eq:RGEMl}). In the SM, the relevant coefficients are given by
\begin{eqnarray}
	\alpha_{\rm u}^{} &=& -\frac{17}{20} g_1^2 -\frac{9}{4}g_2^2 - 8 g_3^2 + {\rm Tr} \left[3\left(Y_{\rm u}^{} Y_{\rm u}^\dagger\right) + 3\left(Y_{\rm d}^{} Y_{\rm d}^\dagger\right) +\left(Y_l^{} Y_l^\dagger\right)\right]\nonumber,\\
	\alpha_{\rm d}^{} &=& -\frac{1}{4} g_1^2 -\frac{9}{4}g_2^2 - 8 g_3^2 + {\rm Tr} \left[3\left(Y_{\rm u}^{} Y_{\rm u}^\dagger\right) + 3\left(Y_{\rm d}^{} Y_{\rm d}^\dagger\right) +\left(Y_l^{} Y_l^\dagger\right)\right].
\end{eqnarray}	

In Fig.~\ref{fig:fig1}, we have presented the numerical solutions to the flavor invariants, running from $\mu = 10_{}^2\,{\rm GeV}$ to $\mu=10_{}^{9}\,{\rm GeV}$. The latter energy scale is chosen for the reason that the quartic Higgs coupling $\lambda$ becomes negative above $\mu = 10^9~{\rm GeV}$, resulting in an unstable electroweak vacuum~\cite{Elias-Miro:2011sqh, Xing:2011aa, Degrassi:2012ry}. The curves refer to normalized flavor invariants $I_j^{\prime}(\mu) \equiv I^{}_j(\mu)/I^{}_j(\mu_0^{})$ for $j=1,2,...,34$. We obtain $I_j^{\prime}(\mu)$ by solving Eqs.~(\ref{eq:I1})-(\ref{eq:I34}) numerically. On the other hand, we also derive $I^{}_j(\mu)/I^{}_j(\mu_0^{})$ by first solving Eqs.~(\ref{eq:RGEMnu})-(\ref{eq:RGEMl}) and Eq.~(\ref{eq:RGEg1}) with the initial values given in Table~\ref{table:initial values} numerically and then substituting the running SM parameters into the flavor invariants. The corresponding results are denoted as $I^{\prime\prime}_j(\mu)$ in Fig.~\ref{fig:fig1}. As one can observe from Fig.~\ref{fig:fig1}, the results obtained by these two different methods match perfectly with each other.

Moreover, we notice that the running behaviors of the invariants at the same degree are very close to each other and no visible differences can be detected from Fig.~\ref{fig:fig1}. Therefore, the flavor invariants at the same degree will be described by one curve and labeled by the common degree. For instance, $I_7^{\prime(\prime \prime)}$ and $I_8^{\prime(\prime \prime)}$ are labeled as $I_{(4,2)}^{\prime (\prime \prime)}$, while $I_{12}^{\prime(\prime \prime)}$, $I_{13}^{\prime(\prime \prime)}$ and  $I_{14}^{\prime(\prime \prime)}$ as $I_{(4,4)}^{\prime (\prime \prime)}$. This feature can be understood as follows. The running behaviors of the invariants are mainly governed by the running masses of charged leptons and neutrinos,\footnote{The running effects of the flavor mixing angles and CP phases are much smaller than those of the lepton masses. More explicitly, from $\mu = 10_{}^2\,{\rm GeV}$ to $\mu=10_{}^{9}\,{\rm GeV}$, the sizes of charged-lepton masses and those of neutrino masses could be changed by 1\% and 10\%, respectively, while those of the mixing angles and CP phases by less than 1\textperthousand. Furthermore, the difference of the running behaviors between different generations can also be neglected.} while the invariants at the same degree are composed of exactly the same power of $M_l^{}$ and $M_\nu^{}$, so their running behaviors are quite similar. Furthermore, we have collected deliberately the curves of the invariants with the same total degree in one plot in order to compare their running effects. For example, the degrees of $I_3^{}$, $I_4^{}$ and $I_5^{}$ are respectively $(4,0)$, $(2,2)$ and $(0,4)$ with the same total degree of 4. From the right panel in the first row of Fig.~\ref{fig:fig1} one can easily see the obvious difference in their running behaviors, which is a direct consequence of the fact that the running effects of neutrino masses are larger than those of the charged-lepton masses.

\section{Summary}
In this article, we have performed a systematic investigation on the flavor invariants in the leptonic sector with massive Majorana neutrinos based on the invariant theory. The physical observables should not depend on the choice of the basis, hence it is useful to study quantities that are invariant under the flavor basis transformations. All the flavor invariants compose a ring in the algebra and for most flavor symmetry groups the ring can be generated by a finite number of basic invariants. Once all the basic generators are found, any invariants in the ring can be written as the polynomials of these basic invariants.

The Hilbert series and plethystic logarithm are powerful tools to explore the algebraic structure of the invariant ring. After calculating the Hilbert series and the plethystic logarithm by using the Molien-Weyl formula, we can obtain the important information about the numbers and degrees of the basic invariants, as well as the relations among the basic invariants, i.e., syzygies. With the help of Hilbert series and plethystic logarithm, one can explicitly construct all the basic invariants. We also propose a practically useful and efficient method to decompose any invariants into the polynomials of the basic invariants and construct all the syzygies at a certain degree. A detailed description of this method is given in Appendix~\ref{appendix:syzygy}.

In the case of two-generation leptons, the ring of flavor invariants is a complete intersection and there are totally 7 basic invariants, where 6 of them are CP-even and 1 is CP-odd. The construction of all the basic invariants are summarized in Table~\ref{table:2 generations} and their RGEs are calculated in Eqs.~(\ref{eq:J1})-(\ref{eq:J7}). In the case of three-generation leptons, the invariant ring is a non-complete intersection whose algebraic structure is much more complicated. Different from the conclusion drawn by reading the plethystic logarithm directly, we find there are totally 34 rather than 33 basic generators, with the addition of one more basic invariant and one more syzygy at the same degree. All the basic invariants are explicitly constructed and collected in Table~\ref{table:3 generations} and their RGEs are calculated in Eqs.~(\ref{eq:I1})-(\ref{eq:I34}). Numerical solutions to the RGEs of basic flavor invariants are shown in Fig.~\ref{fig:fig1}. We have also demonstrated how to extract all the physical observables analytically from the basic invariants, which provides a convenient and basis-independent way to describe the running behaviors of the physical observables.

The invariant theory has proved to be extremely useful in flavor physics. Although only the low-energy effective theory of massive Majorana neutrinos has been considered in the present work, we can easily extend the approach to the realistic seesaw models of neutrino masses. It will be interesting to investigate the flavor invariants in a complete model, and to establish the relationship between the flavor invariants at the low-energy scale and those at high-energy scales. We hope to come back to these issues in the near future.

\label{sec:summary}

\section*{Acknowledgements}
The authors would like to thank Xin Wang, Di Zhang and Jianlong Lu for useful discussions. This work was supported in part by the National Natural Science Foundation of China under
grant No. 11775232 and No. 11835013, by the Key Research Program of the Chinese Academy of
Sciences under grant No. XDPB15, and by the CAS Center for Excellence in Particle Physics.

\begin{appendix}

\section{Cayley-Hamilton Theorem}
\label{appendix:CH theorem}
For any $n \times n$ matrix $A$, its characteristic polynomial is defined as
\begin{eqnarray}
f(\lambda)\equiv {\rm Det}\left(\lambda {\mathbb I}_{n}-A\right)=\lambda^n_{}+a_{n-1}^{}\lambda^{n-1}_{}+...+a_{1}^{}\lambda +a_{0}^{}\;,
\end{eqnarray}
where ${\mathbb I}_n^{}$ is the $n\times n$ identity matrix and $a_{0}^{}, a_{1}^{},..., a_{n-1}^{}$ are some coefficients determined by the matrix $A$ itself. Then the Cayley-Hamilton theorem says
\begin{eqnarray}
\label{eq:CH therorem general}
f(A)=A^n_{}+a_{n-1}^{}A^{n-1}_{}+...+a_{1}^{}A +a_{0}^{}=0\;.
\end{eqnarray}
This is a powerful theorem since it allows $A^n_{}$ to be articulated as a linear combination of the lower matrix powers of $A$. For the special cases of $n=2$ and $n=3$, we have
\begin{eqnarray}
\label{eq:CH therorem 2 generations}
A^2_{}={\rm Tr}(A)A-\frac{1}{2}\left[{\rm Tr}(A)^2_{}-{\rm Tr}(A^2_{})\right]{\mathbb I}_{2}^{}\;,
\end{eqnarray}
and
\begin{eqnarray}
\label{eq:CH therorem 3 generations}
A^3_{}={\rm Tr}(A)A^2_{}-\frac{1}{2}\left[{\rm Tr}(A)^2_{}-{\rm Tr}(A^2_{})\right]A+\frac{1}{6}\left[{\rm Tr}(A)^3_{}-3{\rm Tr}(A){\rm Tr}(A^2_{})+2{\rm Tr}(A^3_{}) \right]{\mathbb I}_3^{}\;.
\end{eqnarray}

\section{Invariant Theory and Hilbert Series}
\label{appendix:Invariant theory}
In this appendix, we give a concise introduction to the invariant theory and Hilbert series (HS) (also known as Poincar{\'e} series or Molien series in some mathematical literature), mainly focusing on their applications in physics. Furthermore, we will introduce the Molien-Weyl (MW) formula, which provides a general and systematic method to calculate HS.

Consider a theory containing $n$ parameters $\vec{x}\equiv\left(x_1^{},x_2^{},...,x_n^{}\right)^{\rm T}$ (not all parameters have to be physical observables) and a symmetry group $G$. For a specific representation $R$ of $G$, any element $g\in G$ could act on the vector space spanned by $\left\{x_1^{},x_2^{},...,x_n^{}\right\}$: $\vec{x}\rightarrow R(g) \vec{x}$. Invariants are such quantities that are polynomial functions of $\vec{x}$ and being invariant under the group action, i.e.,
\begin{eqnarray}
I(\vec{x})=I\left(R(g)\vec{x}\right)\;,\quad
\forall g \in G\;.
\end{eqnarray}

Notice that the addition and multiplication of any two invariants are also invariants, so all the invariants form a \emph{ring} in the sense of algebraic structure, denoted by ${\mathbb C}\left[I_1^{},I_2^{},...\right]$, meaning that the numerical factors in front of the invariants are restricted to be in the complex field ${\mathbb C}$. It can be proved that if $G$ is a reductive group,\footnote{A reductive group is a group whose every representation is completely reducible. Note that there are many equivalent definitions of reductive group. See, e.g., Ref.~\cite{DK2015}, for other more modern but abstract definitions. For the practical purpose, almost all the commonly used groups in physics, including all the semi-simple Lie groups and finite groups, are reductive.} then the ring can be finitely generated\cite{Sturmfels2008, DK2015}. To be more precise, the word ``finitely generated" means that there are a finite number of basic invariants, say $\left\{I_1^{},I_2^{},...,I_m^{}\right\}$, such that any invariant $I_{}^{\prime}$ in the ring can be expressed as the polynomials of these basic invariants
\begin{eqnarray}
I_{}^{\prime}=P_{}^{\prime}\left(I_1^{},I_2^{},...,I_m^{}\right)\;,
\end{eqnarray}
where $P_{}^{\prime}$ denotes a polynomial function. These basic invariants are generators of the ring and $\left\{I_1^{},I_2^{},...,I_m^{}\right\}$ is the generating set. Then we denote the ring as ${\mathbb C}\left[I_1^{},I_2^{},...I_m^{}\right]$ and it is actually a polynomial ring since any element in the ring can be written as the polynomials of the generators.

It is crucial to keep in mind that the generators may \emph{not} be algebraically independent~\cite{Trautner2018, Trautner2020}. This is very different from the case of vector space, where all the basic vectors are also (linearly) independent. Here ``algebraic independence" of a set of invariants, say $\left\{I_1^{}, I_2^{},..., I_r^{}\right\}$, means that there does not exist a polynomial $P$ such that $P\left(I_1^{}, I_2^{},..., I_r^{}\right)=0$, otherwise the set of invariants are algebraically dependent. The maximal number of algebraically-independent invariants, say $r$, is called the Krull dimension of the ring. A significant result which is not difficult to prove is that the Krull dimension of the ring equals the number of physical observables in the theory.

In general, the Krull dimension of the ring is always less than or equal to the dimension of the generating set, i.e.,
\begin{eqnarray}
r \leq m\;,
\end{eqnarray}
and
\begin{itemize}
\item if $r=m$, the ring is free and all the generators of the ring are algebraically independent;
\item if $r<m$, then there exist nontrivial algebraic relations among the generators. These relations are called {\it syzygies} in the mathematical literature.
\end{itemize}

If the number of the generators minus the number of the syzygies is equal to the Krull dimension, the ring is called a complete intersection, otherwise it is a non-complete intersection. We will see later that the algebraic structures are very different for complete and non-complete intersection rings. Given a physical theory and a symmetry group $G$, the three fundamental problems in the invariant theory can be summarized as~\cite{Sturmfels2008}
\begin{itemize}
\item construct all the basic invariants;
\item for any given invariant, find out an algorithm to decompose it into the polynomials of the basic invariants;
\item find out all the syzygies among the basic invariants.
\end{itemize}

These problems have not yet been solved completely in the mathematical literature for the most general case. However, for our cases, i.e., the ring constructed from the flavor invariants in the leptonic sector with massive Majorana neutrinos, we have found indeed effective methods to solve all the above issues.  Before diving into the concrete physical models, let us introduce two very powerful tools in the invariant theory: HS and plethystic logarithm (PL), which provide a convenient way to count the number of basic invariants, as well as the relations among them (i.e., the syzygies).

\subsection{Hilbert Series}
\label{subapp:HS intro}
HS plays the role of the generating function of invariants and is defined as\footnote{The form of HS may remind one of the partition function in statistical physics. Actually, HS is also called partition function in some literature since it encodes almost all information about invariants, just like the partition function in statistical physics, from which we can calculate almost all thermodynamic functions.}
\begin{eqnarray}
\label{eq:HS ungraded def}
{\mathscr H}(q) \equiv \sum_{k=0}^{\infty}c_k^{} q_{}^k\;,
\end{eqnarray}
where $c_k^{}$ (with $c_0^{}\equiv1$) denotes the number of (linearly-) independent invariants at degree $k$  while $q$ is an arbitrary complex variable satisfying $\left|q\right|<1$ and labels the degree of the building blocks of the invariants. A general property is that HS can always be written as the ratio of two polynomial functions~\cite{DK2015}
\begin{eqnarray}
{\mathscr H}(q)=\frac{{\mathscr N(q)}}{{\mathscr D(q)}}\;,
\end{eqnarray}
where ${\mathscr N}(q)$ and ${\mathscr D}(q)$ are both polynomials and satisfy the following properties:
\begin{itemize}
\item The numerator has the palindromic structure, i.e., if we write the numerator as
\begin{eqnarray}
\label{eq:HS property palindromic}
{\mathscr N}(q)=1+a_1^{}q+...+a_{l-1} q_{}^{l-1}+q_{}^l\;,
\end{eqnarray}
then we have $a_k^{}=a_{l-k}^{}$. Especially, ${\mathscr N}(q)=1$ corresponds to the free ring and there are no syzygies at all.
\item The denominator has the general form of
\begin{eqnarray}
\label{eq:HS property Euler form}
{\mathscr D}(q)=\prod_{k=1}^r (1-q^{d_k})\;.
\end{eqnarray}
A highly nontrivial result is that the denominator of HS encodes the information of all algebraically-independent invariants~\cite{Sturmfels2008, DK2015}. The total number of the factors $r$ equals the Krull dimension of the ring, or the maximal number of algebraically-independent invariants, while the power index $d_k^{}$ indicates the degree of each invariant.

\end{itemize}

Let us take a toy model as an example. Consider only one-generation leptons, then the only flavor transformation in the Yukawa sector is the ${\rm U}(1)$ transformation
\begin{eqnarray}
l_{\rm L}^{}\rightarrow e^{{\rm i}\alpha}l_{\rm L}^{}\;,\quad
\nu_{\rm L}^{}\rightarrow e^{{\rm i}\alpha}\nu_{\rm L}^{}\;,\quad
l_{\rm R}^{}\rightarrow e^{{\rm i}\beta}l_{\rm R}^{}\;,
\end{eqnarray}
which corresponds to the transformation rules of masses
\begin{eqnarray}
\label{eq:Toy model transformation rule}
M_l^{}\rightarrow e^{{\rm i}(\alpha-\beta)}M_l^{}\;,\quad
M_\nu^{}\rightarrow e^{2{\rm i}\alpha}M_\nu^{}\;,
\end{eqnarray}
with $\alpha$ and $\beta$ being arbitrary phases. The basic flavor invariants are easy to write down\footnote{Note that in the one-generation case, $M_l^{}$ and $M_\nu^{}$ are complex numbers rather than matrices.}
\begin{eqnarray}
\label{eq:one generation basic invariants}
{\mathscr I}_1^{}=M_l^{}M_l^*\;,\quad
{\mathscr I}_2^{}=M_\nu^{}M_\nu^*\;.\quad
\end{eqnarray}
Labeling the degree of $M_l^{}$ and $M_\nu^{}$ by $q$, then the nonzero coefficients of HS turn out to be
\begin{eqnarray}
c_0^{}&=&1\;,\nonumber\\
c_2^{}&=&2: \left\{{\mathscr I}_1^{}, {\mathscr I}_2^{}\right\}\;,\nonumber\\
c_4^{}&=&3: \left\{{\mathscr I}_1^2,{\mathscr I}_1^{}{\mathscr I}_2^{}, {\mathscr I}_2^2\right\}\;,\nonumber\\
c_6^{}&=&4: \left\{{\mathscr I}_1^3,{\mathscr I}_1^2{\mathscr I}_2, {\mathscr I}_1^{}{\mathscr I}_2^2, {\mathscr I}_2^3\right\}\;,\nonumber\\
&\cdots & \nonumber\\
c_{2k}&=&k+1:\left\{{\mathscr I}_1^k,{\mathscr I}_1^{k-1}{\mathscr I}_2, {\mathscr I}_1^{k-2}{\mathscr I}_2^2,..., {\mathscr I}_2^k\right\}\;,
\end{eqnarray}
which lead to
\begin{eqnarray}
\label{eq:HS ungraded toy model}
{\mathscr H}(q)=\sum_{k=0}^{\infty}(k+1)q_{}^{2k}=\frac{1}{(1-q^2)^2}\;.
\end{eqnarray}
We can see that the denominator of Eq.~(\ref{eq:HS ungraded toy model}) represents two algebraically-independent invariants at degree two, which correspond to ${\mathscr I}_1^{}$ and ${\mathscr I}_2^{}$. The numerator of Eq.~(\ref{eq:HS ungraded toy model}) is trivially one, indicating this is a free ring and there are no syzygies between ${\mathscr I}_1^{}$ and ${\mathscr I}_2^{}$ so they are also the generators of the ring.

The definition of HS in Eq.~(\ref{eq:HS ungraded def}) can be generalized to the multi-graded form. Suppose there are $n$ building blocks in constructing invariants and in order to distinguish them we can label them separately by $(q_1^{},..., q_n^{})$. Let $c_{k_1...k_n}^{}$ (with $c_{0...0}\equiv 1$) denote the number of (linearly-) independent invariants when the $n$ building blocks are at the degree of $(k_1^{},...,k_n^{})$, respectively. Then the multi-graded\footnote{As a comparison, the HS defined in Eq.~(\ref{eq:HS ungraded def}) is called the ungraded HS.} HS is defined as
\begin{eqnarray}
\label{eq:HS multi-graded def}
{\cal H}\left(q_1^{},...,q_n^{}\right)\equiv \sum_{k_1=0}^{\infty}...\sum_{k_n=0}^{\infty}c_{k_1...k_n}q_1^{k_1}...q_n^{k_n}\;.
\end{eqnarray}
Comparing to the ungraded HS, the multi-graded HS does not possess the properties of Eqs.~(\ref{eq:HS property palindromic})-(\ref{eq:HS property Euler form}) in general. However, it owns the huge advantages in constructing invariants since it labels the degree of each building block separately. Clearly the relation between multi-graded and ungraded HS is simply
\begin{eqnarray}
{\mathscr H}(q)={\cal H}(q,q,...,q)\;.
\end{eqnarray}

For the toy model above, let $(q_1^{},q_2^{},q_3^{},q_4^{})$ label the degrees of $(M_l^{},M_l^*,M_\nu^{},M_\nu^*)$ respectively, then the multi-graded HS reads
\begin{eqnarray}
\label{eq:HS multi-graded toy model}
{\cal H}\left(q_1^{},q_2^{},q_3^{},q_4^{}\right)&=&1+\left(q_1^{}q_2^{}+q_3^{}q_4^{}\right)+\left[(q_1^{}q_2^{})^2+(q_1^{} q_2^{})(q_3^{}q_4^{})+(q_3^{}q_4^{})^2 \right]\nonumber\\
&& + \left[(q_1^{}q_2^{})^3+(q_1^{}q_2^{})^2(q_3^{}q_4^{})+(q_1^{}q_2^{})(q_3^{}q_4^{})^2+(q_3^{}q_4^{})^3\right]+\cdots \nonumber\\
&=&1/\left[(1-q_1^{}q_2^{})(1-q_3^{}q_4^{})\right]\;.
\end{eqnarray}

The tool of HS has been widely used in different aspects of particle physics. After firstly introduced in Ref.~\cite{Pouliot1998}
which counts the number of independent gauge-invariant chiral operators in supersymmetric gauge theories, HS has been applied to a wide range of formal theories~\cite{BFHH2007, Romelsberger2005, FHH2007, FHZ2007, Dolan2008, BFHVZ2007, Hanany2007, GHHJM2008, HM2008, HMT2010, Forcella2009, CM2011, HK2014, BP2017, XHM2019}. The construction of all (linearly-) independent gauge- and Lorentz-invariant effective operators at certain mass dimension in the effective field theories is also a standard example in the invariant theory and has been developed a lot by the tool of HS in recent years\cite{LM2015, HLMM2015CMP, LM2016, HLMM2015JHEP, KP2017, HLMM2017, KP2018, ADCP2019, MRW2020, LRSXYZ2020, LM2020, GHLMM2020, KLMT2020, MP2020, Murphy2020,CHMN2021}, which provides a cross-check to the traditional group theory method. In addition, HS also plays a role in some interesting physical processes like the tidal effects~\cite{HH2020, AHH2020}. The calculation of the HS can be automatically carried out with the computer algebra programs such as MATHEMATICA~\cite{Banerjee2020}.

As for our cases in the current paper, we focus on quantities that are invariant under the flavor transformation rather than gauge transformation. Constructing flavor invariants from the couplings in the scalar potential in Two- or Three-Higgs-Doublet Model have been thoroughly studied and the corresponding HS have been calculated~\cite{Bednyakov2018, Trautner2018, Trautner2020, BBST2020, Bento2021}. The HS corresponding to the flavor transformation in the quark and leptonic sector have already been calculated in Refs.~\cite{JM2009, HJMT2011}, while an explicit construction of all basic flavor invariants in the case of three-generation leptons with massive Majorana neutrinos and the calculations of their RGEs are lacking. This is indeed our main purpose in this paper.

\subsection{Plethystic Logarithm}
\label{subapp:PL intro}
For the free ring, it is enough to consider only the denominator of HS since it corresponds to all the algebraically-independent invariants, which are also all generators. However, for more general cases, there are nontrivial relations (syzygies) among generators, which are encoded in the numerator of HS. We need a more convenient method to extract these information from HS and this is what PL does~\cite{BFHH2007}.

Given an arbitrary function $f(x_1^{},...,x_n^{})$, its PL is defined by
\begin{eqnarray}
\label{eq:PL def}
{\rm PL}\left[f(x_1^{},...,x_n^{})\right]\equiv
\sum_{k=1}^{\infty} \frac{\mu(k)}{k}\,{\rm ln}\left[f(x_1^k,...,x_n^k)\right]\;,
\end{eqnarray}
where $\mu(k)$ is the M{\"o}bius function,
\begin{equation}
\mu(k)\equiv
\begin{cases}
0&   \text{$k$ has repeated prime factors} \\
1&   k=1 \\
(-1)^n_{} & \text{$k$ is a product of $n$ distinct primes}
\end{cases}\;.
\end{equation}

The great power of PL is that from it we can read off directly the number and degrees of basic invariants and syzygies~\cite{BFHH2007}: \emph{The leading positive terms of {\rm PL} correspond to the basic invariants while the leading negative terms correspond to the syzygies among these basic invariants.} However, we would like to emphasize here that, for non-complete intersection rings, the number of syzygies may \emph{not} be always read off naively from the leading negative terms and there are actually counter examples~\cite{BFHH2007, JM2009}. This is because the PL for a non-complete intersection is an infinite series with positive and negative terms appearing alternately so the negative terms may count the syzygies from higher-degree invariants.

For the toy model of one-generation leptons above, the PL reads
\begin{eqnarray}
{\rm PL}\left[{\cal H}(q_1^{},q_2^{},q_3^{},q_4^{})\right]=q_1^{}q_2^{}+q_3^{} q_4^{}\;,
\end{eqnarray}
where there are no negative terms, indicating the ring is free. The positive terms give the number and degrees of the basic invariants, i.e., ${\mathscr I}_1^{}$ and ${\mathscr I}_2$ in Eq.~(\ref{eq:one generation basic invariants}). In Sec.~\ref{sec:2g} and Sec.~\ref{sec:3g} we calculate the HS and PL for the cases of two- and three-generation leptons, and they correspond to the scenarios of complete intersection (but not free ring) and non-complete intersection, respectively.

\subsection{Molien-Weyl Formula}
\label{subapp:MW formula}
In Appendix~\ref{subapp:HS intro} we calculate the HS of one-generation case by definition. However, for more complicated cases, it is almost impossible to do this since the number of (linearly-) independent invariants grows very quickly with the degree. Luckily, the MW formula~\cite{Molien1897} provides a systematic method to calculate the HS: As long as the group and the representations of the building blocks are given, the computing of the HS can be reduced to calculating several complex integrals, which can be easily completed via the residue theorem.

For a finite group $G$ and the representation $R$, the HS is given by the following MW formula~\cite{Molien1897}
\begin{eqnarray}
\label{eq:MW formula finite}
{\mathscr H}(q)=\frac{1}{\left|G\right|}\sum_{g\in G}\frac{1}{{\rm Det}\left({\mathbb I}-q R(g)\right)}\;,
\end{eqnarray}
with $\left|G\right|$ denoting the order of $G$, ${\mathbb I}$ denoting the identity matrix and the sum covering the whole group (see page~30 of Ref.~\cite{Sturmfels2008} for a simple and explicit proof). If $G$ is a reductive Lie group, then Eq.~(\ref{eq:MW formula finite}) can be  generalized to
\begin{eqnarray}
\label{eq:MW formula Lie}
{\mathscr H}(q)=\int \left[{\rm d}\mu\right]_G^{}\frac{1}{{\rm Det}\left({\mathbb I}-q R(g)\right)}\;,
\end{eqnarray}
with $\left[{\rm d}\mu\right]_G^{}$ denoting the Haar measure of $G$. If $G$ is compact then this is just a trivial generalization and the integral is performed over $G$. However, if $G$ is non-compact (for example, the Lorentz group), then the integral over it does not make sense. In this case, the integral in Eq.~(\ref{eq:MW formula Lie}) should be performed on the maximum compact subgroup of $G$~\cite{Weyl1926}. Furthermore, if $G$ is also connected, then the integral can be performed on the maximum compact subgroup of its maximum torus, which can be identified with $(S^1_{})^{r_0}_{}$ from the geometrical point of view with $S^{1}_{}$ the unit circle and $r_0^{}$ the rank of $G$~\cite{DK2015}.\footnote{The dimension of the maximum torus of $G$ equals the dimension of the Cartan subalgebra of $G$, which is exactly the rank of $G$.} So for a reductive Lie group with rank $r_0^{}$, the integral region is actually the $r^{}_0$ products of unit circle, which can be calculated via the residues inside the circle.

In order to recast Eq.~(\ref{eq:MW formula Lie}) into a more convenient form for computations, we notice the following identity
\begin{eqnarray}
\left[{\rm Det}\left({\mathbb I}-q R(g) \right)\right]^{-1}_{}=\prod_{j=1}^{d}\left(1-q\lambda_j^{}\right)^{-1}_{}={\rm exp}\left(\sum_{j=1}^{d}\sum_{k=1}^{\infty}\frac{\lambda_j^k q_{}^k}{k} \right)={\rm exp}\left[\sum_{k=1}^{\infty}\frac{q_{}^k\chi_R^{}(\lambda_1^k,...,\lambda_{d}^k)}{k} \right]\;,\nonumber\\
\end{eqnarray}
where $\chi_R^{}\left(\lambda_1,...,\lambda_{d}\right)=\sum_{j=1}^d \lambda_j^{}$ is the character function of $G$ in the representation $R$ with $\lambda_j^{}$ (for $j=1,2,...d$) the eigenvalues of $R(g)$ and $d$ the dimension of the representation. This inspires one to define the plethystic exponential (PE) of an arbitrary function $f(x_1^{},...,x_n^{})$~\cite{BFHH2007}
\begin{eqnarray}
\label{eq:PE def}
{\rm PE}\left[f(x_1^{},...,x_n^{})\right]\equiv {\rm exp}\left[\sum_{k=1}^{\infty}\frac{f\left(x_1^k,...,x_n^k\right)}{k}\right]\;.
\end{eqnarray}
From the definition of PE and PL, it is not difficult to prove that they are inverse operations to each other (see Sec.~2 of Ref.~\cite{FHH2007} for a straightforward proof), i.e.,
\begin{eqnarray}
f(x_1^{},...,x_n^{})={\rm PE}\left[g(x_1^{},...,x_n^{}) \right]\iff
g(x_1^{},...,x_n^{})={\rm PL}\left[f(x_1^{},...,x_n^{}) \right]\;.
\end{eqnarray}

So finally, in the language of PE, for the reductive Lie group $G$ with rank $r_0^{}$, the most general form of the $n$-variable multi-graded HS can be written as
\begin{eqnarray}
\label{eq:Molien-Weyl formula}
{\cal H}\left(q_1^{},...,q_n^{}\right)=\int \left[{\rm d}\mu \right]_G {\rm PE}\left(z_1^{},...,z_{r_0}^{}; q_1^{},...,q_n^{}\right)\;,
\end{eqnarray}
where
\begin{eqnarray}
{\rm PE}\left(z_1^{},...,z_{r_0}^{};q_1^{},...,q_n^{}\right)\equiv\prod_{i=1}^n{\rm PE}\left[\chi_{R_i}^{}\left(z_1^{},...,z_{r_0}^{}\right)q_i^{}\right]={\rm exp}\left[\sum_{k=1}^{\infty}\sum_{i=1}^n \frac{\chi_{R_i}^{}\left(z_1^k,...,z_{r_0}^k\right)q_i^k}{k} \right]\;.\nonumber\\
\end{eqnarray}
Note that we have used $n$ arbitrary complex variables $q_i^{}$ (for $i=1, 2,..., n$)\footnote{In order to guarantee the series in the argument of {\rm PE} to be convergent, we require $\left|q_i^{}\right|<1$.} to label the degree of the $i$-th building block which is assumed to obey the $R_i$ representation of $G$ while $\chi_{R_i}^{}\left(z_1^k,...,z_{r_0}^k\right)$ denotes the character function of $G$ in the $R_i^{}$ representation.\footnote{$z_i^{}$(for $i=1,2,...,r_0^{}$) are coordinates on the maximum torus of $G$.} For the common situation $G={\rm U}(N)$ whose rank is $N$, we have~\cite{CM2011}
\begin{eqnarray}
\int\left[{\rm d}\mu \right]_{{\rm U}(N)}=\frac{1}{N!\left(2\pi {\rm i}\right)^N}\prod_{i=1}^N \oint_{\left|z_i^{}\right|=1}\frac{{\rm d}z_i^{}}{z_i^{}}\left|\Delta(z)\right|^2\;,
\end{eqnarray}
where $\Delta(z)\equiv \prod_{1\leq a < b \leq N}(z_b^{}-z_a^{})$ is the Vandermonde determinant of $N$ integral variables $z_i^{}$ (for $i=1,2,...,N$).
The character functions of the fundamental and anti-fundamental representation of ${\rm U}(N)$ are respectively~\cite{CM2011}
\begin{eqnarray}
\chi_{{\bf N}}^{}=\sum_{i=1}^N z_i^{}\;,\quad
\chi_{{\bf N^*}}^{}=\sum_{i=1}^N z_i^{-1}\;,
\end{eqnarray}
from which one can calculate the character function of any representation via tensor product decomposition. Below we consider the special cases where $N=$1, 2 and 3.

\begin{itemize}
\item $N=1$, the Haar measure is
\begin{eqnarray}
\int \left[d\mu\right]_{{\rm U}(1)}
=\frac{1}{2\pi{\rm i}}\oint_{\left|z\right|=1}\frac{{\rm d}z}{z}\;,
\end{eqnarray}
while the character function of the object carrying charge $Q$ of ${\rm U}(1)$ is
\begin{eqnarray}
\chi_Q^{}=z^Q_{}\;.
\end{eqnarray}

\item $N=2$, the Haar measure is
\begin{eqnarray}
\label{eq:Haar measure U(2)}
\int \left[d\mu\right]_{{\rm U}(2)}
&=&\frac{1}{2!\left(2\pi{\rm i}\right)^2}\oint_{\left|z_1^{}\right|=1}\frac{{\rm d}z_1^{}}{z_1^{}}\oint_{\left|z_2^{}\right|=1}\frac{{\rm d}z_2^{}}{z_2^{}}\prod_{1\leq i<j\leq 2}\left|z_j^{}-z_i^{}\right|^2\nonumber\\
&=&\frac{1}{2\left(2\pi{\rm i}\right)^2}\oint_{\left|z_1^{}\right|=1}\frac{{\rm d}z_1^{}}{z_1^{}}\oint_{\left|z_2^{}\right|=1}\frac{{\rm d}z_2^{}}{z_2^{}}\left(2-\frac{z_2^{}}{z_1^{}}-\frac{z_1^{}}{z_2^{}}\right)\;,
\end{eqnarray}
while the character functions of the fundamental and anti-fundamental representations are
\begin{eqnarray}
\chi_{{\bf 2}^{}}=z_1^{}+z_2^{}\;,\quad
\chi_{{\bf 2^*}^{}}^{}=z_1^{-1}+z_2^{-1}\;.
\end{eqnarray}

\item $N=3$, the Haar measure is
\begin{eqnarray}
\label{eq:Haar measure U(3)}
\int \left[d\mu\right]_{{\rm U}(3)}
&=&\frac{1}{3!\left(2\pi{\rm i}\right)^3}\oint_{\left|z_1^{}\right|=1}\frac{{\rm d}z_1^{}}{z_1^{}}\oint_{\left|z_2^{}\right|=1}\frac{{\rm d}z_2^{}}{z_2^{}}\oint_{\left|z_3^{}\right|=1}\frac{{\rm d}z_3^{}}{z_3^{}}\prod_{1\leq i<j\leq 3}\left|z_j^{}-z_i^{}\right|^2\nonumber\\
&=&\frac{1}{6\left(2\pi{\rm i}\right)^3}\oint_{\left|z_1^{}\right|=1}\frac{{\rm d}z_1^{}}{z_1^{}}\oint_{\left|z_2^{}\right|=1}\frac{{\rm d}z_2^{}}{z_2^{}}\oint_{\left|z_3^{}\right|=1}\frac{{\rm d}z_3^{}}{z_3^{}}\left[-\frac{\left(z_2^{}-z_1^{}\right)^2\left(z_3^{}-z_1^{}\right)^2\left(z_3^{}-z_2^{}\right)^2}{z_1^2z_2^2z_3^2} \right]\;,\nonumber\\
\end{eqnarray}
while the character functions of the fundamental and anti-fundamental representations are
\begin{eqnarray}
\chi_{{\bf 3}}^{}=z_1^{}+z_2^{}+z_3^{}\;,\quad
\chi_{{\bf 3^*_{}}}^{}=z_1^{-1}+z_2^{-1}+z_3^{-1}\;.
\end{eqnarray}

\end{itemize}

We close this appendix by calculating the HS of the toy model of one-generation leptons using the MW formula. From the transformation rules of building blocks under ${\rm U}(1)$ (see Eq.~(\ref{eq:Toy model transformation rule})), one obtains
\begin{eqnarray}
{\cal H}\left(q_1^{},q_2^{},q_3^{},q_4^{}\right)
&=&\left[\frac{1}{2\pi {\rm i}} \oint_{\left|z_1^{}\right|=1}\frac{dz_1^{}}{z_1^{}}\frac{1}{\left(1-q_1^{}z_1^{Q_1}\right)\left(1-q_2^{}z_1^{-Q_1}\right)}\right]\nonumber\\
&& \times \left[\frac{1}{2\pi {\rm i}} \oint_{\left|z_2^{}\right|=1}\frac{dz_2^{}}{z_2^{}}\frac{1}{\left(1-q_3^{}z_2^{Q_2}\right)\left(1-q_4^{}z_2^{-Q_2}\right)}\right]\;,
\end{eqnarray}
where $Q_1^{}=\alpha-\beta$ and $Q_2^{}=2\alpha$ correspond to the charges of $M_l^{}$ and $M_\nu^{}$ under ${\rm U}(1)$. Rescaling the integral variables by $z_1^{\prime}= z_1^{Q_1}$, $z_2^{\prime}= z_2^{Q_2}$ and using the residue theorem one obtains
\begin{eqnarray}
{\cal H}\left(q_1^{},q_2^{},q_3^{},q_4^{}\right)=\frac{1}{Q_1 Q_2}\frac{1}{\left(1-q_1^{}q_2^{}\right)\left(1-q_3^{}q_4^{}\right)}\;,
\end{eqnarray}
which encodes with Eq.~(\ref{eq:HS multi-graded toy model}) after neglecting the unphysical overall factors $\left(Q_1^{}Q_2^{}\right)_{}^{-1}$. In the more complicated situation, such as the cases of two- and three-generation leptons introduced in Secs.~\ref{sec:2g} and \ref{sec:3g}, we will see the great advantages of MW formula because of its generality.

\section{Decomposition Rules and Syzygies}
\label{appendix:syzygy}
In this appendix, we introduce a general method to decompose an arbitrary flavor invariant into the polynomials of the basic invariants, which, though brute-forced, turns out to be effective and efficient for our problem.

This method is based on the observation that each flavor invariant is labeled by the degree of $(q_l^{},q_\nu^{})$, which correspond to the mass power of $M_l$ and $M_\nu$, respectively. Therefore, in order to keep balance of the mass dimension, any flavor invariant at certain degree can be written in the most general form as the linear combination of all the possible monomials at the same degree composed of the power of basic invariants. As a result, to decompose an arbitrary flavor invariant into the polynomials of the basic invariants, we only need the following two steps.
\begin{itemize}
\item The first step: Find out all possible monomials composed of the basic invariants at the same degree as the decomposed invariant. This can be ascribed to solving two multi-variable indefinite linear equations with non-negative integer solutions. Furthermore, the CP parities of the corresponding monomials should match that of the decomposed invariant, which is a stringent constraint and can significantly decrease the number of possible monomials.
\item The second step: Write the decomposed invariant as the linear combination of these possible monomials with undetermined coefficients. These coefficients can be determined efficiently by arbitrarily substituting several lists of values of physical observables then solving a system of linear equations.
\end{itemize}

We take a concrete example as the illustration of our method. All the basic invariants in the case of two-generation leptons together with their corresponding degrees and CP parities have been summarized in Table~\ref{table:2 generations}. Suppose we want to decompose the invariant $J_7^{+}\equiv{\rm Tr}\left(\left\{H_{l}^{}, H_{\nu}^{}\right\} G_{l\nu}^{}\right)$ into the polynomials of the basic invariants. First, the possible monomials have the most general form of
\begin{eqnarray}
M=J_1^{a_1}J_2^{a_2}J_3^{a_3}J_4^{a_4}J_5^{a_5}J_6^{a_6}(J_7^{-})^{a_7}\;,
\end{eqnarray}
where the power indices $a_i^{}$ (for $i=1,2,...,7$) are non-negative integers.\footnote{The requirement of the polynomial form of decomposition excludes the possibility of non-integer power indices.} In order to match with the degree of $J_7^{+}$, the power indices must satisfy
\begin{equation}
\label{eq:indefinite power eq}
\left\{
\begin{aligned}
2a_1^{}+4a_3^{}+2a_4^{}+4a_6^{}+4a_7^{}=4\\
2a_2^{}+2a_4^{}+4a_5^{}+2a_6^{}+4a_7^{}=4
\end{aligned}
\right.
\;.
\end{equation}
The non-negative integer solution set of Eq.~(\ref{eq:indefinite power eq}) gives all the monomials whose degrees match with that of $J_7^{+}$
\begin{eqnarray}
\left\{M_1^{},M_2^{},M_3^{},M_4^{},M_5^{},M_6^{},M_7^{},M_8^{}\right\}
=\left\{J_4^2,J_3^{}J_5^{},J_2^{}J_6^{},J_2^2J_3^{},J_1^{}J_2^{}J_4^{},J_1^2J_5^{},J_1^2J_2^2,J_7^{-} \right\}\;.
\end{eqnarray}
However, $J_7^+$ is CP-even so the CP-odd monomial $M_8^{}=J_7^{-}$ is forbidden. Thus we are only left with 7 possible monomials and the most general form of $J_7^{+}$ can be written as
\begin{eqnarray}
\label{eq:indefinite coe eq}
J_7^{+}=\sum_{i=1}^7 c_i^{}M_i^{}\;,
\end{eqnarray}
where $c_i^{}$ (for $i=1,2,...,7$) are undermined coefficients. Note that both $J_7^+$ and $M_i^{}$ are functions of six physical observables (i.e., two charged-lepton masses, two neutrino masses, one flavor mixing angle and one CP-violating phase), so a direct but efficient way is to substitute into Eq.~(\ref{eq:indefinite coe eq}) arbitrary 7 lists of values of these six physical observables. Then Eq.~(\ref{eq:indefinite coe eq}) will reduce to a system of linear equations about $c_i^{}$, which can easily be solved
\begin{eqnarray}
\left\{c_1^{},c_2^{},c_3^{},c_4^{},c_5^{},c_6^{},c_7^{}\right\}=
\left\{1,0,1,0,0,\frac{1}{2},-\frac{1}{2}\right\}\;,
\end{eqnarray}
so the decomposition is given by
\begin{eqnarray}
J_7^{+}=J_4^2+J_2^{}J_6^{}+\frac{1}{2}J_1^2J_5^{}-\frac{1}{2}J_1^2J_2^2\;,
\end{eqnarray}
which is exactly Eq.~(\ref{eq:J7P decomposition}).

Several remarks about this method are in order before going on:
\begin{itemize}
\item This method provides a general way to decompose an arbitrary invariant into the polynomials of the basic invariants and can be carried out thoroughly with computers. Although brute-forced, it turns out to be effective and efficient for our problem, including the simplification of the right-hand side of Eqs.~(\ref{eq:I1})-(\ref{eq:I34}). For some invariants, manual decomposition acquires some tricks in matrix analysis and it is easy for one to make mistakes because of the tedious expressions after using CH theorem repeatedly. Therefore, an automatically-done and independent method which supplies a cross-check to the manual decomposition will be helpful.
\item This method provides a judgement about the independence and completeness of the basic invariants in the generating set, which is a cross-check to the direct observation from HS and PL. Given a certain invariant to be decomposed, write it into the linear combination of all possible monomials (say, totally $n_0$ possible monomials) with the same degrees and CP parities as this invariant, then substitute arbitrary $n_0$ lists of values of physical observables and solve the system of linear equations to obtain the $n_0$ undermined coefficients. If the substitution of another $n_0$ lists of values of physical observables gives different values of the coefficients, then it is sufficient to assert that the generating set is incomplete and we need more invariants to be generators. This is why we draw the conclusion that the first 33 generators in Table~\ref{table:3 generations}, as what the leading positive terms of PL in Eq.~(\ref{eq:PL 3 generations}) show, are not complete and $I_{34}^{}$ has to be added into the generating list: We find it impossible to decompose $I_{34}^{}$ into the polynomials of $I_1^{}$ to $I_{33}^{}$ using this method. Similarly, the independence of the basic invariants in the generating set can also be judged in this way: If any basic invariant in the generating set can not be decomposed into the polynomials of other basic invariants, then there is no redundancy in the generating set.

\item For invariants with high degrees, it takes much time to obtain the whole solution set of the multi-variable indefinite linear equations with non-negative integer solutions like Eq.~(\ref{eq:indefinite power eq}) because the number of the solutions grows quickly with the degrees. So one should firstly use CH theorem to decrease the degrees of the invariants as soon as possible before using this method.

\end{itemize}

As an additional bonus, we find all the syzygies at a certain degree can be determined by the same methodology. Notice that the syzygies at some degree are indeed the linear relationships among all possible monomials at this degree. Suppose there are $n_0^{}$ monomials at some degree, then any invariant $\tilde{I}$ at the same degree can be written as
\begin{eqnarray}
\tilde{I}\left(\vec{x}\right)=\sum_{i=1}^{n_0}c_i^{}M_i^{}\left(\vec{x}\right)\;,
\end{eqnarray}
where $\vec{x}$ denotes the physical observables that the  invariant and monomials depend on. Substituting arbitrary $n_0^{}$ lists of values of $\vec{x}$ (denoted by $\vec{x}_j^{}$, for $j=1,2,...,n_0^{}$) we get a system of linear equations
\begin{eqnarray}
\label{eq:indefinite coe eq 2}
\sum_{i=1}^{n_0}M_i^{}\left(\vec{x}_j^{}\right)c_i^{}=\tilde{I}\left(\vec{x}_j^{}\right)\;,\quad
j=1,2,...,n_0^{}\;.
\end{eqnarray}
Then the number of the syzygies is determined by the rank of the coefficient matrix $M_{ji}\equiv M_i^{}\left(\vec{x}_j^{}\right)$. If the rank of $M_{ji}$ is $n_0^{}$ (i.e., full rank) then there is no syzygy at this degree, while if the rank of $M_{ji}$ is $n_0^{}-s$ then there are $s$ syzygies at this degree. In the latter case, the solution of Eq.~(\ref{eq:indefinite coe eq 2}) is given by
\begin{eqnarray}
c_i^{}=c_{i0}^{}+\sum_{j=n_0-s+1}^{n_0}k_i^j c_j^{}\;,\quad
i=1,2,...,n_0^{}-s\;,
\end{eqnarray}
with $s$ coefficients $c_j^{}$ (for $j=n_0^{}-s+1,n_0^{}-s+2,...,n_0^{}$) undetermined. Note $c_{i0}^{}$ are numbers determined by the decomposed invariant $\tilde{I}$ while $k_i^j$ are independent of the decomposed invariant and encode the relations among the monomials. So we can rewrite $\tilde{I}$ as
\begin{eqnarray}
\label{eq:decompose Itilde}
\tilde{I}&=&\sum_{i=1}^{n_0}c_i^{}M_i^{}=\sum_{i=1}^{n_0-s}\left[c_{i0}^{}+\sum_{j=n_0-s+1}^{n_0}k_i^jc_j^{}\right]M_i^{}+\sum_{j=n_0-s+1}^{n_0}c_j^{}M_j^{}\nonumber\\
&&=\sum_{i=1}^{n_0-s}c_{i0}^{}M_i^{}+\sum_{j=n_0-s+1}^{n_0}c_j^{}\left[ M_j^{}+\sum_{i=1}^{n_0-s}k_i^jM_i\right]\;.
\end{eqnarray}
Since the undetermined coefficient $c_j^{}$ can take any values, in order to guarantee Eq.~(\ref{eq:decompose Itilde}) to hold, we must have
\begin{eqnarray}
\label{eq:syzygy equation}
M_j^{}+\sum_{i=1}^{n_0-s}k_i^jM_i=0\;,\quad
j=n_0^{}-s+1,n_0^{}-s+2,...,n_0^{}\;,
\end{eqnarray}
which are just the $s$ syzygies at this degree. We emphasize that although we choose a specific invariant $\tilde{I}$ during the derivation of syzygies, they are indeed independent of the invariant to be decomposed and any invariant at the same degree works.

We take the case of two-generation leptons as an illustration.
Eq.~(\ref{eq:PL 2 generations}) tells us that there is one syzygy at the degree $(8,8)$ and we show below how to derive it explicitly.  The non-negative integer solutions of the following equation
\begin{equation}
\left\{
\begin{aligned}
2a_1^{}+4a_3^{}+2a_4^{}+4a_6^{}+4a_7^{}=8\\
2a_2^{}+2a_4^{}+4a_5^{}+2a_6^{}+4a_7^{}=8
\end{aligned}
\right.
\;,
\end{equation}
give all possible monomials at this degree
\begin{eqnarray}
\left\{M_1^{},M_2^{},...,M_{36}^{}\right\}
&=&\left\{\left(J_7^{-}\right)^2, J_5 J_6^2, J_4^2 J_7^{-}, J_4^4, J_3 J_5 J_7^{-}, J_3 J_4^2 J_5, J_3^2 J_5^2, J_2 J_6 J_7^{-}, J_2 J_4^2 J_6, J_2 J_3 J_5 J_6, \right.\nonumber\\
&&\left. J_2^2 J_6^2, J_2^2 J_3 J_7^{-}, J_2^2 J_3 J_4^2, J_2^2 J_3^2 J_5, J_2^3 J_3 J_6, J_2^4 J_3^2, J_1 J_4 J_5 J_6, J_1 J_2 J_4 J_7^{-}, J_1 J_2 J_4^3,\right.\nonumber\\
&& \left. J_1 J_2 J_3 J_4 J_5, J_1 J_2^2 J_4 J_6, J_1 J_2^3 J_3 J_4, J_1^2 J_5 J_7^{-}, J_1^2 J_4^2 J_5, J_1^2 J_3 J_5^2, J_1^2 J_2 J_5 J_6, \right.\nonumber\\
&&\left. J_1^2 J_2^2 J_7^{-}, J_1^2 J_2^2 J_4^2, J_1^2 J_2^2 J_3 J_5, J_1^2 J_2^3 J_6, J_1^2 J_2^4 J_3, J_1^3 J_2 J_4 J_5, J_1^3 J_2^3 J_4, J_1^4 J_5^2,\right.\nonumber\\
&&\left.J_1^4 J_2^2 J_5, J_1^4 J_2^4\right\}\;.
\end{eqnarray}
Then we choose $\left(J_7^{-}\right)^2_{}$ to be the decomposed invariant, whose degree is also $(8,8)$,
\begin{eqnarray}
\label{eq:linear equations}
\left(J_7^{-}\right)^2_{}=\sum_{i=1}^{36}c_i^{}M_i^{}\;.
\end{eqnarray}
These coefficients $c_i^{}$ (for $i=1,2,...,36$) can be solved by substituting arbitrary 36 lists of values of physical observables and solving a system of linear equations,
\begin{eqnarray}
\left\{c_1^{},c_2^{},...,c_{35}^{},c_{36}^{}\right\}=
c_{36}^{}&&\left\{1/c_{36}+ 4, -8, 0, -4, 0, 8, -4, 0, 8, 0, 4, 0, -8 , 6 , 0, -2 , 16 , 0, 0,\right.\nonumber\\
&& \left. -8, -16 , 8 , 0, -12 , 4 , -4 , 0, 12 , -4, 4 , 0, 8 , -8 , -1, 0 , 1\right\}\;.
\end{eqnarray}
From the above values of $c_i^{}$, we can read off the syzygy according to Eq.~(\ref{eq:syzygy equation})
\begin{eqnarray}
&&4\left(J_7^{-}\right)^2-8J_5^{}J_6^2-4J_4^4+8J_3^{}J_4^2J_5^{}-4J_3^2J_5^2+8J_2^{}J_4^2J_6^{}+4J_2^2J_6^2-8J_2^2J_3^{}J_4^2+6J_2^2J_3^2J_5^{} -2J_2^4J_3^2\nonumber\\
&&+16J_1^{}J_4^{}J_5^{}J_6^{}-8J_1^{}J_2^{}J_3^{}J_4^{}J_5^{}-16J_1^{}J_2^2J_4^{}J_6^{}+8J_1^{}J_2^3J_3^{}J_4^{}-12J_1^2J_4^2J_5^{}+4J_1^2J_3^{}J_5^2-4J_1^2J_2^{}J_5^{}J_6^{}\nonumber\\
&&+12 J_1^2J_2^2J_4^2-4J_1^2J_2^2J_3^{}J_5^{}+4J_1^2J_2^3J_6^{}+8J_1^3J_2^{}J_4^{}J_5^{}-8J_1^3J_2^3J_4^{}-J_1^4J_5^2+J_1^4J_2^4=0\;,
\end{eqnarray}
which is exactly Eq.~(\ref{eq:syzygy 2g}).

\end{appendix}


\begin{thebibliography}{99}
\bibitem{Xing2020}
Z.~z.~Xing,
``Flavor structures of charged fermions and massive neutrinos,''
Phys. Rept. \textbf{854}, 1-147 (2020)
[arXiv:1909.09610 [hep-ph]].

\bibitem{PDG2020}
P.~A.~Zyla \textit{et al.} [Particle Data Group],
``Review of Particle Physics,''
PTEP \textbf{2020}, no.8, 083C01 (2020).


\bibitem{Majorana1937}
E.~Majorana,
``Teoria simmetrica dell\textquoteright{}elettrone e del positrone,''
Nuovo Cim. \textbf{14}, 171-184 (1937).

\bibitem{Racah1937}
G.~Racah,
``On the symmetry of particle and antiparticle,''
Nuovo Cim. \textbf{14}, 322-328 (1937).


\bibitem{Pontecorvo1957}
  B.~Pontecorvo,
  ``Mesonium and anti-mesonium,''
  Sov.\ Phys.\ JETP {\bf 6}, 429 (1957)
  [Zh.\ Eksp.\ Teor.\ Fiz.\  {\bf 33}, 549 (1957)].

\bibitem{MNS1962}
  Z.~Maki, M.~Nakagawa and S.~Sakata,
  ``Remarks on the unified model of elementary particles,''
  Prog.\ Theor.\ Phys.\  {\bf 28}, 870 (1962).

\bibitem{Jarlskog:1985ht}
C.~Jarlskog,
``Commutator of the Quark Mass Matrices in the Standard Electroweak Model and a Measure of Maximal CP Violation,''
Phys. Rev. Lett. \textbf{55}, 1039 (1985)

\bibitem{Jarlskog:1985cw}
C.~Jarlskog,
``A Basis Independent Formulation of the Connection Between Quark Mass Matrices, CP Violation and Experiment,''
Z. Phys. C \textbf{29}, 491-497 (1985)

\bibitem{Branco1986quark}
J.~Bernabeu, G.~C.~Branco and M.~Gronau,
``CP Restrictions on Quark Mass Matrices,''
Phys. Lett. B \textbf{169}, 243-247 (1986).

\bibitem{Branco1986lepton}
G.~C.~Branco, L.~Lavoura and M.~N.~Rebelo,
``Majorana Neutrinos and CP Violation in the Leptonic Sector,''
Phys. Lett. B \textbf{180}, 264-268 (1986).

\bibitem{Yu2019PLB}
B.~Yu and S.~Zhou,
``The number of sufficient and necessary conditions for CP conservation with Majorana neutrinos: three or four?,''
Phys. Lett. B \textbf{800}, 135085 (2020)
[arXiv:1908.09306 [hep-ph]].


\bibitem{Yu2020ICHEP}
B.~Yu and S.~Zhou,
``Weak-basis invariants and CP conservation in the leptonic sector with Majorana neutrinos,''
[arXiv:2010.08758 [hep-ph]].


\bibitem{Yu2020PRD}
B.~Yu and S.~Zhou,
``Sufficient and Necessary Conditions for CP Conservation in the Case of Degenerate Majorana Neutrino Masses,''
Phys. Rev. D \textbf{103}, no.3, 035017 (2021)
[arXiv:2009.12347 [hep-ph]].


\bibitem{Pilaftsis1997}
A.~Pilaftsis,
``CP violation and baryogenesis due to heavy Majorana neutrinos,''
Phys. Rev. D \textbf{56}, 5431-5451 (1997)
[arXiv:hep-ph/9707235].


\bibitem{Branco2001}
  G.~C.~Branco, T.~Morozumi, B.~M.~Nobre and M.~N.~Rebelo,
  ``A Bridge between CP violation at low-energies and leptogenesis,''
  Nucl.\ Phys.\ B {\bf 617}, 475 (2001)
  [hep-ph/0107164].


\bibitem{CIP2006}
  V.~Cirigliano, G.~Isidori and V.~Porretti,
  ``CP violation and Leptogenesis in models with Minimal Lepton Flavour Violation,''
  Nucl.\ Phys.\ B {\bf 763}, 228 (2007)
  [hep-ph/0607068].

\bibitem{FMS2015}
T.~Feldmann, T.~Mannel and S.~Schwertfeger,
``Renormalization Group Evolution of Flavour Invariants,''
JHEP \textbf{10}, 007 (2015)
[arXiv:1507.00328 [hep-ph]].

\bibitem{Talbert:2021iqn}
J.~Talbert and M.~Trott,
``Dirac Masses and Mixings in the (geo)SM(EFT) and Beyond,''
[arXiv:2107.03951 [hep-ph]].

\bibitem{Sturmfels2008}
B.~Sturmfels
``Algorithms in Invariant Theory,''
  Springer-Verlag, Wien (2008).

\bibitem{DK2015}
H.~Derksen, G.~Kemper, V.~L.~Popov and N.~A’~Campo,
``Computational invariant theory,''
 Springer-Verlag, Berlin Heidelberg (2015).

\bibitem{Processi76}
C.~Processi,
``The Invariant Theory of $n\times n$ Matrices,"
Adv. in Math. \textbf{19}, 306 (1976).

\bibitem{Formanek84}
E.~Formanek,
``Invariants and the Ring of Generic Matrices,"
J. Alg. \textbf{89}, 178 (1984).

\bibitem{Chankowski:1993tx}
P.~H.~Chankowski and Z.~Pluciennik,
``Renormalization group equations for seesaw neutrino masses,''
Phys. Lett. B \textbf{316} (1993), 312-317
[arXiv:hep-ph/9306333 [hep-ph]].

\bibitem{Babu:1993qv}
K.~S.~Babu, C.~N.~Leung and J.~T.~Pantaleone,
``Renormalization of the neutrino mass operator,''
Phys. Lett. B \textbf{319} (1993), 191-198
[arXiv:hep-ph/9309223 [hep-ph]].

\bibitem{Antusch:2001ck}
S.~Antusch, M.~Drees, J.~Kersten, M.~Lindner and M.~Ratz,
``Neutrino mass operator renormalization revisited,''
Phys. Lett. B \textbf{519} (2001), 238-242
[arXiv:hep-ph/0108005 [hep-ph]].

\bibitem{XZ2011}
  Z.~z.~Xing and S.~Zhou,
  ``Neutrinos in particle physics, astronomy and cosmology,''
  Springer-Verlag, Berlin Heidelberg (2011).

\bibitem{Ohlsson:2013xva}
T.~Ohlsson and S.~Zhou,
``Renormalization group running of neutrino parameters,''
Nature Commun. \textbf{5}, 5153 (2014)
[arXiv:1311.3846 [hep-ph]].

\bibitem{JM2009}
  E.~E.~Jenkins and A.~V.~Manohar,
  ``Algebraic Structure of Lepton and Quark Flavor Invariants and CP Violation,''
  JHEP {\bf 0910}, 094 (2009)
  [arXiv:0907.4763 [hep-ph]].

\bibitem{Wallach2009}
	A.~Garsia, N.~Wallach, G.~Xin and M.~Zabrocki,
	``Hilbert series of invariants, constant terms and Kostka–Foulkes polynomials,''
		Discrete Mathematics {\bf 309}, 5206-5230 (2009).


\bibitem{NuFit2020}
I.~Esteban, M.~C.~Gonzalez-Garcia, M.~Maltoni, T.~Schwetz and A.~Zhou,
``The fate of hints: updated global analysis of three-flavor neutrino oscillations,''
JHEP \textbf{09}, 178 (2020)
[arXiv:2007.14792 [hep-ph]].

\bibitem{HZ2020}
G.~y.~Huang and S.~Zhou,
``Precise Values of Running Quark and Lepton Masses in the Standard Model,''
Phys. Rev. D \textbf{103}, no.1, 016010 (2021)
[arXiv:2009.04851 [hep-ph]].

\bibitem{Elias-Miro:2011sqh}
J.~Elias-Miro, J.~R.~Espinosa, G.~F.~Giudice, G.~Isidori, A.~Riotto and A.~Strumia,
``Higgs mass implications on the stability of the electroweak vacuum,''
Phys. Lett. B \textbf{709}, 222-228 (2012)
[arXiv:1112.3022 [hep-ph]].

\bibitem{Xing:2011aa}
Z.~z.~Xing, H.~Zhang and S.~Zhou,
``Impacts of the Higgs mass on vacuum stability, running fermion masses and two-body Higgs decays,''
Phys. Rev. D \textbf{86}, 013013 (2012)
[arXiv:1112.3112 [hep-ph]].

\bibitem{Degrassi:2012ry}
G.~Degrassi, S.~Di Vita, J.~Elias-Miro, J.~R.~Espinosa, G.~F.~Giudice, G.~Isidori and A.~Strumia,
``Higgs mass and vacuum stability in the Standard Model at NNLO,''
JHEP \textbf{08}, 098 (2012)
[arXiv:1205.6497 [hep-ph]].

\bibitem{Trautner2018}
A.~Trautner,
``Systematic construction of basis invariants in the 2HDM,''
JHEP \textbf{05}, 208 (2019)
[arXiv:1812.02614 [hep-ph]].


\bibitem{Trautner2020}
A.~Trautner,
``On the systematic construction of basis invariants,''
J. Phys. Conf. Ser. \textbf{1586}, no.1, 012005 (2020)
[arXiv:2002.12244 [hep-ph]].



\bibitem{Pouliot1998}
P.~Pouliot,
``Molien function for duality,''
JHEP \textbf{01}, 021 (1999)
[arXiv:hep-th/9812015 [hep-th]].

\bibitem{BFHH2007}
S.~Benvenuti, B.~Feng, A.~Hanany and Y.~H.~He,
``Counting BPS Operators in Gauge Theories: Quivers, Syzygies and Plethystics,''
JHEP \textbf{11}, 050 (2007)
[arXiv:hep-th/0608050 [hep-th]].

\bibitem{Romelsberger2005}
C.~Romelsberger,
``Counting chiral primaries in N = 1, d=4 superconformal field theories,''
Nucl. Phys. B \textbf{747}, 329-353 (2006)
[arXiv:hep-th/0510060 [hep-th]].


\bibitem{FHH2007}
B.~Feng, A.~Hanany and Y.~H.~He,
``Counting gauge invariants: The Plethystic program,''
JHEP \textbf{03}, 090 (2007)
[arXiv:hep-th/0701063 [hep-th]].



\bibitem{FHZ2007}
D.~Forcella, A.~Hanany and A.~Zaffaroni,
``Baryonic Generating Functions,''
JHEP \textbf{12}, 022 (2007)
[arXiv:hep-th/0701236 [hep-th]].



\bibitem{Dolan2008}
F.~A.~Dolan,
``Counting BPS operators in N=4 SYM,''
Nucl. Phys. B \textbf{790}, 432-464 (2008)
[arXiv:0704.1038 [hep-th]].



\bibitem{BFHVZ2007}
A.~Butti, D.~Forcella, A.~Hanany, D.~Vegh and A.~Zaffaroni,
``Counting Chiral Operators in Quiver Gauge Theories,''
JHEP \textbf{11}, 092 (2007)
[arXiv:0705.2771 [hep-th]].



\bibitem{Hanany2007}
A.~Hanany,
``Counting BPS operators in the chiral ring: The plethystic story,''
AIP Conf. Proc. \textbf{939}, no.1, 165-175 (2007)




\bibitem{GHHJM2008}
J.~Gray, A.~Hanany, Y.~H.~He, V.~Jejjala and N.~Mekareeya,
``SQCD: A Geometric Apercu,''
JHEP \textbf{05}, 099 (2008)
[arXiv:0803.4257 [hep-th]].

\bibitem{HM2008}
A.~Hanany and N.~Mekareeya,
``Counting Gauge Invariant Operators in SQCD with Classical Gauge Groups,''
JHEP \textbf{10}, 012 (2008)
[arXiv:0805.3728 [hep-th]].

\bibitem{HMT2010}
A.~Hanany, N.~Mekareeya and G.~Torri,
``The Hilbert Series of Adjoint SQCD,''
Nucl. Phys. B \textbf{825}, 52-97 (2010)
[arXiv:0812.2315 [hep-th]].


\bibitem{Forcella2009}
D.~Forcella,
``Master Space and Hilbert Series for N=1 Field Theories,''
[arXiv:0902.2109 [hep-th]].


\bibitem{CM2011}
Y.~Chen and N.~Mekareeya,
``The Hilbert series of U/SU SQCD and Toeplitz Determinants,''
Nucl. Phys. B \textbf{850}, 553-593 (2011)
[arXiv:1104.2045 [hep-th]].

\bibitem{HK2014}
A.~Hanany and R.~Kalveks,
``Highest Weight Generating Functions for Hilbert Series,''
JHEP \textbf{10}, 152 (2014)
[arXiv:1408.4690 [hep-th]].

\bibitem{BP2017}
A.~Bourget and A.~Pini,
``Non-Connected Gauge Groups and the Plethystic Program,''
JHEP \textbf{10}, 033 (2017)
[arXiv:1706.03781 [hep-th]].

\bibitem{XHM2019}
Y.~Xiao, Y.~H.~He and C.~Matti,
``Standard Model Plethystics,''
Phys. Rev. D \textbf{100}, no.7, 076001 (2019)
[arXiv:1902.10550 [hep-th]].


\bibitem{LM2015}
L.~Lehman and A.~Martin,
``Hilbert Series for Constructing Lagrangians: expanding the phenomenologist's toolbox,''
Phys. Rev. D \textbf{91}, 105014 (2015)
[arXiv:1503.07537 [hep-ph]].


\bibitem{HLMM2015CMP}
B.~Henning, X.~Lu, T.~Melia and H.~Murayama,
``Hilbert series and operator bases with derivatives in effective field theories,''
Commun. Math. Phys. \textbf{347}, no.2, 363-388 (2016)
[arXiv:1507.07240 [hep-th]].

\bibitem{LM2016}
L.~Lehman and A.~Martin,
``Low-derivative operators of the Standard Model effective field theory via Hilbert series methods,''
JHEP \textbf{02}, 081 (2016)
[arXiv:1510.00372 [hep-ph]].


\bibitem{HLMM2015JHEP}
B.~Henning, X.~Lu, T.~Melia and H.~Murayama,
``2, 84, 30, 993, 560, 15456, 11962, 261485, ...: Higher dimension operators in the SM EFT,''
JHEP \textbf{08}, 016 (2017)
[erratum: JHEP \textbf{09}, 019 (2019)]
[arXiv:1512.03433 [hep-ph]].


\bibitem{KP2017}
A.~Kobach and S.~Pal,
``Hilbert Series and Operator Basis for NRQED and NRQCD/HQET,''
Phys. Lett. B \textbf{772}, 225-231 (2017)
[arXiv:1704.00008 [hep-ph]].

\bibitem{HLMM2017}
B.~Henning, X.~Lu, T.~Melia and H.~Murayama,
``Operator bases, $S$-matrices, and their partition functions,''
JHEP \textbf{10}, 199 (2017)
[arXiv:1706.08520 [hep-th]].

\bibitem{KP2018}
A.~Kobach and S.~Pal,
``Reparameterization Invariant Operator Basis for NRQED and HQET,''
JHEP \textbf{11}, 012 (2019)
[arXiv:1810.02356 [hep-ph]].

\bibitem{ADCP2019}
Anisha, S.~Das Bakshi, J.~Chakrabortty and S.~Prakash,
``Hilbert Series and Plethystics: Paving the path towards 2HDM- and MLRSM-EFT,''
JHEP \textbf{09}, 035 (2019)
[arXiv:1905.11047 [hep-ph]].

\bibitem{MRW2020}
C.~B.~Marinissen, R.~Rahn and W.~J.~Waalewijn,
``..., 83106786, 114382724, 1509048322, 2343463290, 27410087742, ... efficient Hilbert series for effective theories,''
Phys. Lett. B \textbf{808}, 135632 (2020)
[arXiv:2004.09521 [hep-ph]].

\bibitem{LRSXYZ2020}
H.~L.~Li, Z.~Ren, J.~Shu, M.~L.~Xiao, J.~H.~Yu and Y.~H.~Zheng,
``Complete Set of Dimension-8 Operators in the Standard Model Effective Field Theory,''
[arXiv:2005.00008 [hep-ph]].

\bibitem{LM2020}
Y.~Liao and X.~D.~Ma,
``An explicit construction of the dimension-9 operator basis in the standard model effective field theory,''
JHEP \textbf{11}, 152 (2020)
[arXiv:2007.08125 [hep-ph]].

\bibitem{GHLMM2020}
L.~Graf, B.~Henning, X.~Lu, T.~Melia and H.~Murayama,
``2, 12, 117, 1959, 45171, 1170086, \textellipsis{}: a Hilbert series for the QCD chiral Lagrangian,''
JHEP \textbf{01}, 142 (2021)
[arXiv:2009.01239 [hep-ph]].

\bibitem{KLMT2020}
G.~D.~Kribs, X.~Lu, A.~Martin and T.~Tong,
``Custodial Symmetry (Violation) in SMEFT,''
[arXiv:2009.10725 [hep-ph]].

\bibitem{MP2020}
T.~Melia and S.~Pal,
``EFT Asymptotics: the Growth of Operator Degeneracy,''
[arXiv:2010.08560 [hep-th]].


\bibitem{Murphy2020}
C.~W.~Murphy,
``Low-Energy Effective Field Theory below the Electroweak Scale: Dimension-8 Operators,''
[arXiv:2012.13291 [hep-ph]].

\bibitem{CHMN2021}
W.~Cao, F.~Herzog, T.~Melia and J.~R.~Nepveu,
``Renormalization and non-renormalization of scalar EFTs at higher orders,''
[arXiv:2105.12742 [hep-ph]].


\bibitem{HH2020}
K.~Haddad and A.~Helset,
``Tidal effects in quantum field theory,''
JHEP \textbf{12}, 024 (2020)
[arXiv:2008.04920 [hep-th]].

\bibitem{AHH2020}
R.~Aoude, K.~Haddad and A.~Helset,
``Tidal effects for spinning particles,''
[arXiv:2012.05256 [hep-th]].

\bibitem{Banerjee2020}
U.~Banerjee, J.~Chakrabortty, S.~Prakash and S.~U.~Rahaman,
``Characters and group invariant polynomials of (super)fields: road to \textquotedblleft{}Lagrangian\textquotedblright{},''
Eur. Phys. J. C \textbf{80}, no.10, 938 (2020)
[arXiv:2004.12830 [hep-ph]].


\bibitem{Bednyakov2018}
A.~V.~Bednyakov,
``On three-loop RGE for the Higgs sector of 2HDM,''
JHEP \textbf{11}, 154 (2018)
[arXiv:1809.04527 [hep-ph]].


\bibitem{BBST2020}
M.~P.~Bento, R.~Boto, J.~P.~Silva and A.~Trautner,
``A fully basis invariant Symmetry Map of the 2HDM,''
JHEP \textbf{21}, 229 (2020)
[arXiv:2009.01264 [hep-ph]].


\bibitem{Bento2021}
M.~P.~Bento,
``The invariant space of multi-Higgs doublet models,''
[arXiv:2102.13120 [hep-ph]].



\bibitem{HJMT2011}
A.~Hanany, E.~E.~Jenkins, A.~V.~Manohar and G.~Torri,
``Hilbert Series for Flavor Invariants of the Standard Model,''
JHEP \textbf{03}, 096 (2011)
[arXiv:1010.3161 [hep-ph]].


\bibitem{Molien1897}
T.~Molien,
``{\"U}ber die Invarianten der linearen Substitutionsgruppe,''
Sitzungber. K{\"o}nig. Preuss. Akad. Wiss. (J. Berl. Ber.). 52: 1152–1156.

\bibitem{Weyl1926}
H.~Weyl,
``Zur Darstellungstheorie und Invariantenabzählung der projektiven, der Komplex-und der Drehungsgruppe,''
Acta Mathematica 48.3-4 (1926): 255-278.






\end{thebibliography}
\end{document}